\documentclass[apj,twocolumn]{openjournal}
\usepackage{natbib} 
\usepackage[dvipsnames]{xcolor}
\usepackage{aas_macros} 
\usepackage{amssymb}
\usepackage{amsmath}
\usepackage[title]{appendix}
\usepackage{hyperref}	
\hypersetup{colorlinks=true,linkcolor=blue,citecolor=blue,filecolor=blue,urlcolor=blue}
\usepackage[caption=false]{subfig}
\usepackage{soul}                          
\usepackage{xspace}
\usepackage{booktabs}
\usepackage{savesym}
\savesymbol{tablenum}
\usepackage{siunitx}
\usepackage{pifont}
\restoresymbol{SIX}{tablenum}

\DeclareSIUnit\Msun{\ensuremath{M_\odot}}
\DeclareSIUnit\Zsun{\ensuremath{Z_\odot}}
\DeclareSIUnit\hred{\ensuremath{\textit{h}}}

\begin{document}
\title[]{MEGATRON: Reproducing the Diversity of High-Redshift Galaxy Spectra with Cosmological Radiation Hydrodynamics Simulations\vspace{-15mm}}
\author{Harley Katz$^{1,2*}$, Martin P. Rey$^{3}$, Corentin Cadiou$^4$, \\ Oscar Agertz$^5$, Jeremy Blaizot$^{6}$, Alex J. Cameron$^{7}$, Nicholas Choustikov$^{7}$, Julien Devriendt$^{7}$, Uliana Hauk$^{1}$, Gareth C. Jones$^{8,9}$, Taysun Kimm$^{10}$, Isaac Laseter$^{11}$, Sergio Martin-Alvarez$^{12}$, Kosei Matsumoto$^{13}$, Autumn Pearce$^{1}$, Francisco Rodríguez Montero$^{1,2}$, Joki Rosdahl$^{6}$, Mahsa Sanati$^{7}$, Aayush Saxena$^{7}$, Adrianne Slyz$^{7}$, Richard Stiskalek$^{7}$, Anatole Storck$^{7}$, Oscar Veenema$^{7}$, and Wonjae Yee$^{1}$}
\thanks{$^*$E-mail: \href{mailto:harleykatz@uchicago.edu}{harleykatz@uchicago.edu}}

\affiliation{$^{1}$Department of Astronomy \& Astrophysics, University of Chicago, 5640 S Ellis Avenue, Chicago, IL 60637, USA}
\affiliation{$^{2}$Kavli Institute for Cosmological Physics, University of Chicago, Chicago IL 60637, USA}
\affiliation{$^{3}$University of Bath, Department of Physics, Claverton Down, Bath, BA2 7AY, UK}
\affiliation{$^{4}$Institut d'Astrophysique de Paris, Sorbonne Université, CNRS, UMR 7095, 98 bis bd Arago, 75014 Paris, France}
\affiliation{$^{5}$Division of Astrophysics, Department of Physics, Lund University, Box 118, SE-221 00 Lund, Sweden}
\affiliation{$^{6}$Universite Claude Bernard Lyon 1, CRAL UMR5574, ENS de Lyon, CNRS, Villeurbanne, F-69622, France}
\affiliation{$^{7}$Sub-department of Astrophysics, University of Oxford, Keble Road, Oxford OX1 3RH, United Kingdom}
\affiliation{$^{8}$Kavli Institute for Cosmology, University of Cambridge, Madingley Road, Cambridge CB3 0HA, UK}
\affiliation{$^{9}$Cavendish Laboratory, University of Cambridge, 19 JJ Thomson Avenue, Cambridge CB3 0HE, UK}
\affiliation{$^{10}$Department of Astronomy, Yonsei University, 50 Yonsei-ro, Seodaemun-gu, Seoul 03722, Republic of Korea}
\affiliation{$^{11}$Department of Astronomy, University of Wisconsin-Madison, Madison, WI 53706, USA}
\affiliation{$^{12}$Kavli Institute for Particle Astrophysics \& Cosmology (KIPAC), Stanford University, Stanford, CA 94305, USA}
\affiliation{$^{13}$Sterrenkundig Observatorium Department of Physics and Astronomy Universiteit Gent, Krijgslaan 281 S9, B-9000 Gent, Belgium}

\begin{abstract}
We present the {\small MEGATRON} suite of cosmological radiation hydrodynamics simulations following the formation of Milky Way-mass galaxies from the earliest cosmic epochs when Population III stars form to Cosmic Noon. The suite represents the first set of cosmological simulations that couples a vast non-equilibrium thermochemistry network of primordial species, metals, and molecules to multifrequency, on-the-fly radiation transport, allowing us to directly predict the spectral properties of early galaxies. By initializing the simulations at zero metallicity, resolving haloes well below the atomic cooling threshold, reaching parsec-scale resolution, and modeling a Milky Way-mass environment, we aim to address four key science themes: 1) Star formation at cosmic dawn, 2) Galaxy formation and the interstellar medium in the epoch of reionization, 3) The circumgalactic medium towards cosmic noon, and 4) Reionization in a local volume environment and near-field cosmology. In this introductory work, we present an overview of the physical characteristics of high-redshift {\small MEGATRON} galaxies and their environment at $z>8$. We present a library of $>175,000$ simulated galaxy spectra and demonstrate how much of the diversity of galaxy spectra seen by JWST is naturally reproduced in the context of a $\Lambda$CDM cosmology. Caveats are discussed, such as the lack of AGN in our simulations and the limitations of our adopted stellar population and chemical yield models. This project represents a step towards making more direct comparisons between simulations and observations and is particularly applicable for optimizing methods to infer galaxy properties from existing high-redshift JWST spectra and imaging data.
\end{abstract}

\section{Introduction}
Prior to the 2020s, few constraints existed on the properties of the high-redshift ($z>6$) interstellar medium (ISM). The sample of spectroscopic detections was limited to a few bright sources detected with the Hubble Space Telescope (HST) that could be targeted by the Atacama Large Millimeter Array \citep[e.g.,][]{Pentericci2016,Bradac2017,Hashimoto2019,Harikane2020,Carniani2020} at mm wavelengths, or UV line detections from other large ground-based facilities \citep[e.g.,][]{Stark2017,Mainali2017}. 

The launch of the James Webb Space Telescope (JWST) catalyzed a paradigm shift in our understanding of galaxy formation in the first billion years. Designed with a large collecting area (allowing it to observe faint objects), spectroscopic capabilities (needed for precision measurements of ISM properties such as density, temperature, metallicity), and a low operating temperature and mirror coatings (needed to detect infrared photons that correspond to the rest-frame UV and optical in the epoch of reionization), JWST is optimal for studies of the ISM in the first galaxies \citep{Gardner2006,Gardner2023}.

The thousands of high-quality public JWST spectra\footnote{More specifically, there are nearly 9,000 grade 3 spectra in version 4.3 of the DAWN JWST Archive \citep{deGraaff2024,Heintz2024} at $z>3$ with a median S/N $>1$. Such a cut returns mostly PRISM spectra ($R\sim100$) with robust and high S/N emission lines and continuum detections in the UV and optical.} of early galaxies are beginning to reveal population-level trends of the high-redshift ISM \citep[e.g.,][]{Sanders2023,Cameron2023,Hu2024,Topping2025,Shapley2025}. In general, the high-redshift ISM exists at higher gas density, higher excitation, and lower metallicity compared to the low-redshift Universe. While these broad trends had been expected based on observations at intermediate redshifts \citep[e.g.,][]{Steidel2016,Strom2017,Reddy2018,Shapley2019} and from low-redshift ``analog'' galaxies \citep[e.g.,][]{Cardamone2009,Izotov2011,Izotov2021,Mingozzi2022}, many surprising aspects of early galaxies have also emerged. 

For example, multiple high-redshift galaxy surveys conducted with JWST have reported an overabundance of bright galaxies at $z\gtrsim9$, a finding that contrasts with earlier HST-based constraints and many theoretical models \citep[e.g.,][although c.f. \citealt{Willott2023}]{Finkelstein2023_b,Harikane2024,Leung2023,Chemerynska2023}. Several early galaxies also exhibit chemical abundance patterns that have never been observed in the gas-phase \citep[e.g.,][]{Cameron2023_nitrogen,Senchyna2024,Isobe2023_nitrogen,Topping2024}. Some of the high-redshift spectra appear to require exotic stellar populations or a top-heavy stellar initial mass function \citep[e.g.,][]{Cameron2023_NDG,Katz2024_BJ,Cullen2025}. Certain gas temperature measurements may be at odds with standard ISM heating mechanisms \citep[e.g.,][]{Katz2023_jwst,Laseter2024}. The star formation histories of massive quiescent galaxies may place their formation at very early epochs \citep[e.g.,][]{deGraaff2024,Glazebrook2024}. Finally, a new class of compact red galaxies with peculiar ``v''-shaped spectra (``Little Red Dots'') has been discovered whose exact nature remains unknown \citep[e.g.,][]{Matthee2024}.

These findings highlight the need for updated models of high-redshift galaxy formation and the ISM. This is a significant challenge, as the thermodynamic state of the ISM for which many of the observed emission lines are exponentially sensitive is a complex interplay between heating and cooling thermochemistry processes, stellar feedback in the form of supernova (SN) explosions, stellar winds, and ionizing radiation, and large-scale galactic processes such as turbulence and inflows/outflows \citep[e.g.,][]{Field1969,McKee1977,Tielens2005}. Modifications to, for example, chemical abundance patterns or modes of star formation compared to what occurs in the solar neighborhood can lead to a vastly different ISM pressure-density distribution \citep[e.g.,][]{Bialy2019,Katz2022_prism,Kim2023}.

Historically, inferring and interpreting the physical properties of the ISM from spectra relies on line ratio measurements and photoionization modeling (see \citealt{Kewley2019} for a review). While many inferred properties from line ratios are based on quantum physics with (reasonably) well measured atomic data, photoionization modeling is subject to significant uncertainties. For example, simply changing the structure of an illuminated cloud from a 1D sphere or slab geometry to a 3D turbulent medium with the same mean density can have non-negligible impacts on the observed line ratios \citep[e.g.,][]{Grey2017,Jin2022}. Likewise, the underlying stellar population \citep[e.g.][]{Byler2017,Xiao2018} and included physics \citep[cosmic rays, X-ray heating, etc, e.g.,][]{Katz2023_jwst} can affect the observed spectrum, as can galactic-scale effects (e.g., non-linear averaging over different geometries and emission regions; \citealt{Cameron2023}). Finally, there is no guarantee that the ISM is in an equilibrium state \citep[e.g.][]{Richings2022}, which is an implicit assumption of most photoionization models. However, photoionization modeling is computationally inexpensive, allowing for large parameter spaces to be explored. Such wide parameter spaces are necessary for interpreting the diversity of spectra seen in the real Universe.

3D numerical simulations of high-redshift galaxies represent a complementary approach to interpreting JWST spectra and images. There is no shortage of numerical simulations of early galaxy formation that predict line emission, full spectra, or imaging \citep[e.g.,][]{Ceverino2017,Barrow2017,Katz2019_ism,spdrv1,Vogelsberger2020,Arata2020,Lupi2020,Lovell2021,Trebitsch2021,Pallottini2022,Hirschmann2023,Yang2023,Kannan2022,Nakazato2023,Kannan2025,Garg2024,Schimek2024,Bhagwat2024,Nakazato2026}. However, all these simulations lack the detailed physics present in photoionization models that is needed to robustly predict spectra. More specifically, most of these simulations do not perform a detailed modeling of the ISM \citep[e.g.,][]{Vogelsberger2020,Lovell2021,Hirschmann2023} or include radiation transport. Reducing physical fidelity allows for simulations of much larger cosmological volumes that probe more massive galaxies, and crucially, the bright end of the UV luminosity function. This tradeoff is necessary to model massive galaxies in their environments. For the subset of simulations that attempt ISM modeling of lower mass objects \citep[e.g.,][]{Barrow2017,Katz2019_ism,spdrv1,Trebitsch2021,Yajima2022,Pallottini2022,Bhagwat2024,Kannan2025,Garcia2025O,Hirai2025}, the non-equilibrium chemistry is typically limited to only primordial species (e.g., H, He, H$_2$, e$^-$); hence, emission from only hydrogen and helium can be directly predicted from the simulation. None of these simulations include a detailed, non-equilibrium metal or molecular chemical network, nor have the ISM models been benchmarked against local constraints \citep{Kim2023}. 

Simulations without non-equilibrium metal chemistry must be post-processed with photoionization models such as {\small CLOUDY} \citep{Ferland2017} or Monte Carlo radiation transfer \citep[e.g.,][]{McClymont2024} in order to model their intrinsic (before dust attenuation) spectra. Such an approach has been successful in creating realistic looking galaxies that reproduce many of the observed properties of the high-redshift galaxy population \citep[e.g.,][]{Hirschmann2023b,spdrv1,Wilkins2023, Nyhagen2024}. However, numerous assumptions must be made in post-processing that limit the predictive power of the simulations. Different (reasonable) sets of assumptions can lead to significantly different mock observations. For this reason, there is strong motivation for the development of more self-consistent and predictive forward models, especially in the context of dwarf galaxies where such an approach is numerically tractable. 

There are now a minority of cosmological simulation codes that can model the detailed physics from photoionization models, in non-equilibrium, in a 3D galaxy formation setting \citep[e.g.,][]{Richings2014,RTZ,Chan2025}. {\small RAMSES-RTZ} is unique among these as it is connected to on-the-fly, multi-frequency radiation transport and has been extensively benchmarked to reproduce equilibrium results computed with 1D photoionization codes \citep{RTZ,Katz2024_meg} and the properties of the Milky Way ISM at solar metallicity \citep{Katz2022_prism}. The non-equilibrium aspect of the code allows one to drop a major equilibrium assumption that is ubiquitous across post-processing methods and was recently shown to fail in certain regimes \citep{Richings2022,Ploeckinger2025}. Moreover, the 3D nature of {\small RAMSES-RTZ} enables it to capture the geometric effects of H~{\small II} region structure on emission lines \citep{Grey2017,Jin2022}. Hence {\small RAMSES-RTZ} can better predict the intrinsic spectra\footnote{I.e.\ the observed spectra prior to dust attenuation. Note that we have adopted an empirical model for dust in the simulation that interacts with the radiation and the gas thermo-chemistry, but we do not sample enough frequency bins (due to memory constraints) in the code to accurately predict the attenuated spectra on-the-fly, nor do we consider scattering at UV and optical wavelengths and, therefore, these must be computed in post-processing. Note also that when the resolution is too low, Strömgren spheres become unresolved and corrections to the temperature and ionization state must be made in post-processing (see e.g.\ \citealt{spdrv1} and further details below). However, this only occurs in very dense gas which we empirically find is close to equilibrium.} of simulated galaxies from cosmological initial conditions, making it uniquely suited for studying the spectral evolution of early galaxies.

\begin{figure*}
    \centering
    \includegraphics[width=0.98\textwidth,trim={0cm 0cm 0cm 1.3cm},clip]{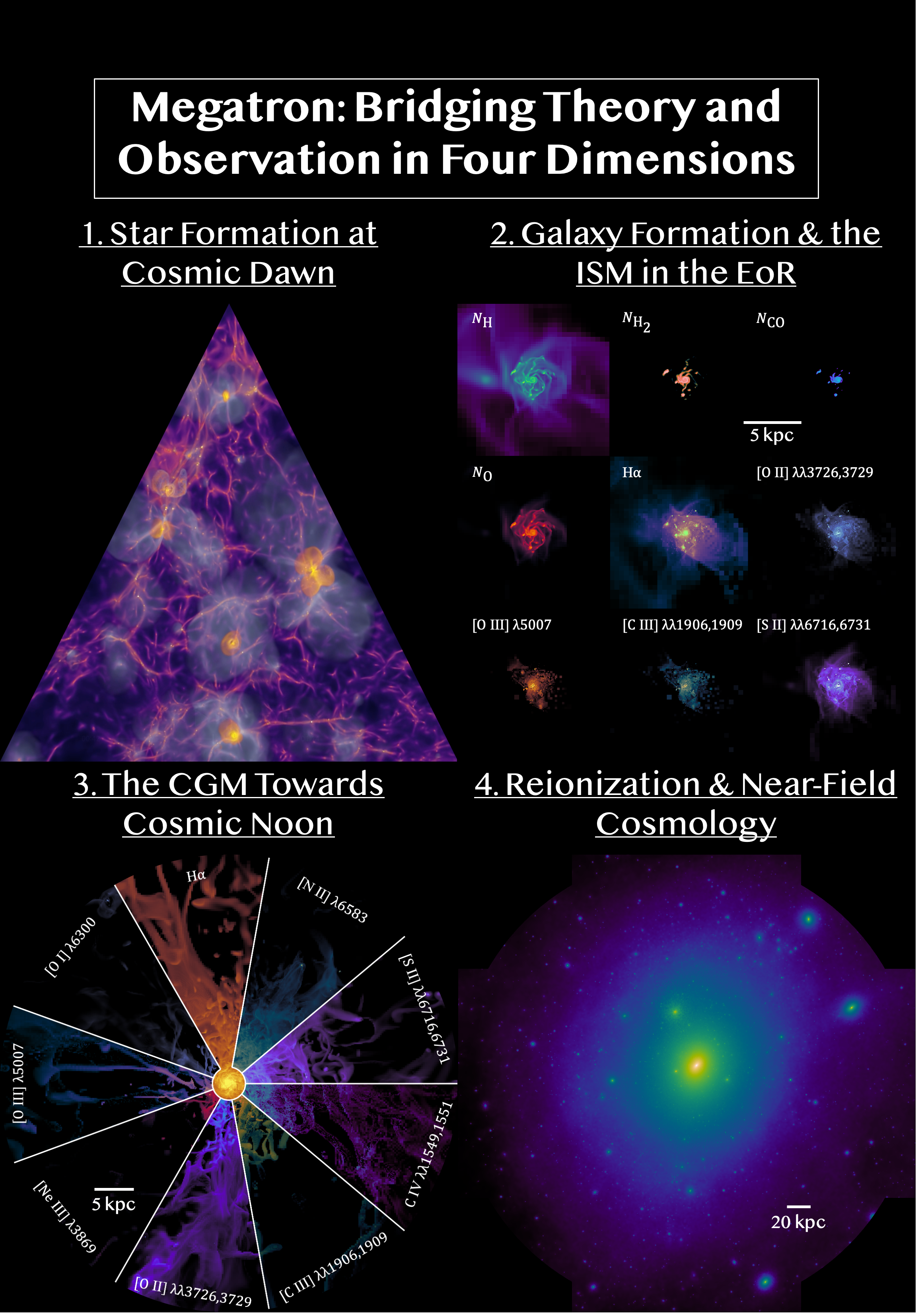}
    \caption{Summary of the four key science drivers of the {\small MEGATRON} simulations. {\bf Top left}: A $z\sim15$ density map with H$\beta$ (white) and [O~{\small III}]~$\lambda$5007 (yellow) shows radiation-driven H$\beta$ emission extending into the IGM, while metals remain near galaxies. {\bf Top right}: A rotationally-supported $z=10$ galaxy reveals early H$_2$ and CO formation and complex emission line morphology. {\bf Bottom left}: The CGM of a massive $z\sim4$ star-forming galaxy highlights how different ions trace distinct gas phases. {\bf Bottom right}: The $z=0$ dark matter distribution connects high- and low-redshift structure formation. }
    \label{fig:hero}
\end{figure*}

In this work, we introduce the {\small MEGATRON} suite of simulations, that, for the first time, employ a detailed thermochemistry network of primordial species, metals, and molecules, coupled to on-the-fly radiation transport and a state-of-the-art galaxy formation model. Our work focuses on star-forming galaxies --- AGN will be considered in future work. The simulations represent a large zoom region around an object that collapses to a Milky Way-mass galaxy at $z=0$ (although the simulations are stopped at much higher redshift). The {\small MEGATRON} project is designed to advance our understanding in four key science areas (Figure~\ref{fig:hero}), in chronological order: 
\begin{enumerate} 
    \item Predicting the demographics and spectral properties of the first generation of Population III (Pop.~III) stars.
    \item Quantifying how observable ISM properties respond to different galaxy and stellar evolution models at low metallicity.  
    \item Studying how non-equilibrium chemistry, local radiation fields, and mass assembly affect the emission and absorption observables of the circumgalactic medium (CGM) at cosmic noon.
    \item Deciphering the archaeological traces left by high-redshift galaxy formation physics in the local volume environment of a Milky Way-mass galaxy.
\end{enumerate} 

\begin{figure*}
    \centering
    \includegraphics[width=\textwidth,trim={0cm 0cm 0cm 0cm},clip]{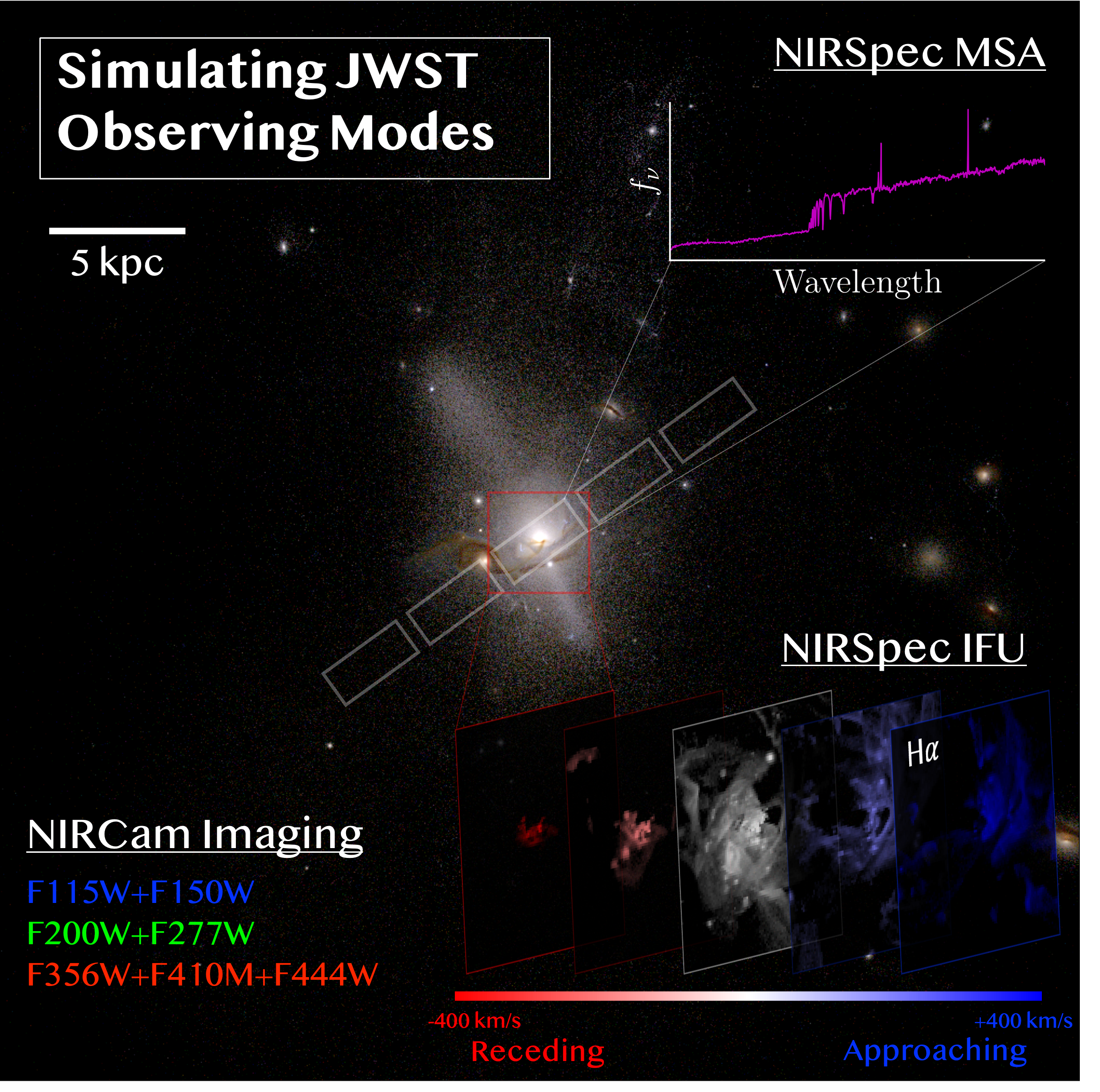}
    \caption{Demonstration of how {\small MEGATRON} galaxies can be mock observed in multiple JWST observing modes. The example galaxy is a dusty spiral at $z=4$ from the Cosmic Noon suite. The background RGB image combines JWST NIRCam filters F115W and F150W, in the blue channel, F200W and F277W, in the green channel, and F356W, F410M, and F444W in the red channel. Prominent dust lanes are visible as are large plumes of stars that are a remnant of a merger. To mock the NIRSpec micro shutter array (MSA), we overlay slitlets and show the spectrum of the central region of the galaxy. The object has a very red UV slope from dust attenuation and a clear Balmer break due to an aging stellar population, and some weak emission lines that are remnants of previous and ongoing star formation. Finally, we mock the central region of the NIRSpec IFU focusing on the velocity-resolved H$\alpha$ line. We split the velocity channels in to bins of $v<-400~{\rm km/s}$, $-400<v<-200~{\rm km/s}$, $-200<v<200~{\rm km/s}$, $200<v<400~{\rm km/s}$, and $v>400~{\rm km/s}$. Note how the morphology of the H$\alpha$ line strongly depends on velocity. All images, spectra, and IFU data were created with the Monte Carlo radiation transfer code RASCAS \protect\citep{Leo2020}.}
    \label{fig:hero2}
\end{figure*}

To achieve this, the {\small MEGATRON} simulations bring together several technical improvements that build upon previous numerical efforts to model early galaxy formation. In particular:
\begin{itemize}
    \item The explicit modeling of Pop.~III star formation from zero-metallicity initial conditions facilitated by non-equilibrium H$_2$ cooling and a high spatial resolution ($\approx$pc). This mode of star formation is common in targeted studies of minihaloes and a select few larger-scale cosmological simulations \citep[e.g.,][]{Wise2012,Oshea2015,Kimm2017,Brauer2025}, but remains absent in most high-redshift simulations attempting to resolve the ISM, e.g., {\small SPHINX} \citep{Rosdahl2018}, {\small FIRE} \citep{Ma2018}, {\small THESAN-ZOOM} \citep{Kannan2025}.
    
    \item An energetic feedback model from the {\small VINTERGATAN} that was calibrated on Milky Way-mass galaxies at $z=0$ (with only minor modifications here), and shown to produce a realistic stellar mass-halo mass relation at $z=0$ \citep{Agertz2021, Rey2023Vintergatan}.

    \item A non-equilibrium chemistry model for primordial species, metals, and molecules. This is not generally captured by high-redshift simulations\footnote{Recent state-of-the-art simulations do have non-equilibrium H and He thermochemistry, but assume equilibrium for all other elements. Note also that a small minority of simulations include the low ionization states of species like carbon and oxygen \citep[e.g.,][]{Lupi2020,Khatri2025}.}. Non-equilibrium chemistry can be important for accurately predicting ISM emission lines depending on the density and ionizing conditions \citep[e.g.,][]{Richings2022,Katz2022_prism}, and its importance is expected to increase in the CGM where lower gas density leads to longer equilibrium timescales\footnote{For example, the recombination timescale scales inversely with electron density.} \citep{Gnat2007,Oppenheimer2013,Kumar2025}. 
    
    \item An ISM thermochemistry model \citep{Katz2022_prism} that reproduces the conditions of the local ISM. The importance of recovering the inferred gas pressure as a function of density under such conditions was recently discussed in \cite{Kim2023}.

    \item An emphasis on mock observations (see Figure~\ref{fig:hero2}), where we have aimed to reproduce (with as few additional assumptions as possible) the observing modes of NIRCam and NIRSpec on JWST. 

    \item A grid of simulations studying the response of observables to both (i) variations in fundamental subgrid models; and (ii) controlled variations in the mass growth history of an object using the `genetic modification' approach \citep{Roth2016, Rey2018}. 
        
\end{itemize}

In this paper, we introduce the suite of {\small MEGATRON} simulations. Most of the physical ingredients of the simulations are presented in the methods paper \citep{Katz2024_meg} and we outline here the numerics of two simulation suites targeting high-redshift galaxy formation and the CGM towards cosmic noon. As a validation of our approach and model, we show how the {\small MEGATRON} simulations are able to reproduce the spectral diversity seen at high redshift with JWST.

\section{Numerical Methods}
We present seven cosmological radiation hydrodynamics simulations of the Lagrange region around a Milky Way-mass galaxy. The simulations are all run with the {\small MEGATRON} galaxy formation model presented in \cite{Katz2024_meg} that was developed within the {\small RAMSES-RTZ} adaptive mesh refinement (AMR) code \citep{RTZ}. {\small RAMSES-RTZ} is a fork of {\small RAMSES} \citep{Teyssier2002} and {\small RAMSES-RT} \citep{Rosdahl2013,Rosdahl2015}. Four simulations comprise the ``high-redshift'' suite, where the resolution is primarily focused on resolving the ISM in the early Universe. The physical inputs are loosely motivated by various physical mechanisms that may help explain the excess numbers of bright galaxies at high redshift \citep[e.g.,][]{Chemerynska2023,Leung2023,Harikane2024,Finkelstein2023_b}, while achieving high-enough resolution to predict the spectral properties of the ISM that can be directly compared to existing JWST and ALMA data. Three simulations run at slightly lower resolution are designed to capture the non-equilibrium physics in the CGM towards cosmic noon, with the aim of studying the observable properties of galaxies that can be used to probe the cosmic baryon cycle. The CGM simulations will be presented in detail in a companion paper \citep{Cadiou2025tmp}. 

Details of the seven simulations can be found in Table~\ref{tab:sim_desc} and a summary of the methods employed for galaxy formation physics, halo finding, and dust radiation transport are described in Appendix~\ref{app:gf}. In this work, we focus primarily on the high-redshift simulation suite which is stopped at $z\sim8.5$, while the cosmic noon simulations (which end closer to $z\sim3$) is presented in \cite{Cadiou2025tmp}. Furthermore, the Milky Way mass initial conditions of these simulations are designed to allow us to make connections with Milky Way satellite population in the low-redshift Universe and the properties of Milky Way progenitors. A dark matter-only simulation using the same initial conditions has been run to $z=0$ and the connection between high and low redshift is presented in a second companion paper \citep{Rey2025tmp}.

\subsection{Initial Conditions}
We construct cosmological, zoomed initial conditions (ICs) for a Milky-Way-mass halo (with virial mass of $\approx \SI{e12}{\Msun}$ at $z=0$). This halo is based on a reference object first presented in \citet{Rey2022} and evolved to $z=0$ for the \textsc{vintergatan-gm} project \citep{Agertz2021, Rey2023Vintergatan}. 

\begin{figure}
  \centering
    \includegraphics[width=\columnwidth]{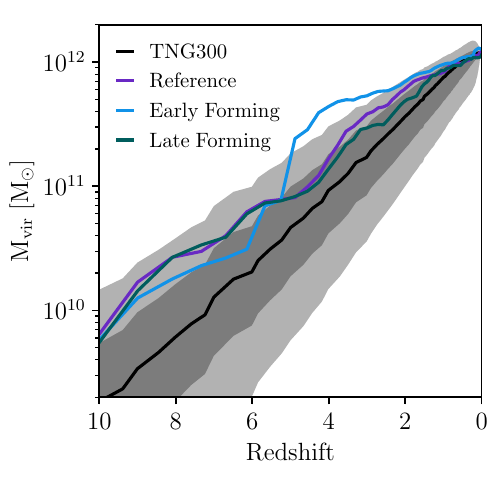}
    \caption{Mass growth histories of the main progenitor in the three sets of initial conditions compared to the typical mass growth histories of 28,475 similar mass haloes from TNG300. The dark and light shaded regions represent the $1\sigma$ and $2\sigma$ results from TNG300 with the black line representing the median relation. The early forming halo (purple) is used for the high-redshift suite as it results in more numerous massive progenitors at high redshift rather than a single dominant object. The reference and late forming initial conditions (dashed blue and green lines, respectively) are only used for the CGM suite.}
    \label{fig:mahs}
\end{figure}

\begin{table*}
    \centering
    \caption{Details of the simulations in the MEGATRON suite. We list the simulation names, the suite they belong to, the mass of the dark matter particles, minimum mass of the stellar particles, minimum physical cell size at $z=8.5$ or the target constant physical resolution, initial metallicity, the current redshift of the simulation, the virial mass of the primary halo at the current redshift, the Jeans length below which star formation is triggered, the efficiency of star formation per free-fall time, the energy per SN, and whether the simulation included hypernovae (HN). A star formation efficiency denoted as $f(\sigma_V)$ indicates that the model uses a variable efficiency related to the local turbulent properties of the gas.}
    \begin{tabular}{llccccccccccc}
         Name & Suite & $m_{\rm DM}$ & $m_{*}$ & $\Delta x_{\rm min}$ & $Z_{\rm initial}$ & $z_{\rm current}$ & ${\rm M_{vir,max}}$ & $\lambda_{\rm J,SF}$ & $\epsilon_{\rm ff}$ & $E_{\rm SN}$ & HN \\
         \hline
          & & (M$_{\odot}\ h^{-1}$) & (M$_{\odot}$) & (pc~$h^{-1}$) & $Z_{\odot}$ & & $\log_{10}({\rm M_{\odot}})$ & & & $(10^{51}$\ erg) & \\
         \hline
         Efficient SF & High-Redshift & $1.67\times10^4$ & 500 & $\sim 5$ & 0 & $8.5$ & 10.30 & $\Delta x$ & $f(\sigma_V)$ & 1.0 & \ding{55} \\
         Bursty SF & High-Redshift & $1.67\times10^4$ & 500 & $\sim 5$ & 0 & $8.5$ & 10.27  & $\Delta x$ & $f(\sigma_V)$ & 5.0 & \ding{55}  \\
         Variable IMF & High-Redshift & $1.67\times10^4$ & 500 & $\sim 5$ & 0 & $8.5$ & 10.30  & $\Delta x$ & $f(\sigma_V)$ & 1.0 & \ding{51}  \\
         HN, High $\epsilon_{\rm ff}$ & High-Redshift & $1.67\times10^4$ & 2,000 & $\sim 5$ & 0 & $8.5$ & 10.30  & $4\Delta x$ & 100\% & 1.0 & \ding{51}  \\
         \hline
         Early Collapse & Cosmic Noon & $1.67\times10^4$ & 4,600 & $\sim 20$ & $10^{-4}$ & $2.94$ & 11.67 & $\Delta x$ & $f(\sigma_V)$ & 1.0 & \ding{55}  \\
         Fiducial Collapse & Cosmic Noon & $1.67\times10^4$ & 4,600 & $\sim 20$ & $10^{-4}$ & $3.60$ & 11.26 & $\Delta x$ & $f(\sigma_V)$ & 1.0 & \ding{55}  \\
         Late Collapse & Cosmic Noon & $1.67\times10^4$  & 4,600 & $\sim 20$ & $10^{-4}$ & $3.32$ & 11.29  & $\Delta x$ & $f(\sigma_V)$ & 1.0 & \ding{55}  \\
         \hline
         & 
    \end{tabular}
    \label{tab:sim_desc}
\end{table*}

We generate ICs using the {\small genetIC} software \citep{Stopyra2021} and a flat \citet{Planck2016Cosmo} cosmology with $\Omega_\mathrm{M}=0.3139$, $h=0.6727$, $\sigma_8=0.8440$, $n_\mathrm{s}=0.9645$ and $\Omega_\mathrm{b}=0.04916$. From a dark matter-only cosmological volume with a box size of $50\ {\rm Mpc}\ h^{-1} = \SI{73}{Mpc}$ and mass resolution $m_\text{DM} = 1.2 \times 10^8 \, {\rm M_{\odot}}$, we identify an isolated halo with Milky-Way virial mass ($M_{\text{200}} \approx \SI{e12}{M_{\odot}}$ and no more massive neighbors within 5$r_{\mathrm{200c}}$, where $r_{\mathrm{200c}}$ is the radius enclosing 200 times the critical density of the Universe; see \citealt{Rey2022} for further details). 

We trace the region enclosing 3$r_{\mathrm{200c}}$ at $z=0$ around this object to the low-resolution ICs and refine the mass resolution within this Lagrange region to $m_\text{DM} = \SI{1.67e4}{\Msun}$~$h^{-1}$ (effective resolution $8912^3$). Importantly, since this region is selected from the $z=0$ halo, it represents a large cosmic volume at high redshift (initially $18$ comoving Mpc$^3$~$h^{-3}$) and ensures several thousands of high-redshift galaxies are adequately resolved without contamination.

The mass growth history of our reference object is shown as the blue dashed line in Figure~\ref{fig:mahs}. To further maximize the size of the high-redshift galaxy population, we use a quadratic genetic modification technique, as outlined in \cite{Rey2018, Rey2019}, to modify the collapse of progenitors within the Lagrangian patch. More specifically, we increase the variance by 10\% on scales of $0.30\ {\rm Mpc}\ h^{-1}$. This ensures that all progenitor haloes with masses $\approx \SI{2e10}\Msun$ collapse earlier, generating a larger sample of more massive high-redshift galaxies, which is useful for comparing with JWST data. The result of this modification is to initially delay the mass accretion onto the main progenitor (purple line in Figure~\ref{fig:mahs}) since mass is spread across multiple, more massive progenitors early on. However, the modification promotes a rapid assembly at $z\sim6-4$ when the multiple, more massive progenitors merge. To ensure a fixed total mass at $z=0$, we reduce the mean overdensity of the Lagrangian region by \SI{12}{\percent} (see also \citealt{Roth2016, Pontzen2017} for further details).

Finally to study the impact of the timescale of the collapse of the Lagrange region on the CGM towards cosmic noon, we create a third IC (green dashed line in Figure~\ref{fig:mahs}) that assembles slightly faster at early times and is delayed at later times by reducing the variance by 5\% on scales of $0.30\ {\rm Mpc}\ h^{-1}$. 

To assess the likelihood of the formation scenarios we engineered with the genetic modification approach, we first compute the likelihood difference between the reference and modified initial conditions. We find $\Delta \chi^2 = -2630$ and $\Delta \chi^2 = 9530$ for the `Early' and `Late' ICs, respectively. Both $\Delta \chi^2$ are low compared to the number of degrees of freedom available in our Lagrangian region which contains $\approx10^8$ modes, ensuring that our ICs are compatible with the $\Lambda$CDM power spectrum. We also compare in Figure~\ref{fig:mahs} our mass assembly histories to a large sample of $\approx 30,000$ mass assembly histories for haloes with a similar $z=0$ mass in TNG300 (\citealt{Pillepich2018,Nelson2018}, see \citealt{Rey2022} for more details of the comparison). All of our ICs assemble more rapidly $z\geq 4$ compared to a typical Milky-Way-mass halo, with the `Early Forming' IC used for the high-redshift suite having an excursion above the $2\sigma$ contour at intermediate redshift ($z\sim5-3$). 

We emphasize that our chosen ICs are not typical of haloes of similar mass at $z=0$, but rather engineered for specific purposes --- having a Milky Way mass galaxy at $z=0$ while still producing a significant number of high-redshift galaxies. While this is ideal for our science goals focused on studying the internal properties of galaxies (e.g., the structure of the ISM and CGM, how different modes of star formation are realized in observable properties, etc.), comparing with population statistics (e.g., luminosity functions, stellar mass functions, etc.) will require more care given the biased growth history. 

All ICs are evolved to $z=149$ using first-order linear theory \citep{Zeldovich1970}. For the high-redshift suite simulations, we assume a primordial composition of \SI{76}{\percent} hydrogen and \SI{24}{\percent} helium, and no metals. For the cosmic noon suite, we adopt a metallicity of $12+\log_{10}({\rm O/H})=4.69$ ($10^{-4}Z_{\odot}$) for the oxygen abundance and follow \cite{Ramambason2022} to compute the abundances of other elements\footnote{Note that we assume a solar metallicity of $12+\log_{10}({\rm O/H})=8.69$ \citep{Asplund2009}.}.

To reduce advection errors and increase the accuracy of high-redshift hydrodynamics, we `genetically' modify the initial conditions to ensure that the Lagrangian region is at rest compared to the numerical grid following the procedure described in \citet{Pontzen2021}. 

\begin{figure}
    \centering
    \includegraphics[width=\columnwidth]{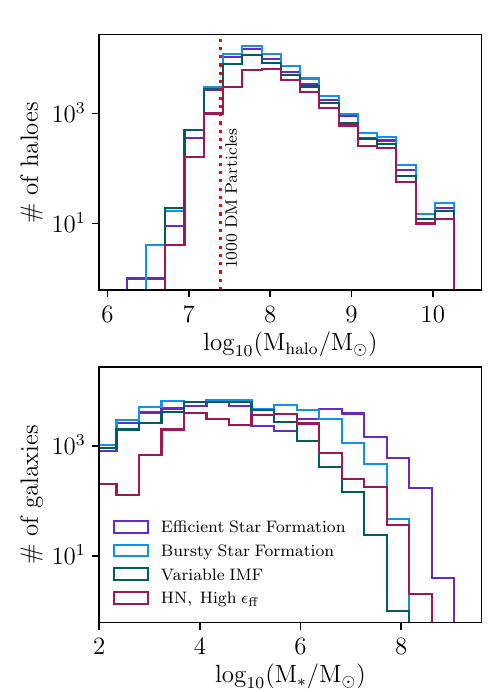}
    \caption{Histograms of halo mass (top) and stellar mass (bottom) across all snapshots for each simulation in the high-redshift suite. In the top panel we indicate the halo mass that corresponds to 1,000 DM particles, which represents the minimum required to be included in our spectroscopic sample.}
    \label{fig:smhf}
\end{figure}

\begin{figure*}
    \centering
    \includegraphics[width=\textwidth]{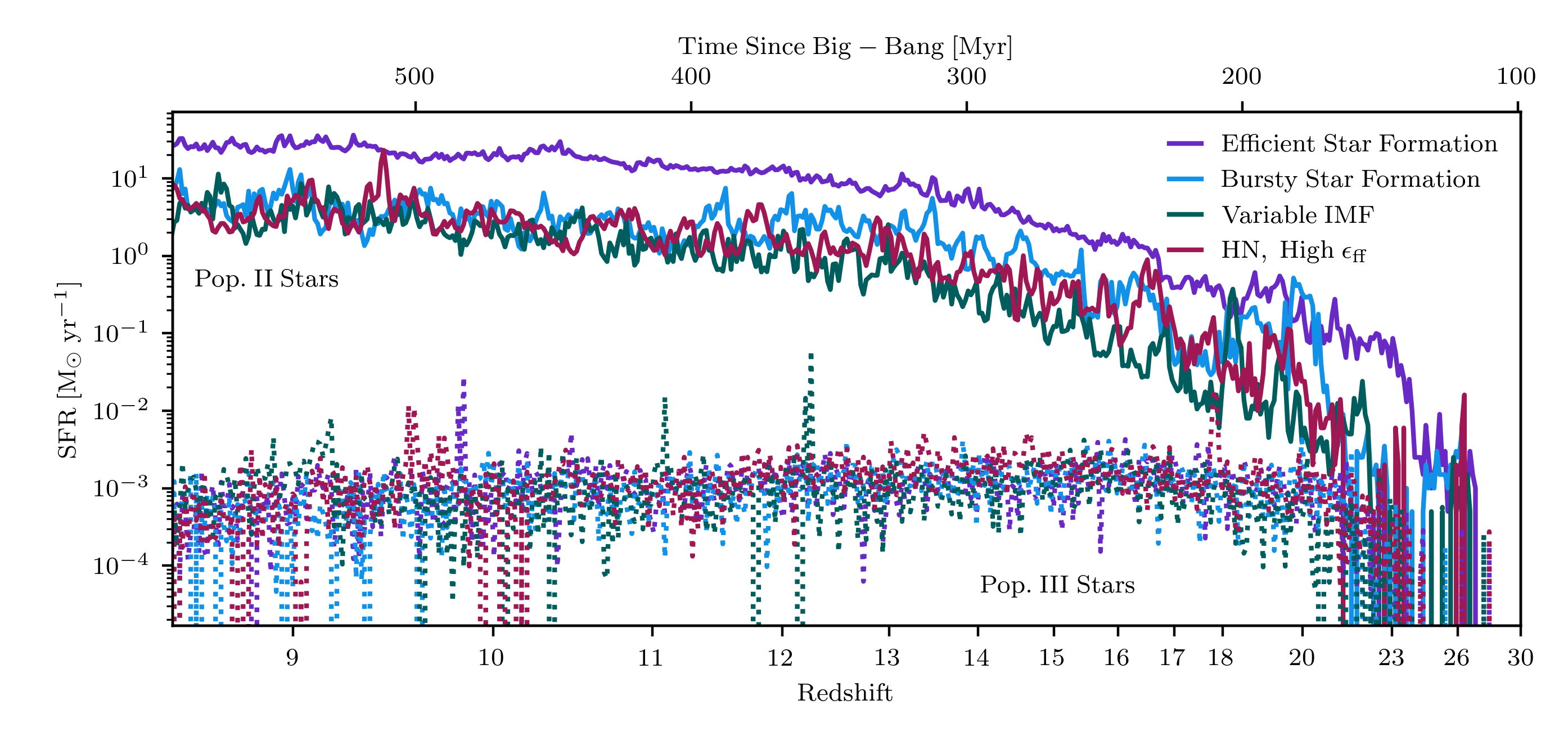}
    \caption{Star formation histories of Pop.~II stars (solid) and Pop.~III stars (dotted) across the entire Lagrange volume for each of the high-redshift suite simulations. We adopt a critical metallicity of $Z\geq2\times10^{-8}$ for Pop.~II star formation.}
    \label{fig:SFH}
\end{figure*}

\subsection{Summary of Simulations}
The key differences between the seven simulations are summarized in this section. The high-redshift suite uses a constant comoving resolution which results in extremely high-resolution (up to $\sim1$~pc physical resolution at the time when the first Pop.~III stars form) and is thus better suited for studying the details of star formation and the ISM at high redshift. The cosmic noon suite employs a constant physical resolution of $\approx20~{\rm pc}~h^{-1}$. This is computationally less expensive, allowing us to run to lower redshift. However, the sacrifice in spatial resolution makes these simulations more ideal for studying more massive galaxies and the CGM. 

Within the high-redshift suite, we consider variations to feedback, IMF, and star formation. In the Bursty star formation simulation, the energy injected per SNII is increased to $5\times10^{51}~{\rm erg}$, which results in large fluctuations in galaxy star formation rates. The variable IMF simulation allows for hypernova and a density and metallicity-dependent IMF. Finally, in the HN, High $\epsilon_{\rm ff}$ simulation, we require the Jeans length to be better resolved for star formation to occur, form stars at 100\% efficiency per free-fall time, and allow for hypernova. 

The galaxy formation physics in the cosmic noon suite is identical to the efficient star formation model as this is closest to what was used in {\small VINTERGATAN}. The only difference between the three cosmic noon simulations is the ICs. These ICs are systematically designed to study the impact of an earlier and later formation history on the properties of the lower-redshift CGM. 

For all high-redshift simulations, we output full simulation snapshots at a cadence of 5~Myr after $z=30$\footnote{Note that we store many more snapshots at irregular time intervals, as the simulation is checkpointed just before reaching the wall-clock time limit.} This cadence is increased to 20~Myr for the CGM suite. 

\subsection{Mock Spectra}
The primary advantage of using {\small RAMSES-RTZ} and a detailed non-equilibrium thermochemistry and star formation model is that the intrinsic spectra of the simulated galaxies are self-consistently predicted by the code. We consider the stellar continuum, nebular continuum, and nebular emission lines. 

For Pop.~II stellar continua, we adopt the same SEDs that are used when computing the on-the-fly RT --- either {\small BPASS} (default) or {\small STARBURST99} (Variable IMF). Pop.~III stellar continua are from \cite{Larkin2023}\footnote{Note that in the simulation, the ionizing photon rates for Pop.~III stars from from \cite{Schaerer2002}; however, because they do not provide the full SEDs, we defer to those from \cite{Larkin2023} when computing the full spectrum of each Pop.~III galaxy. There are minimal differences between these models.}. 

Nebular line emission is calculated for both recombination lines and collisionally excited lines. For each cell, we have an electron density, temperature, and ion number density, which allow us to compute the luminosity of each line. Emissivities for each line are computed with {\small PyNeb} \citep{Luridiana2015} assuming a 15-level atom. We adopted the atomic data from {\small CHIANTI} \citep{Dere2019}. For Balmer and Lyman lines of H and He, recombination is typically the dominant component of the emission; however, collisional excitation is non-negligible. In these cases, {\small CHIANTI} is used to compute the collisional component. In certain cases, the $\mathrm{H}\,{\textsc{ii}}$ regions around star particles are numerically unresolved. This typically happens for older star particles that have low ionizing output (and thus rarely contribute meaningfully to the spectrum), or very young stars that are embedded in dense gas clouds. When this occurs, the local cell properties become a mix between the neutral and ionized phases and must be corrected. We thus replace the emission of cells with unresolved Strömgren spheres by a spherical {\small CLOUDY} \citep{Ferland2017} calculation adopting the gas properties of the host cell and SED of the star particle. We emphasize that in the spherical Strömgren sphere limit, {\small CLOUDY} and {\small RAMSES-RTZ} predict very similar emission line luminosities (to better than 10\% on average). When there are multiple star particles within the same gas cell with unresolved Strömgren spheres, we calculate whether the Strömgren spheres overlap and create a graph of connected regions. We then group each set of connected Strömgren spheres into a single {\small CLOUDY} model (i.e.\ a single gas cell can host multiple unresolved Strömgren spheres when the cell is extremely dense).

Nebular continuum emission is computed very similarly to nebular line emission. {\small PyNeb} is used to calculate the free-free, free-bound, and two-photon emission (considering only H and He, which dominate) and similar corrections are made for unresolved Strömgren spheres using {\small CLOUDY}.

\section{Star \& Galaxy Formation in a Milky Way mass Lagrangian Region}

We begin by describing the basic physical properties of the dark matter, gas, and stars in the Lagrange region as a function of time for the simulations that are part of the high-redshift suite. The top panel of Figure~\ref{fig:smhf} shows the distribution of dark matter halo masses in our simulations, combining results from all snapshots up to $z=8.5$. By genetically modifying the reference ICs, we resolve thousands of haloes in the epoch of reionization, up to masses $>10^{10}\ {\rm M_{\odot}}$, as well as those well below the atomic cooling threshold, leading to a statistical sample size that is ideal for comparing to galaxies at the faint-end of the UV luminosity function at high redshift. In the $z=8.5$ snapshots, we have slightly more than 5,000 haloes resolved by at least 300 dark matter particles in each simulation. Across all snapshots, considering only those haloes with at least one star particle and resolved by $>$1,000 dark matter particles and $>$1,000 gas cells, which meet our criteria to be included in our spectroscopic sample, the simulations produce a data set of $>175,000$ spectra. This represents a nearly 100$\times$ increase compared to the number of individual galaxies with spectra computed for {\small SPHINX$^{20}$} \citep{spdrv1} and {\small RENAISSANCE} \citep{Barrow2017} simulations\footnote{Note that both of these simulations have more resolved galaxies compared to MEGATRON; however, post-processing the simulations to compute spectra from {\small SPHINX$^{20}$} or {\small RENAISSANCE} is much more time-consuming than what is needed for MEGATRON, which is why our sample is so much larger.}.

\begin{figure}
    \centering
    \includegraphics[width=\columnwidth]{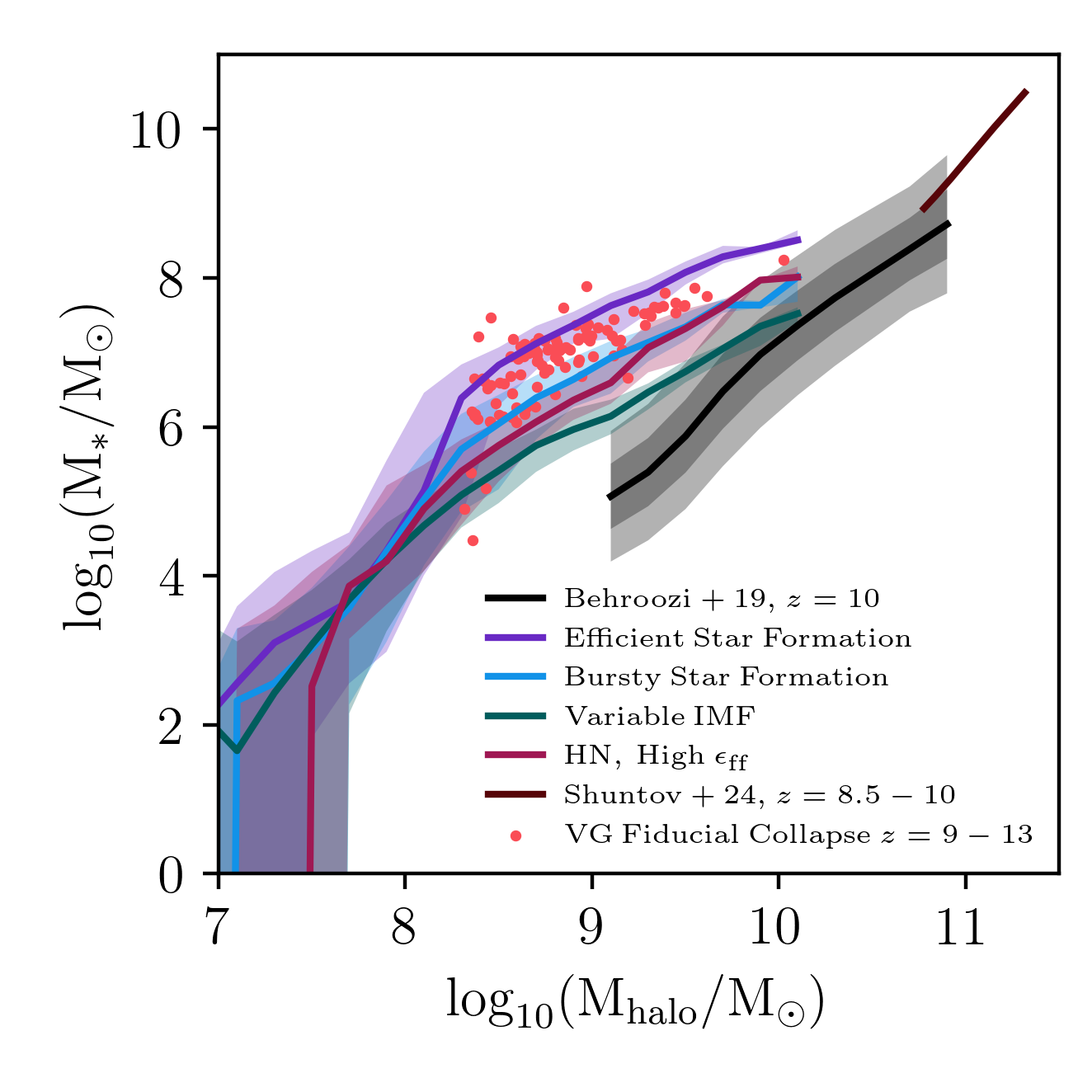}
    \caption{Stellar mass - halo mass relation for each simulation, including all galaxies at every redshift. We show the median and $1\sigma$ scatter about the relation. For comparison, the empirical model of \protect\cite{Behroozi2019} at $z=10$, the median redshift of our spectroscopic sample (black) and its $1$ and $2\sigma$ scatter as well as inferences from high-redshift JWST observations from \protect\cite{Shuntov2025} (brown line). We also show the results from the VINTERGATAN simulation (salmon points) of the fiducial collapse initial conditions, which serve as our benchmark.}
    \label{fig:smhm}
\end{figure}

\begin{figure}
    \centering
    \includegraphics[width=\columnwidth]{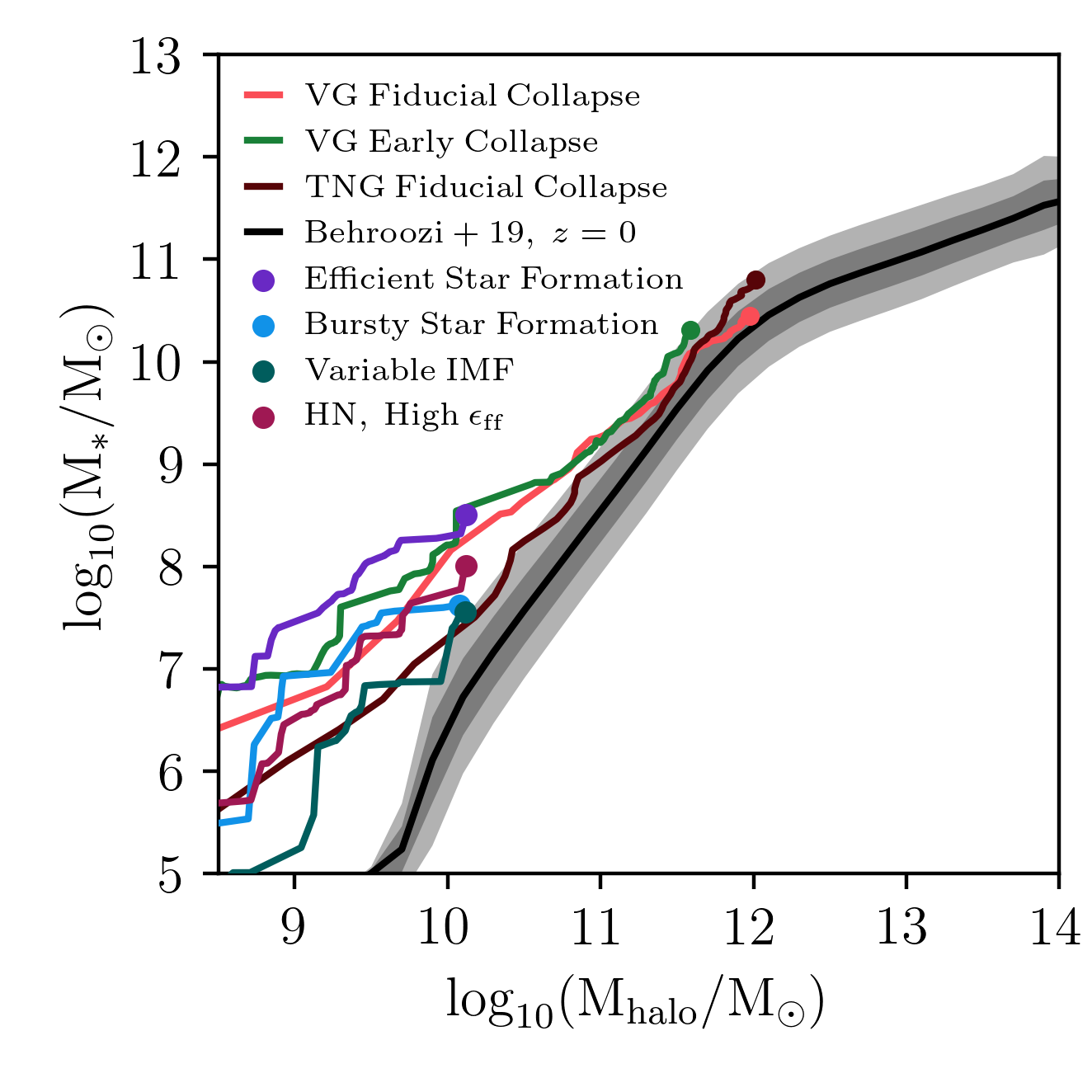}
    \caption{Stellar mass - halo mass relation for VINTERGATAN simulations of the fiducial (salmon line) and early collapse (green line) compared with the $z=0$ empirical constraints from \protect\cite{Behroozi2019} at $z=0$ (black). Note that there are more simulation outputs for the early collapse model, which explains why the line is less smooth and the simulation has not been run to $z=0$. The brown line shows the fiducial initial conditions run with the IllustrisTNG model \protect\cite{Joshi2025}. Here we see that the benchmark halo is in good agreement with empirical constraints at low redshift, despite being a more than $2\sigma$ outlier at $z\sim9-13$. Coloured points represent the values for the most massive progenitor halo at $z=8.5$ in each of the high-redshift suite simulations.}
    \label{fig:smhm2}
\end{figure}

Star formation first occurs in the simulations at $z\sim27$. In our model, a single Pop.~III explosion, regardless of mass, can provide enough heavy elements to reach the critical metallicity ($Z\geq2\times10^{-8}$) for Pop.~II star formation \citep[see also e.g.,][]{Wise2012,Brauer2025}. By $z=26$, Pop.~II star formation is already dominant across the Lagrange region. This is demonstrated in Figure~\ref{fig:SFH} where we show the star formation history of the zoom region for all high-redshift simulations. By $z=20$ the Pop.~II SFR is typically an order of magnitude or more greater than that of Pop.~III stars. The effect is strongest when weaker feedback allows more efficient Pop.~II star formation (purple). Despite the dominance of Pop.~II star formation in all models at $z\lesssim26$, a perhaps surprising result is that Pop.~III stars are still forming in the simulation at $z=8.5$. Late Pop.~III star formation has been observed in previous simulations \citep[e.g.,][]{Pallottini2014,Xu2016,Jaacks2018,Sarmento2018}, although this typically occurs in low-density regions, and not necessarily in the Lagrange region around a Milky Way mass object. 

The bottom panel of Figure~\ref{fig:smhf} shows a histogram of the stellar masses of the galaxies in our spectroscopic sample (after accounting for mass loss from stellar evolutionary processes). Depending on the simulations, we find stellar masses up to $10^8-10^9\ {\rm M_{\odot}}$. The efficient star formation model produces the highest stellar masses, again due to the weaker feedback, while in the variable IMF model, the galaxy stellar masses are much better regulated. The stellar masses are better controlled in the variable IMF model partially because the energetic feedback budget is increased due to the inclusion of HN and the increase in number of SN due to a shallower upper-mass IMF slope. In addition, the increased number of massive stars in the variable IMF model leads to the star particles losing a greater fraction of their initial mass. As can be seen in Figure~\ref{fig:SFH}, by $z=8.5$, the SFR in the Lagrange region is similar in the bursty star formation and variable IMF models and significantly lower than in the efficient star formation simulation. However, the extra mass loss from the top-heavy IMF reduces the stellar mass of the variable IMF simulation with respect to the bursty star formation model.

The regulation of star formation is explored in Figure~\ref{fig:smhm} where we show the stellar mass-halo mass relation for all high-redshift simulations. We do not find a significant evolution with redshift for this relation. The benchmark for our simulation are the results from {\small VINTERGATAN} (shown as salmon points) for the set of initial conditions with the fiducial collapse. The {\small VINTERGATAN} simulations have been run to $z=0$ and good agreement was found with the local stellar mass-halo mass relation \citep[see Figure~\ref{fig:smhm2} and also ][]{Agertz2021}. The efficient star formation model uses a feedback scheme very similar to {\small VINTERGATAN} (by design) and we find a small offset such that this {\small MEGATRON} model forms $\sim50\%$ more stars compared to our benchmark. This is partially due to the initial condition modification for early collapse but also likely caused by the higher DM resolution, the adoption of constant physical versus constant comoving resolution, and the non-equilibrium cooling used in {\small MEGATRON}. Note that in Figure~\ref{fig:smhm2} the stellar mass-halo mass relation for the main progenitor in the early collapse initial conditions is consistently higher than for the fiducial collapse. Nevertheless, the agreement between our simulations and  {\small VINTERGATAN} is very acceptable given the minor changes. The other three models form fewer stars than the benchmark {\small VINTERGATAN} simulation due to the enhanced feedback, with the variable IMF simulation having the lowest stellar masses. At $z=8.5$, our four high-redshift simulations bracket the stellar masses expected for the same set of early collapsing initial conditions from the {\small PARADIGM} project \citep{Joshi2025}. In that work, the early collapse initial conditions were simulated with both the {\small VINTERGATAN} and {\small IllustrisTNG} models. Interestingly, they found that the {\small IllustrisTNG} model forms fewer stars at high redshift compared to the {\small VINTERGATAN} model but then falls higher on the stellar mass-halo mass relation at $z=0$. Our results thus span reasonable predictions of the high redshift Universe based on models calibrated at $z=0$. 

For further comparison, we show the inferred stellar mass-halo mass relation from recent JWST survey data at high redshift \citep{Shuntov2025}, which unfortunately does not overlap in mass range with our models as well as empirical constraints from \cite{Behroozi2019}. We caution comparisons with \cite{Behroozi2019} because this model is known to significantly underpredict galaxy number counts at high redshift, even at $z<9$ \citep[e.g.][]{Finkelstein2023_b}. Nevertheless, this comparison contextualizes our results. From this figure, we emphasize two key points: 1) as discussed in \cite{Katz2024_meg}, agreement with the stellar mass-halo mass relation at $z=0$ does not imply significant regulation at high redshift. Most of the progenitor haloes in our model collapse into a single object at $z=0$ and thus the mild offset we see for the efficient star formation run compared to the benchmark is not concerning. In fact, if we were able to run to $z=0$, the three models with stronger feedback may over-regulate the galaxy because the benchmark halo falls well within the $1\sigma$ scatter of the \cite{Behroozi2019} empirical model at $z=0$. 2) Our initial conditions represent a highly biased environment due to the engineered formation history. Compared to the other {\small VINTERGATAN} haloes, this particular object has a higher conversion rate of gas into stars at high redshift due to the early collapse. Again, our results do not preclude agreement with the $z=0$ stellar mass-halo mass relation (as shown in Figure~\ref{fig:smhm2}).

\begin{figure}
    \centering
    \includegraphics[width=\columnwidth]{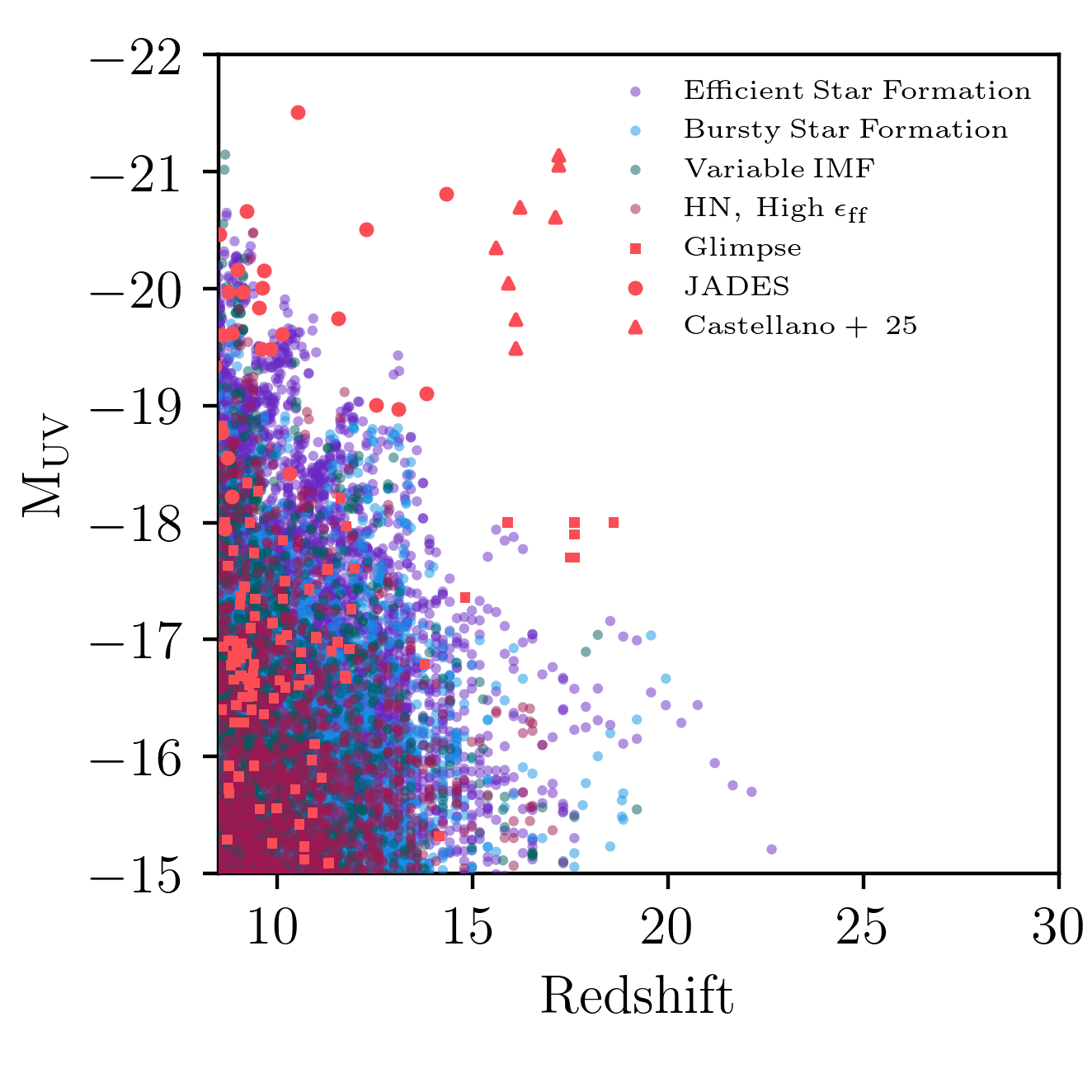}
    \caption{Intrinsic UV magnitude as a function of redshift for galaxies in the high-redshift suite simulations. For context, we also show spectroscopically confirmed high-redshift galaxies from the JADES survey \protect\citep{Carniani2024,Bunker2024,DEugenio2025} as well as the photometrically selected high-redshift candidate galaxies from the GLIMPSE survey \protect\citep{Kokorev2024,Chemerynska2025} and by \protect\cite{Castellano2025}. Note that we have excluded uncertainties on the photometric data points for clarity.}
    \label{fig:mag_z}
\end{figure}

Because our simulated Lagrange region represents that of a Milky Way mass halo, one particularly interesting question is when the progenitors first become observable. In Figure~\ref{fig:mag_z}, we show the UV magnitudes of bright resolved galaxies as a function of redshift for all high-redshift suite simulations. Note that here and throughout this work, the UV magnitude represents the intrinsic value at $1500$~\AA, considering both the stellar and nebular continuum\footnote{For such low mass galaxies, dust attenuation is not expected to be significant \citep[e.g.][]{Ma2018,Rosdahl2018}, but it will be considered in future work.}. For context, we also show spectroscopically confirmed high-redshift galaxies from the JADES survey \citep{Carniani2024,Bunker2024,DEugenio2025} as well as the photometrically selected high-redshift galaxy candidates from the GLIMPSE survey \citep{Kokorev2024,Chemerynska2025}, and by \cite{Castellano2025}. There is significant overlap in M$_{\rm UV}$-redshift space between the high-redshift photometric candidates from GLIMPSE and the simulated {\small MEGATRON} galaxies. All simulations except for the HN, High~$\epsilon_{\rm ff}$ run predict that certain Milky Way progenitors can reach UV magnitudes brighter than $-17$ at $z>17$, which is now within reach of the deepest JWST imaging surveys around lensing clusters. Thus, we may be able to detect galaxies that evolve into those like our own at a redshift representing $\sim1.5\%$ of the current age of the Universe. While this is extremely promising for the opportunity to constrain the origins of our own galaxy, it is clear from Figure~\ref{fig:mag_z} that JWST is currently detecting the absolute brightest progenitors that represent the tail end of the distribution for what is typical in such an environment at these redshifts.

\pagebreak
\section{The Spectral Diversity of High-Redshift Galaxies}
A key aspect of {\small MEGATRON} is the ability to predict the intrinsic spectra of galaxies. Across all four high-redshift simulations, we produce $>175,000$ spectra, accounting for the stellar continuum, nebular line emission, and nebular continuum emission. In this work, we ignore the role of dust absorption, re-emission, and scattering, all of which which will be considered in subsequent work (see Appendix~\ref{app:gf}). This is likely a reasonable assumption since we focus primarily on low-mass objects and most of the observed high-redshift galaxies exhibit $\beta$ slopes that are suggestive of minimal dust attenuation\footnote{Although blue UV slopes could also be due to a gray attenuation curve \citep[e.g.][]{McKinney2025}.} \citep[e.g.,][]{Cullen2024,Topping2024,Saxena2024}. 

\begin{figure*}
    \centering
    \includegraphics[width=\textwidth,trim={0.0cm 0cm 0.0cm 1cm},clip]{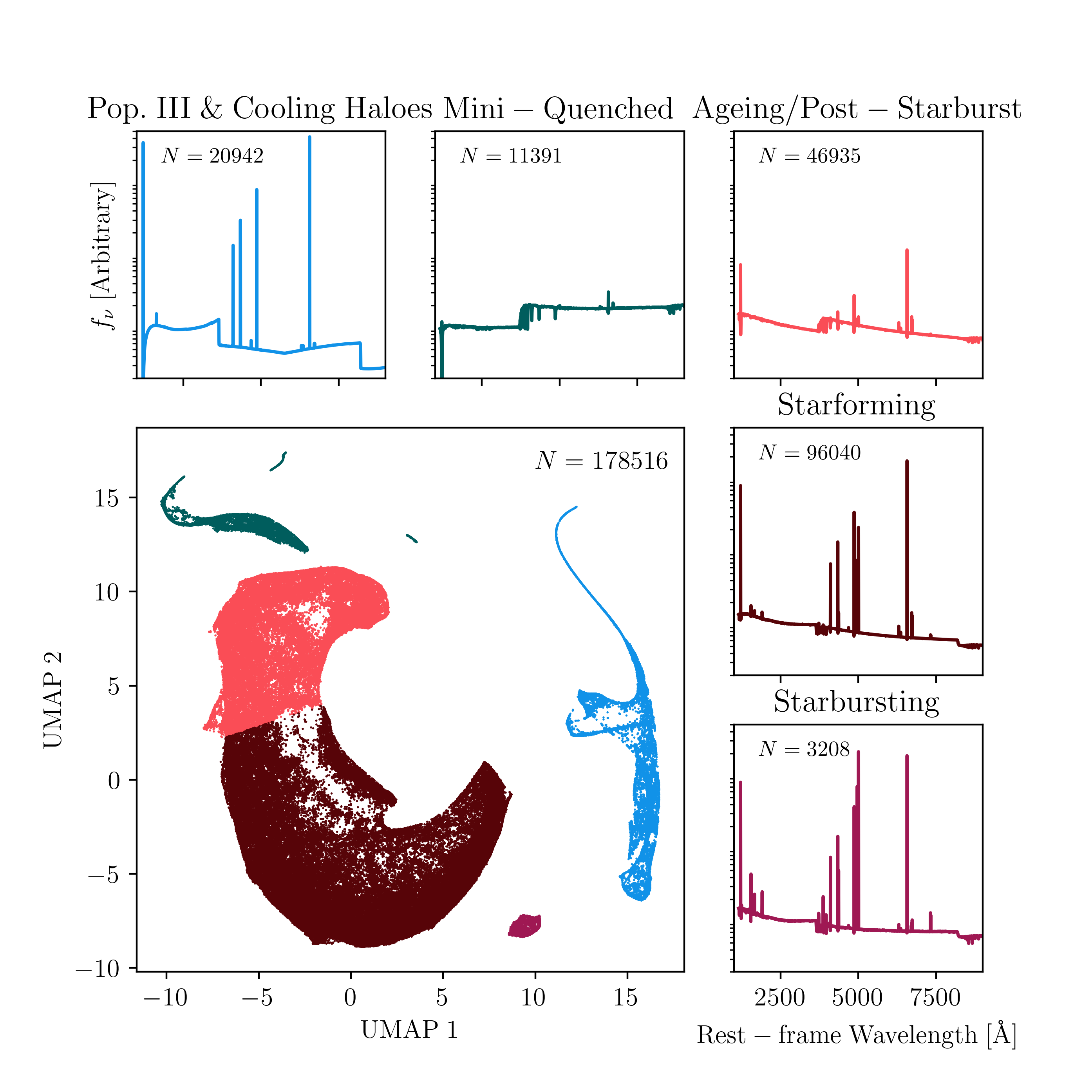}
    \caption{UMAP embeddings for all 178,516 spectra from the high-redshift suite. For each visually selected sample of points, we show the median stacked spectrum and identify the galaxy type. The number of galaxies in each sample is listed in each panel.}
    \label{fig:umap}
\end{figure*}

To qualitatively demonstrate the types of spectra that naturally emerge from our simulations, we perform dimensionality reduction and compute a 2D embedding for all $>175,000$ spectra in the suite using {\small UMAP}\footnote{We set the number of neighbours to 100, and a minimum distance of 0 for the UMAP decomposition.} \citep{umap1e,umap2}. This exercise aims to reveal the variety of spectral types in our simulations by reducing their high-dimensional representations to a 2D plane, preserving both local relationships and global structure. Spectra are all normalized to their value at rest-frame 3,000~\AA~in $f_{\nu}$ and embeddings are computed in the rest-frame. In Figure~\ref{fig:umap}, we show this 2D embedding and plot the median stacked spectra of visually selected regions. The types of galaxies we identify are:
\begin{enumerate}
    \item Pop.~III galaxies and cooling haloes that have strong emission H and He emission lines, no (or very weak) metal lines, and a dominant nebular continuum.
    \item Mini-quenched galaxies with very weak or no emission lines and a strong Balmer break.
    \item Post-starburst galaxies with weak emission lines and a Balmer break.
    \item Starforming galaxies with a blue UV continuum, and strong H, He, and metal emission lines.
    \item Starbursting galaxies with extremely strong H, He, and metal emission lines, a blue UV continuum, and a Balmer jump.
\end{enumerate}
The majority of galaxies in our simulations are normal starforming galaxies. These objects form a smooth continuum with the aging/post-starburst galaxies that represent the next most populous group. Various clusters of objects separate themselves from this connected continuum due to their peculiar spectral properties. In this section, we explore this diversity of spectral types in detail and place them in the context of recent JWST observations. 

\vspace{1cm}
\subsection{Population III Galaxies}
The first, clearly-isolated cluster in Figure~\ref{fig:umap} maps onto the spectra of galaxies dominated by Pop.~III stars (blue). This includes haloes with both live Pop.~III star formation and haloes that are cooling after the Pop.~III stars have died. The lack of metal lines, strong He~{\small II} emission, and dominant nebular continuum are clear signatures of Pop.~III physics.

\begin{figure}
    \centering
    \includegraphics[width=0.43\textwidth,trim={0cm 0.3cm 0cm 0.4cm},clip]{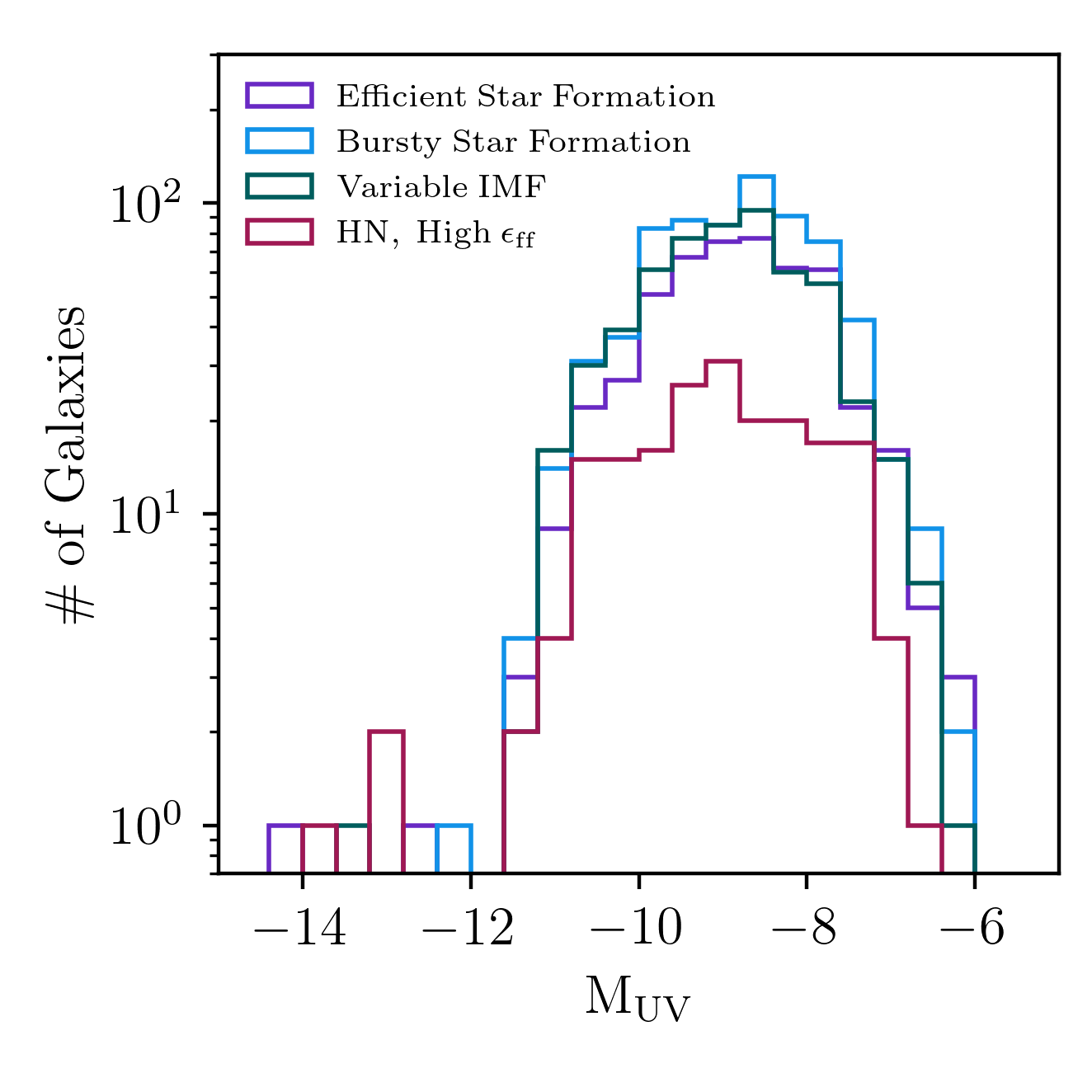}
    \includegraphics[width=0.43\textwidth,trim={0cm 0.3cm 0cm 0.4cm},clip]{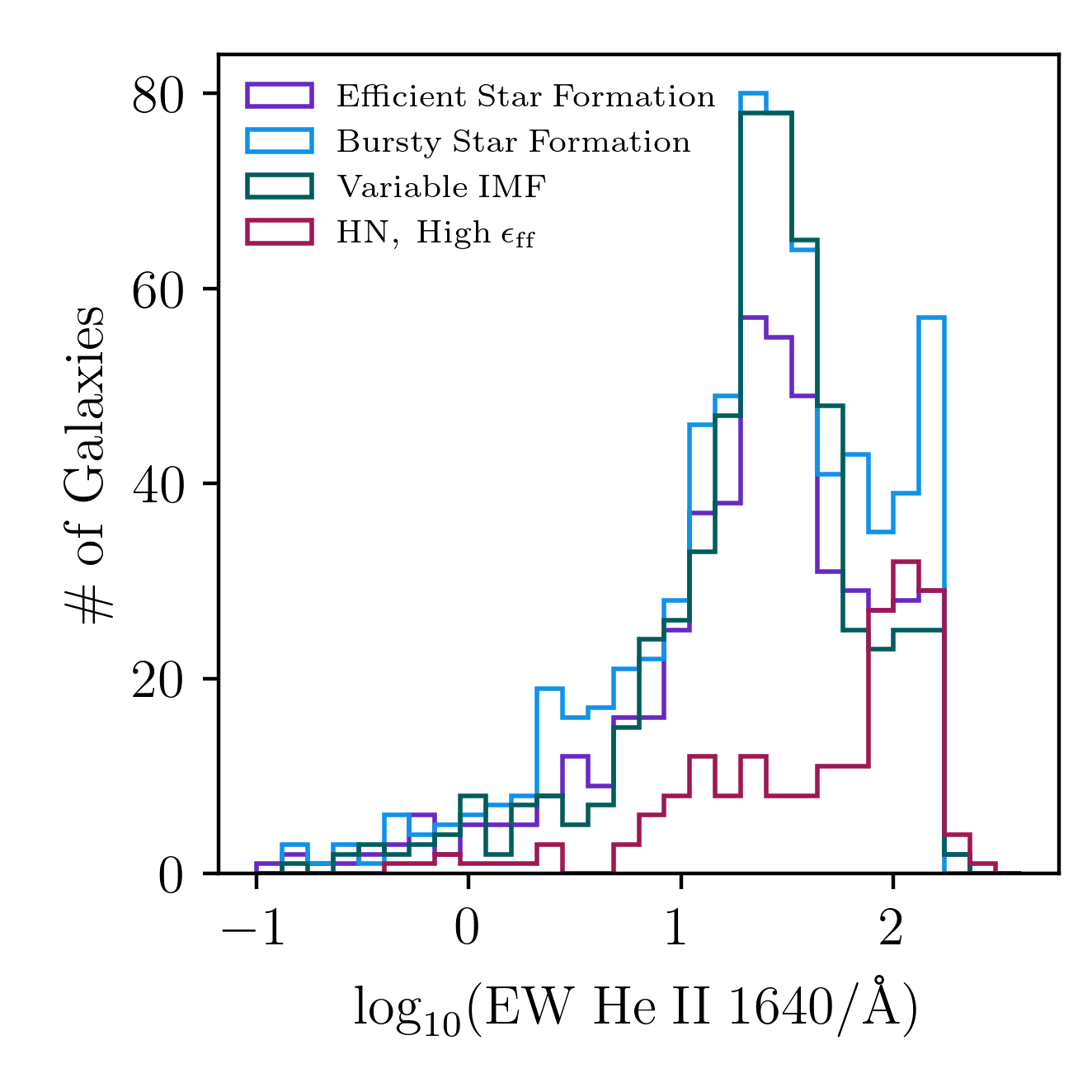}
    \includegraphics[width=0.43\textwidth,trim={0cm 0.3cm 0cm 0.4cm},clip]{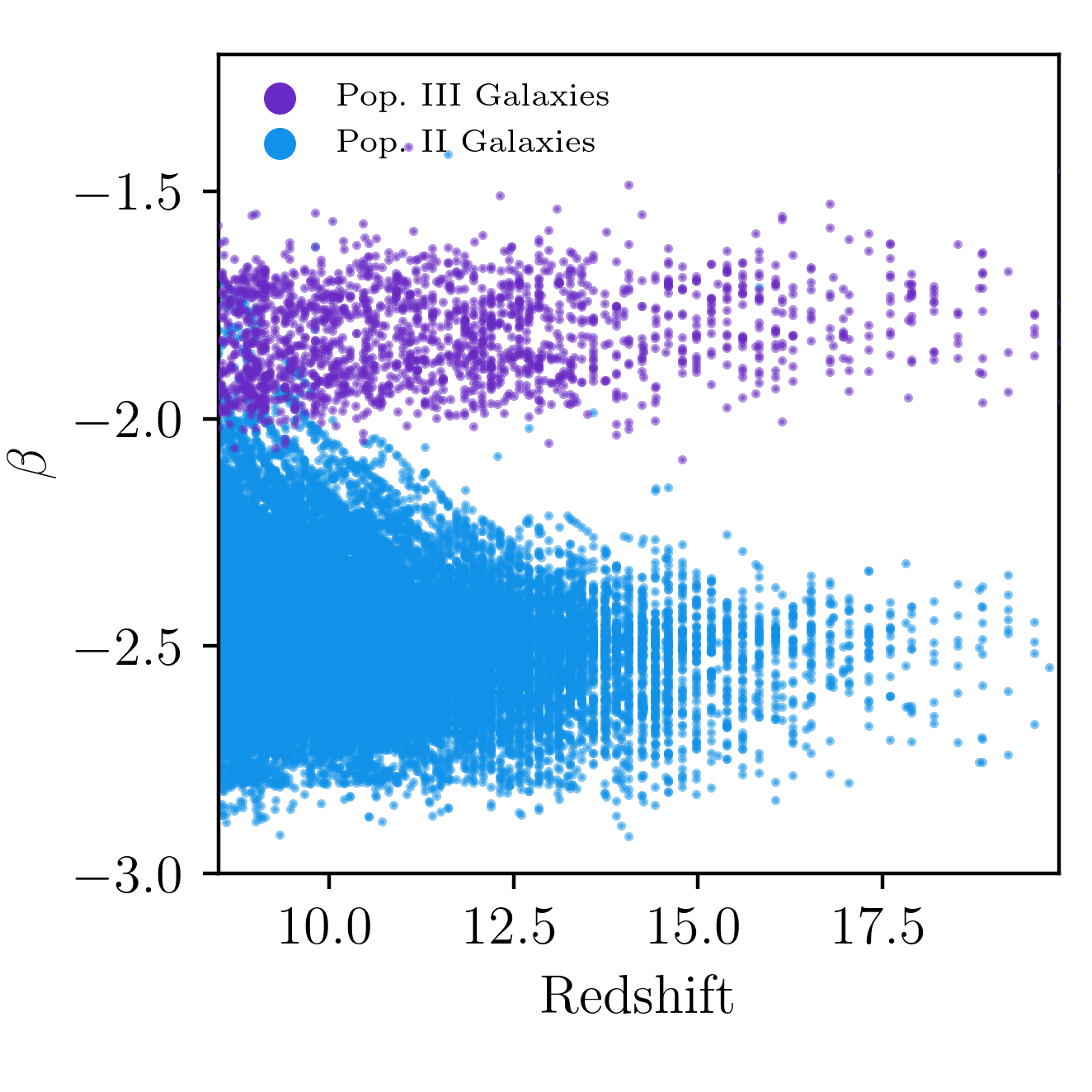}
    \caption{(Top) Histogram of the 1500~\AA{} magnitude of Pop.~III galaxies. (Middle) Histogram of the He~{\small II} EW of Pop.~III galaxies. (Bottom) Intrinsic UV slopes of Pop.~III (purple) and Pop.~II (blue) galaxies as a function of redshift. We only consider galaxies where Pop.~III stars have formed in the previous 1~Myr.}
    \label{fig:pop3_beta}
\end{figure}

The UV magnitudes of Pop.~III galaxies in our simulations tend to be very faint. In the top panel of Figure~\ref{fig:pop3_beta}, we show a histogram of 1500~\AA{} UV magnitudes of star-forming Pop.~III galaxies. The vast majority of Pop.~III stars in the simulations form in haloes close to the atomic cooling limit, either individually or in small groups, leading to UV magnitudes fainter than $-11$. Objects of this magnitude have yet to be observed at high redshift, with the faintest strongly-lensed object in the reionization era having an M$_{\rm UV}$ of $-12.3$ \citep{Vanzella2024,Nakajima2025b}.

Among all of the simulations, there are seven Pop.~III galaxies that break the M$_{\rm UV}<-12$ barrier, reaching magnitudes as bright as $-14.4$, well within the reach of JWST surveys leveraging gravitational lensing \citep[e.g.][]{Fujimoto2025}. This is highly encouraging for the prospects of detection, even if such systems are extremely rare (7/20,942). \cite{Storck2026-meg} explore in detail the physical conditions required to produce bright Pop.~III galaxies and their sensitivity to galaxy formation assumptions. 

However, differentiating Pop.~III galaxies from their low-metallicity Pop.~II counterparts remains a key concern for robustly detecting them. This is especially true at high redshifts (e.g., at $z\gtrsim10$) where strong rest-frame optical metal lines drop out of the NIRSpec PRISM and one must rely on lower equivalent width UV lines. He~{\small II}~1640~\AA\ has long-been considered one of the key signatures of massive Pop.~III stars \citep[e.g.,][]{Tumlinson2000,Oh2001,Schaerer2002}, but as we show in the middle panel of Figure~\ref{fig:pop3_beta}, not all Pop.~III galaxies in {\small MEGATRON} have high He~{\small II} EWs. A lack of He~{\small II}~1640~\AA\ does not imply that a galaxy is not a Pop.~III system. This is a direct result of the fact that not all Pop.~III stars are expected to be massive and have extreme surface temperatures $\gtrsim10^5$~K \citep[e.g.,][]{Clark2011,Hirano2014,Stacy2016}. While the EWs are primarily sensitive to the input SED, which depends on our chosen IMF, star formation history and ISM properties also impact this quantity. It should also be noted that strong, narrow He~{\small II}~1640~\AA\ can also be generated by sources other than Pop.~III stars, for example AGN or Wolf-Rayet stars with weak winds \citep[e.g.,][]{Grafener2015}, and thus this particular diagnostic is not necessarily definitive.

One of the key differentiating factors between Pop.~III and Pop.~II galaxies that we find in the simulations is that Pop.~III galaxies are generically redder. This is counterintuitive because the intrinsic UV slopes of massive Pop.~III stars are extremely blue, with values much steeper than $-3$ \citep[e.g.,][]{Schaerer2002,Tumlinson2003,Larkin2023}. However, Pop.~III galaxies represent a rare class of systems where the nebular continuum can outshine the stars in the UV which leads to redder slopes. This has been demonstrated previously using photoionization models \citep[e.g.,][]{Raiter2010,Trussler2023,Katz2024_BJ}, where the effect highly depends on the electron density of the nebula due to the relatively low critical density of H~{\small I} two-photon emission. As we show in the bottom panel of Figure~\ref{fig:pop3_beta}, the UV slopes of Pop.~III galaxies are typically $\gtrsim-2$, which is much redder than the standard Pop.~II starforming galaxies in {\small MEGATRON} at the same redshift. In fact, the intrinsic Pop.~II galaxy slopes only become this red at lower redshifts when the stars have aged enough. For this reason, photometric selection cuts for high-redshift galaxies should not exclude systems with UV slopes $\gtrsim-2$.

For the same reason that Pop.~III galaxies appear redder than Pop.~II galaxies --- i.e.\ due to their strong nebular continuum emission --- they also exhibit a downturn in the UV, blueward of Ly$\alpha$, which is a signature of intense two-photon emission. Such a feature has rarely been discussed in the literature \citep[although see][]{Dijkstra2009,Raiter2010,Katz2024_BJ}, but has gained prominence more recently due to its possible detection at high redshift \citep[][albeit with significant metal line emission]{Cameron2024,Katz2024_BJ,Witstok2024}.

\begin{figure}
    \centering
    \includegraphics[width=0.45\textwidth]{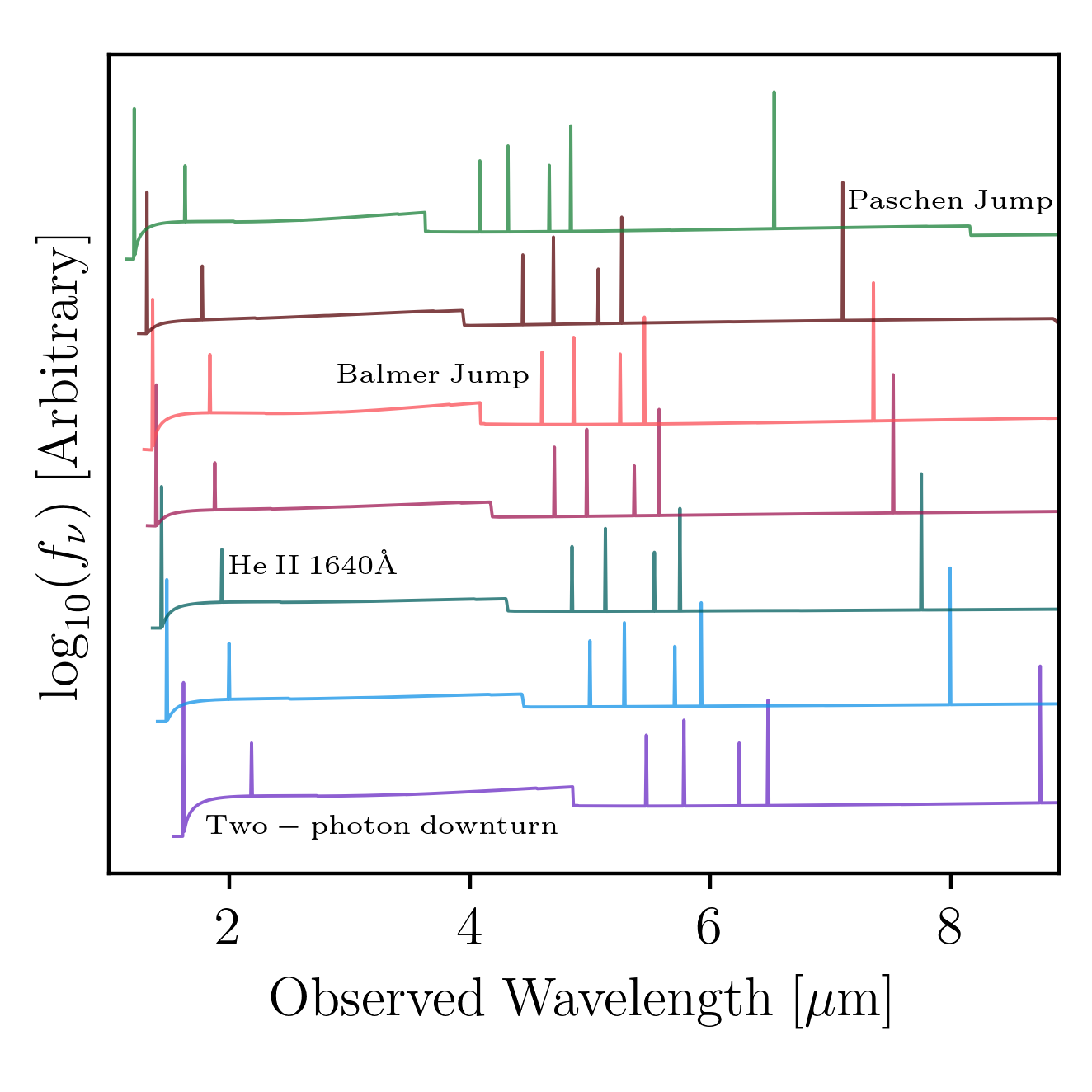}
    \caption{Seven example bright Pop.~III galaxy spectra. Spectra are shown in $f_{\nu}$ and have been renormalized onto an arbitrary scale for clarity. The brightest Pop.~III systems all show strong He~{\small II} emission in the UV and optical, a Balmer (and Paschen) jump, and a downturn in the UV just redward of Ly$\alpha$ from two-photon emission.}
    \label{fig:pop3_spec}
\end{figure}

Many of the spectral features discussed in this section can be seen in Figure~\ref{fig:pop3_spec} where we show the spectra of seven example bright Pop.~III galaxies. These systems all show strong He~{\small II} emission in the UV and optical, a Balmer (and Paschen) jump, and a downturn in the UV from two-photon emission. The different strengths of the UV downturn depend on the electron density near the star particles and the fact that the spectra appear flat or slightly rising in $f_{\nu}$ indicates that the spectral slopes are redder than $\sim-2$.

\begin{figure}
    \centering
    \includegraphics[width=0.45\textwidth,trim={0cm 0.0cm 0cm 0.0cm},clip]{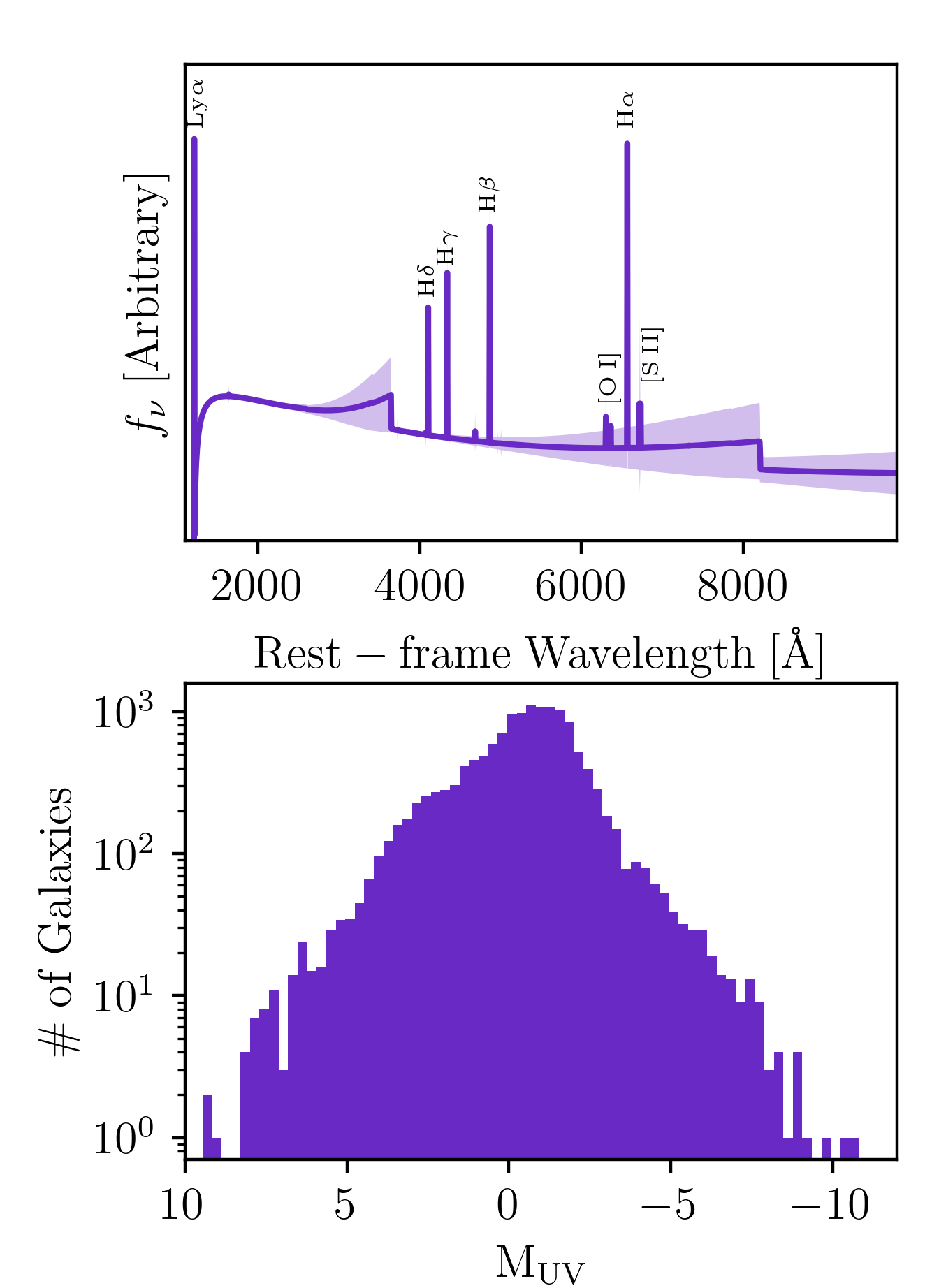}
    \caption{(Top) Median stacked spectra of cooling haloes (pure line emitters) across all simulations. The shaded region represents the $1\sigma$ standard deviation of the population. (Bottom) Histogram of the UV magnitudes of the cooling haloes.}
    \label{fig:ples}
\end{figure}

If the energy injection from the Pop.~III radiation and SNe is high enough, star formation can be quenched and cooling radiation from the hot gas may also be detectable. This has been considered previously in the more general case of cooling clouds \citep[e.g.,][]{Dijkstra2009} and also in the case of Pop.~III remnant galaxies \citep{Katz2023_Pop3}. In Figure~\ref{fig:ples}, we show a stacked spectrum of $>10,000$ galaxies (across all high-redshift simulations) that are cooling after having a Pop.~III star formation event (pure line emitters). Their spectra are characterized by having no stellar emission and are thus purely nebular line and continuum emitters. This means that the equivalent widths of the lines are extremely strong (thousands of \AA\ for H$\alpha$ and H$\beta$ depending on temperature, see Appendix~A of \citealt{Katz2024_BJ}) and the nebular two-photon and Balmer/Paschen jump features are highly visible. The strength of the Balmer/Paschen jump and the line EWs provides a strong constraint on the temperature of the cooling halo. Since a subset of the pure line emitters were heated by SN, there are also metals present in the system. For some galaxies we see faint O~{\small I}, O~{\small II}, O~{\small III}, Ne~{\small III}, and S~{\small II} lines. The lowest ionization state lines tend the be the brightest of the metal lines and we highlight the notable lack of He~{\small II} emission. 

In most cases, these cooling haloes have ${\rm M_{UV}}=0$ (see the bottom panel of Figure~\ref{fig:ples}), well beyond current capabilities. But the extreme tail of the UV magnitude distribution reaches $-11$ which is within reach of JWST when coupled with strong gravitational lensing.

\subsection{Starbursting, Extreme Emission Line Galaxies}
As the star formation rates within galaxies rise after the transition from Pop.~III to Pop.~II star formation due to the enhanced cooling from metals compared to H$_2$, a small fraction of galaxies undergo catastrophic bursts of star formation. Such bursts manifest spectroscopically as extreme emission line galaxies (EELGs). A stacked example of EELGs is shown as the red clump in the bottom right panel of Figure~\ref{fig:umap}. While EELGs have been observed at all redshifts, for example blueberries at $z\sim0$ \citep{Yang2017}, green peas at $z\sim0.3$ \citep{Cardamone2009}, as well as others up to $z\sim1$ \citep{Amorin2015}, they are certainly more common at high redshifts \citep{Boyett2024,Davis2024,Llerena2024}.

\begin{figure}
    \centering
    \includegraphics[width=0.45\textwidth,trim={0cm 0.3cm 0cm 0.4cm},clip]{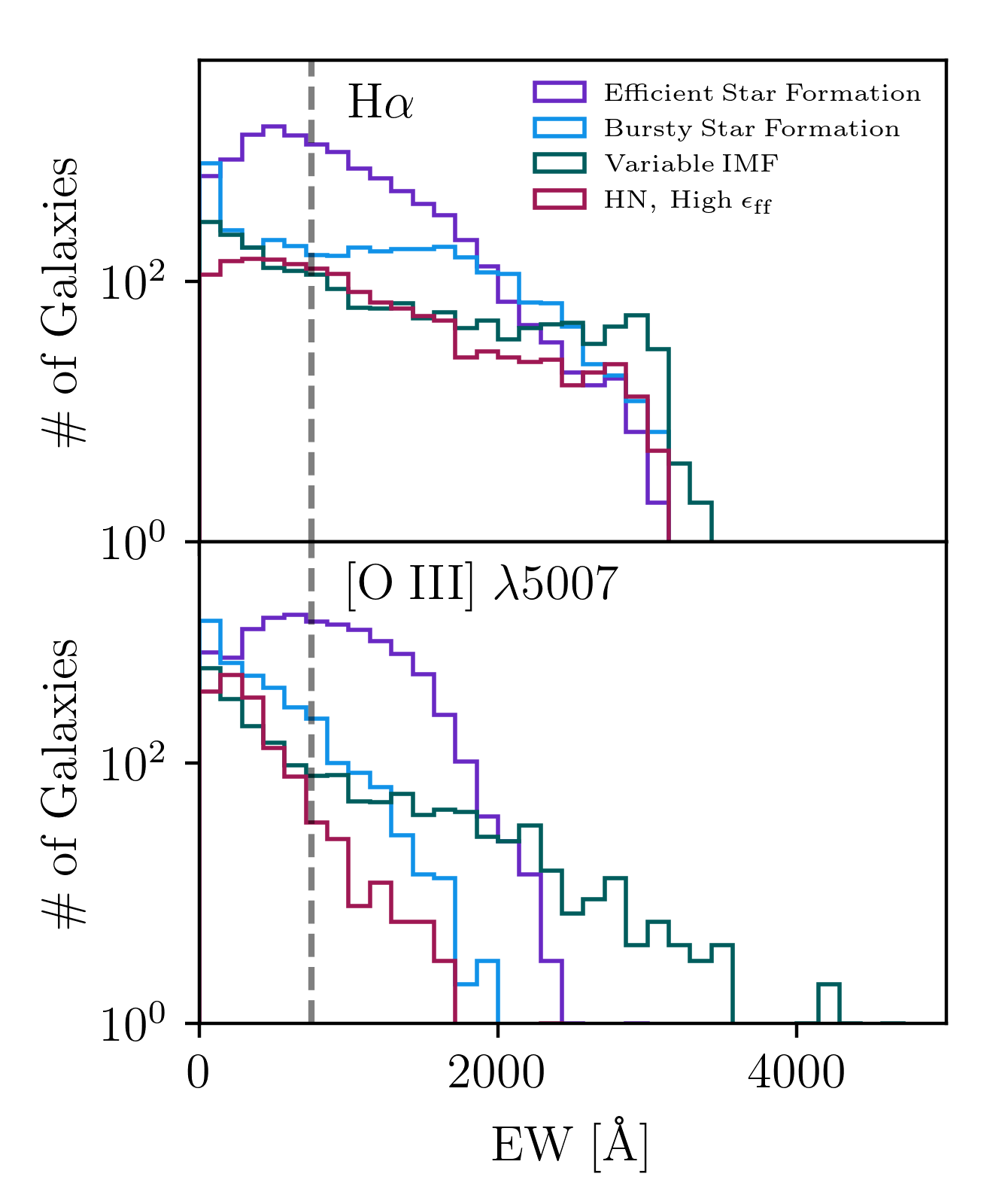}
    \caption{Histograms of H$\alpha$ (top) and [O~{\small III}]~$\lambda$5007 (bottom) equivalent width for galaxies with UV magnitudes brighter than $-15$. The dashed vertical line represents an EW of 750~\AA~which is our threshold for an EELG.}
    \label{fig:EELG_EW}
\end{figure}

\begin{figure*}
    \centering
    \includegraphics[width=\textwidth,trim={0cm 0.0cm 0cm 0.0cm},clip]{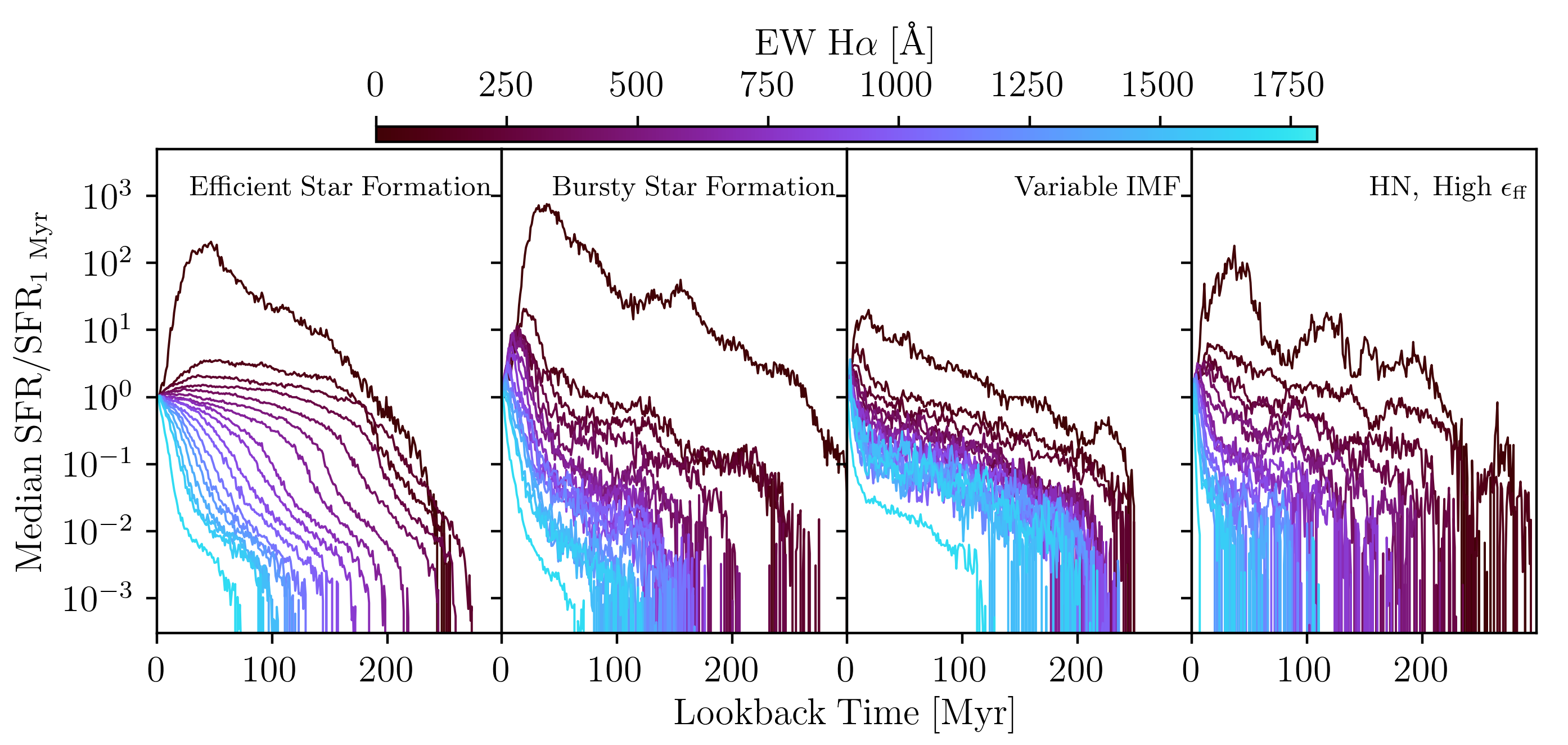}
    \includegraphics[width=\textwidth,trim={0cm 0.0cm 0cm 0.0cm},clip]{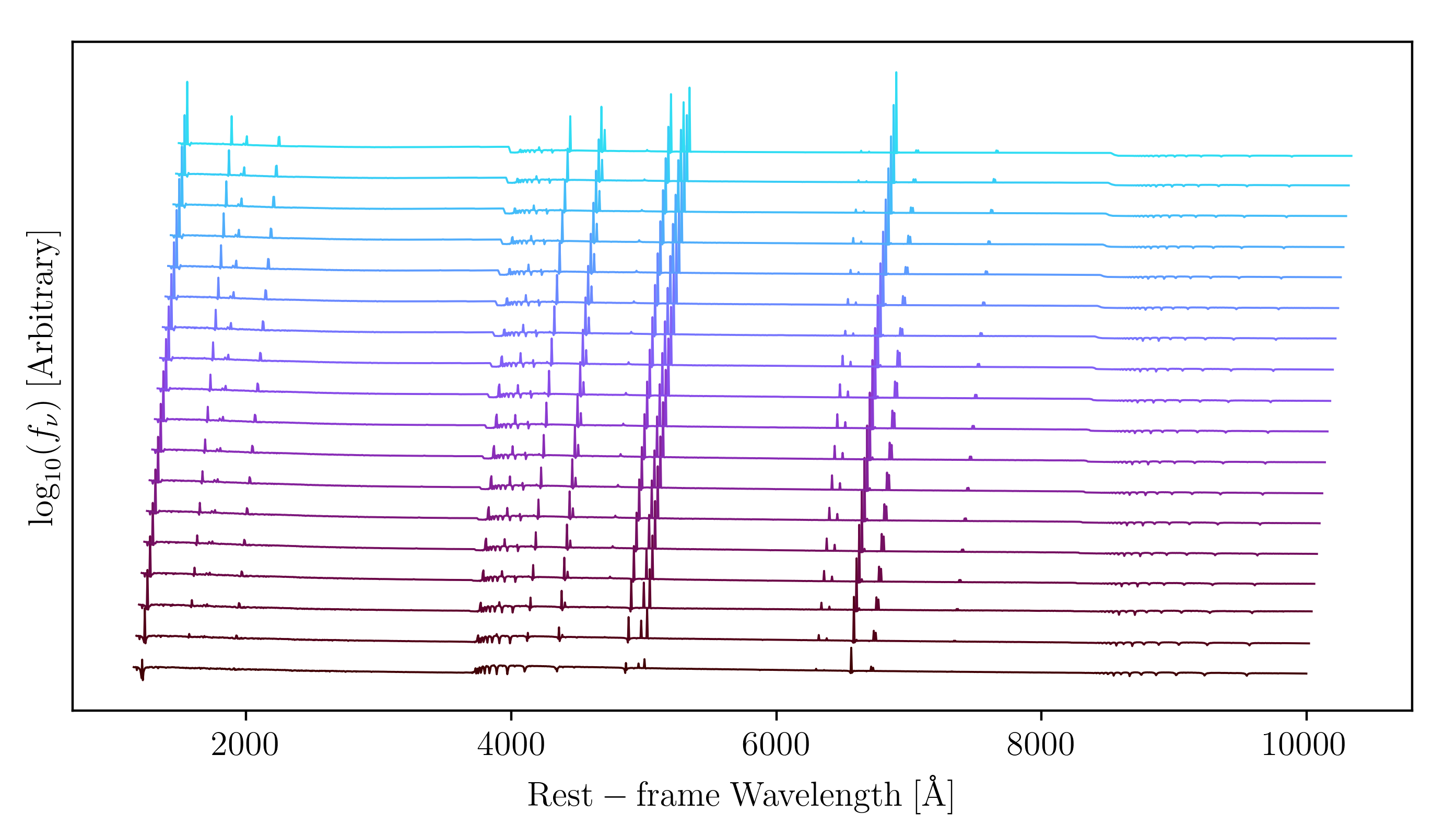}
    \caption{(Top) Median normalized star formation histories of bright galaxies with different H$\alpha$ EWs in each simulation. The lines are coloured by the minimum H$\alpha$ EW in the bin. (Bottom) Normalized stacked spectra for galaxies in each H$\alpha$ EW bin in the bursty star formation simulation. For clarity, we have offset each spectrum by 20~\AA~from the previous. In both panels, we consider only galaxies with M$_{\rm UV}<-15$. Over a 50~Myr period, in the most extreme cases, the median stacked star formation histories can differ by a factor of $>10^4$ between the different EW bins. Such drastic changes in SFR are reflected in the shape of the continuum and in the strength of the emission lines.}
    \label{fig:EELG_SFH}
\end{figure*}

\vspace{5mm}
\subsubsection{Emission Line Properties of EELGs}

The {\small MEGATRON} simulations produce numerous EELGs, as shown in Figure~\ref{fig:EELG_EW}, where we plot histograms of H$\alpha$ (top) and [O~{\small III}]~$\lambda$5007 (bottom) equivalent widths for galaxies with UV magnitudes brighter than $-15$. Following \cite{Boyett2024}, we define an EELG as having an H$\alpha$ or [O~{\small III}]~$\lambda$5007 EW $>750$~\AA. The fraction of bright galaxies that would be considered EELGs are similar across all runs at $\sim50\%$\footnote{Note that the EELG fraction is highly sensitive to the stellar mass distribution.}; however, the distributions of EWs are very different. For example, the efficient star formation H$\alpha$ distribution is peaked at close to 750~\AA\ while the bursty star formation and variable IMF models have much flatter distributions. Because the efficient star formation simulation forms more stars overall compared to the other two models it produces the most EELGs. 

A key result is that in the variable IMF simulation, [O~{\small III}] EWs can reach $>4000$~\AA. This occurs for galaxies in the magnitude range $\lesssim-19$ undergoing extreme bursts of star formation. Simulations historically struggle to reproduce the tail of high [O~{\small III}] EWs observed at high redshift \citep[e.g.,][]{Ceverino2021,Wilkins2023}. However, two physical effects differentiate the variable IMF model from the other simulations. First, HN in the variable IMF simulation inject a significant amount of oxygen which can increase the EW in scenarios where the oxygen emission is limited by metallicity\footnote{One may assume that the same physics would lead to higher [O~{\small III}] EWs in the HN, High~$\epsilon_{\rm ff}$ simulation; however, the stars form at much lower density in this run. Extreme densities are a requirement for forming the most extreme EELGs.} and not by strong cooling. Second, the stellar populations in the variable IMF simulation have higher mass loss which suppresses the stellar continuum at older ages. Both of these effects combined may help explain the high [O~{\small III}] EW galaxies seen in observations.

\subsubsection{Star Formation Histories of EELGs}

Analyzing the star formation histories of EELGs demonstrates how extreme they are. In the top panel of Figure~\ref{fig:EELG_SFH} we show the median, normalized (by their most recent 1~Myr-averaged SFR), stacked star formation histories for galaxies with different H$\alpha$ EW for three of the high-redshift suite simulations. For this exercise, we only consider galaxies with M$_{\rm UV}<-15$. While the EELG population with H$\alpha$~EW~$>1,800$~\AA~sees a rapid rise in SFR where the current value is typically $>100$ times more than what it was 20~Myr ago, an ordinary bright galaxy with an H$\alpha$ EW~$\sim400$~\AA~sees almost no difference in SFR over the past 20~Myr. In contrast, the lowest H$\alpha$ EW galaxies have significant downturns in their recent star formation of up to a factor of 1,000 in the past 50~Myr in the most extreme examples. From this figure, we see how the H$\alpha$ EW is an extremely strong predictor for the typical SFH and sSFR of a galaxy (see also e.g., \citealt{Fumagalli2012,MQ2016,Khostovan2024}). While the details of the SFH depend on the feedback and star formation model (compare the SFHs in the three panels of Figure~\ref{fig:EELG_SFH}), in all cases EELGs require an extreme recent burst of star formation.

\subsubsection{Continuum Properties of EELGs}

The differences in the emission line properties of EELGs are also reflected in the continuum features of the spectra. In the bottom panel of Figure~\ref{fig:EELG_SFH} we show the normalized, stacked spectra for galaxies in bins of H$\alpha$ EW for the bursty star formation model\footnote{The other models are not noticeably different.}. The most noticeable difference of the EELGs compared to lower EW galaxies (apart from the strength of the emission lines) occurs near H$_{\infty}$ at $\sim3645$~\AA. 

As discussed in \cite{Raiter2010,Byler2017,Katz2024_BJ}, a unique property of EELGs, and more generally galaxies with very young stellar populations, is that the nebular continuum emission can contribute significantly in both the rest-frame UV and optical. The nebular continuum consists of three components: free-free, free-bound, and two-photon emission. While free-free emission tends to be subdominant, free-bound and two-photon emission can have substantial impacts on galaxy spectra. More specifically, free-bound emission imparts a characteristic jump at 3645~\AA\ due to recombination directly to the $n=2$ state of hydrogen. Two-photon emission, caused by $2s\rightarrow1s$ transitions in neutral hydrogen peaks in the UV and has a characteristic downturn towards Ly$\alpha$, which looks similar to a damped Ly$\alpha$ system. In general, the nebular continuum causes the UV magnitude to increase while the spectra reddens \citep[e.g.,][]{Bouwens2010,Dunlop2013}. This is because the H~{\small I} free-bound increases towards redder wavelengths, while the two-photon continuum peaks in the rest-frame UV. The nebular continuum is amplified when allowing the mass of the most massive star in the IMF to increase \citep{Raiter2010,Katz2024_BJ,Schaerer2025}.

\begin{figure}
    \centering
    \includegraphics[width=0.45\textwidth,trim={0cm 0.0cm 0cm 0.0cm},clip]{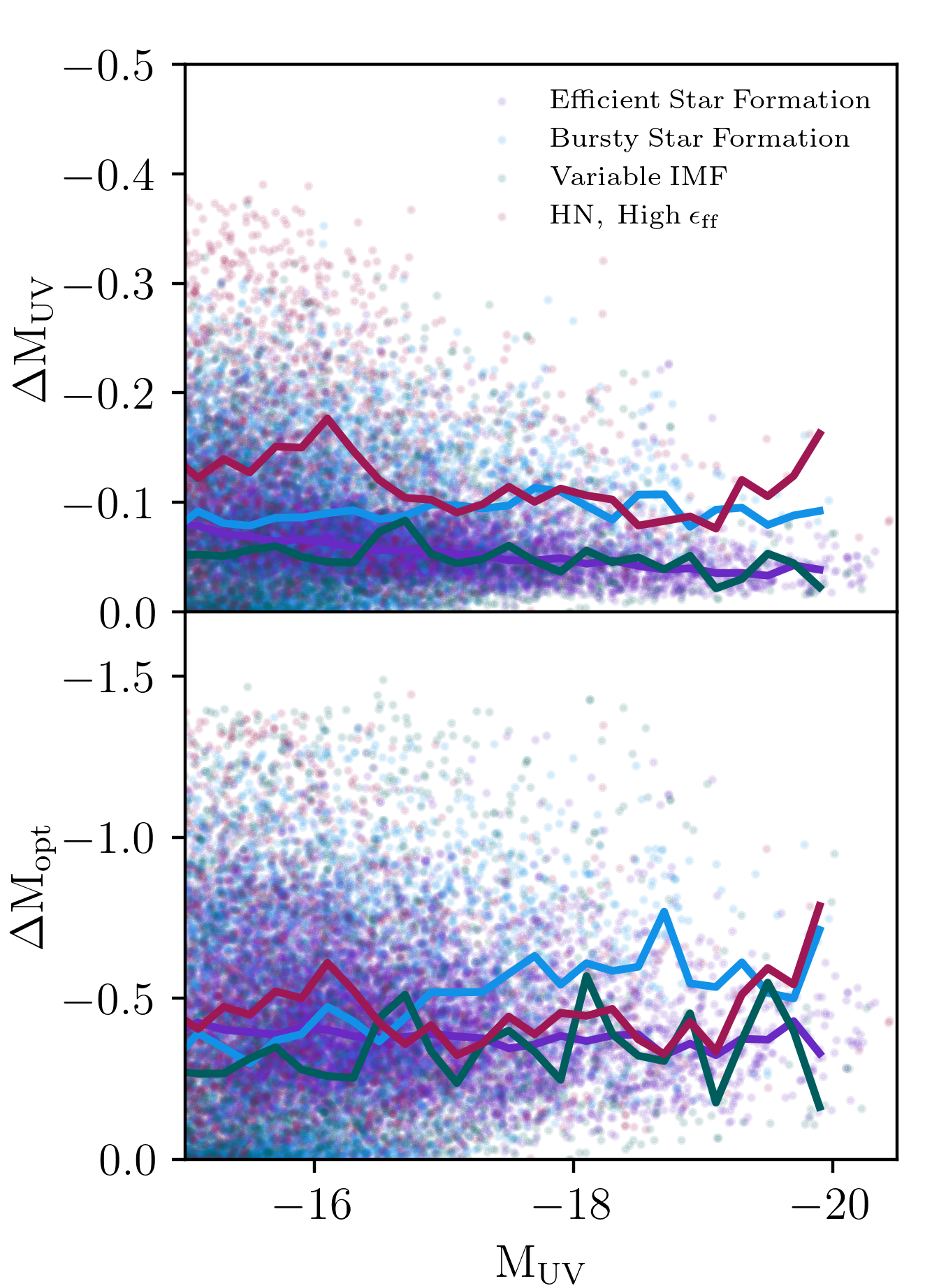}
    \caption{Contribution of the nebular continuum to the UV magnitude at 1500~\AA\ (top) or the optical magnitude at 3640~\AA\ (bottom) as a function of intrinsic UV magnitude. We consider all galaxies with M$_{\rm UV}<-15$ at all redshifts. In both panels, solid lines represent the median trends as a function of UV magnitude.}
    \label{fig:neb_cont}
\end{figure}

The change in magnitude of the spectrum of each galaxy at 1500~\AA\ and 3640~\AA\ due to the addition of the nebular continuum as a function of intrinsic UV magnitude is shown in Figure~\ref{fig:neb_cont}. Focusing on a UV magnitude of $\sim-18$, there is a wide diversity in nebular contributions. For the typical bright ($M_{\rm UV}<-15$) galaxy in the efficient star formation and variable IMF models, the change in $M_{\rm UV}$ is $\sim0.05$ magnitudes which would have no impact on the UV luminosity. This increases to $\sim0.1$ magnitudes for the other two models. However, for the galaxies undergoing the most extreme starbursts, the effect can be as high as 0.4 magnitudes for our assumed SSP models. The impact of the nebular continuum on the UV magnitudes of high-redshift galaxies therefore must not be ignored, especially for the the most extreme systems which are typically the first to be detected.

The strength\footnote{Here we refer to both the overall normalization and with respect to the stellar continuum} and shape of the nebular contribution to the overall spectrum is primarily sensitive to three parameters: the specific star formation rate, the density of the ISM, and $\xi_{\rm ion}$. The nebular continuum is strongest when $\xi_{\rm ion}$ is maximized (when there are more ionizing photons per unit 1500~\AA{} luminosity), the LyC escape fraction is low (such that all ionizing photons are converted into nebular continuum), and the density is low ($l$-changing collisions shift electrons from the $2s$ to $2p$ state, reducing two-photon emission). In our models, $\xi_{\rm ion}$ is limited by our adopted SSP models; however, if high-redshift stars are hotter either due to lower metallicity and being more massive \citep{Schaerer2002}, or due to peculiar chemical abundance patterns \citep{Katz2024_He}, the nebular contribution, particularly in the UV, can become much more significant. Most stars form in embedded in dense clouds so LyC leakage rarely impacts the nebular continuum. However, the ISM in our models can reach densities of $\gtrsim10^5\ {\rm cm^{-3}}$, particularly in galaxies undergoing extreme bursts of star formation, which is above the critical density for H~{\small I} two-photon emission. This latter effect leads to a wide diversity in spectral slope. If we consider the strongest EELGs with H$\alpha$ EW~$>1500$~\AA{}, we find a trend such that intrinsic UV slope scales with the mean gas density of a galaxy as $\beta=0.038\log_{10}({\rm n_H/cm^{-1}})-2.58$, i.e.\ galaxies with denser gas have redder slopes. Our simulations clearly demonstrate how the detailed properties of the ISM impact not only the emission lines but also the shape of the observed continuum.

\subsection{Starforming Galaxies}
The overwhelming majority (more than half) of galaxies in our simulations can be described as being starforming\footnote{This definition is qualitative an entirely based on the UMAP decomposition}. A stacked spectra of these galaxies is shown in brown in Figure~\ref{fig:umap}. These galaxies have strong emission lines and blue UV slopes, and represent the less-extreme counterparts to the galaxies described in the previous section.

\begin{figure}
    \centering
    \includegraphics[width=0.45\textwidth,trim={0cm 0.0cm 0cm 0.0cm},clip]{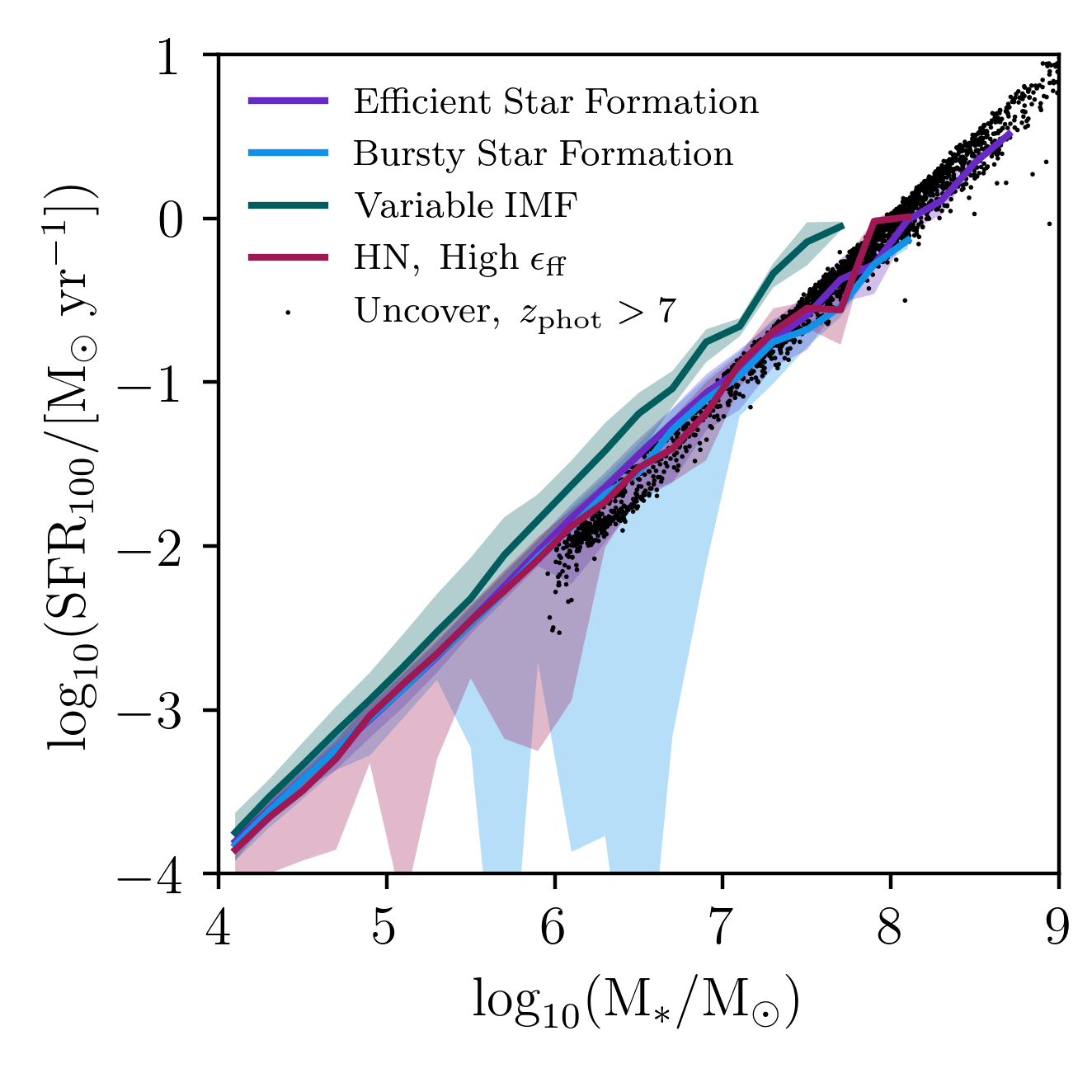}
    \caption{Star formation main-sequence averaged over 100~Myr for all galaxies in each high-redshift suite simulation. We show the median and $1\sigma$ scatter for each simulation. For comparison, we show SED fitting results from photometrically selected $z>7$ galaxies from the {\small UNCOVER} survey. }
    \label{fig:sfms_100}
\end{figure}

The starforming galaxy population exhibits less extreme scatter in their star formation rate as a function of stellar mass, which maintains a blue $\beta$, but lowers the EWs. Because of the less extreme changes in SFR, these galaxies form a tight main-sequence \citep[e.g.][]{Speagle2014,Schreiber2015}. This main-sequence has now been characterized up to $z\sim10$ with JWST data \citep[e.g.][]{RB2024}.

{\small MEGATRON} galaxies also exhibit a tight main-sequence, especially when average over long ($\sim100$~Myr) time scales. We show this main-sequence and its $1\sigma$ scatter for each high-redshift simulation in Figure~\ref{fig:sfms_100}. Nearly independent of feedback model and star formation criteria, the galaxies all follow the same main sequence, which is in excellent agreement with SED fitting results from photometrically selected $z>7$ galaxies from the {\small UNCOVER} survey \citep{Wang2024,Weaver2024,Suess2024,Furtak2023}. The exception to this trend is the variable IMF simulation where the enhanced mass loss causes an offset towards lower stellar mass. The SED fitting codes typically assume a fixed IMF and thus it is expected that the variable IMF simulation is offset.

Even though the median star formation main sequence are consistent across all runs with a fixed stellar IMF, the scatter in the relation depends substantially on the star formation and feedback model. For example, at stellar masses below $10^7\ {\rm M_{\odot}}$, the scatter becomes extremely large for the bursty star formation simulation. This is because the strong feedback can suppress star formation for time scales longer than 100~Myr. Hence the ratio of galaxies on and off the main-sequence is sensitive to the physics, even if the location of the main-sequence is not. The galaxies that have suppressed SFRs will be discussed in the next sections.

\subsection{``Mini-Quenched'' Galaxies}
The star formation rates required to power an EELG are unsustainable for long periods of time. This is because these galaxies rapidly deplete their ISM gas reservoirs through star formation and intense, temporally coincident feedback can expel the residual ISM gas from the central parts of the galaxy. A sample of ``Mini-Quenched'' galaxies that have seen a relatively rapid and sustained drop in their SFR have been identified by JWST for having blue UV slopes, no emission lines, and Balmer absorption features \citep[e.g.,][]{Looser2024,Trussler2025}. The {\small MEGATRON} simulations contain a small population of such systems and their stacked spectra are shown in green in Figure~\ref{fig:umap}. 

\begin{figure}
    \centering
    \includegraphics[width=0.45\textwidth,trim={0cm 0.0cm 0cm 1.0cm},clip]{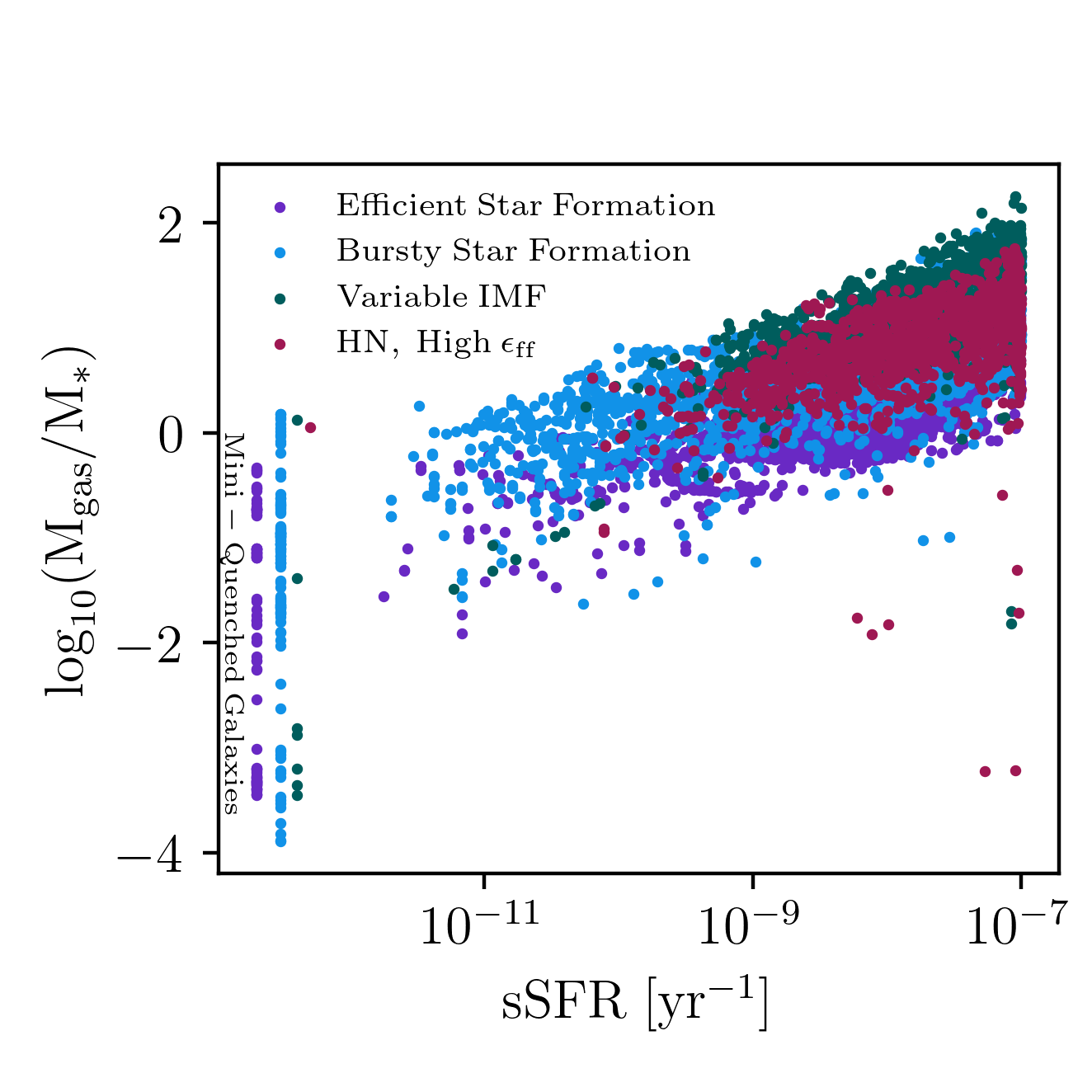}
    \caption{Ratio of gas to stellar mass as a function of sSFR for bright galaxies (M$_{\rm UV}<-15$) in the high-redshift simulations. We have put a floor in sSFR at $10^{-13}\ {\rm yr^{-1}}$ (and slightly separated different simulations for visibility) so that the quenched systems are visible in this diagram. To calculate sSFR, we use the 10~Myr-averaged SFR.}
    \label{fig:miniq}
\end{figure}

Mini-quenched galaxies can be selected by having a low  ratio of gas to stellar mass and a near-zero sSFR as shown in Figure~\ref{fig:miniq}. Low values of sSFR indicate that the galaxies are not currently undergoing any star formation and a low ratio of gas to stellar mass selects systems that have expelled all of their gas. Indeed, a population of galaxies emerges with extremely low sSFR (consistent with being 0) and very low gas fractions. 

\begin{figure}
    \centering
    \includegraphics[width=0.45\textwidth,trim={0cm 0.0cm 0cm 0.0cm},clip]{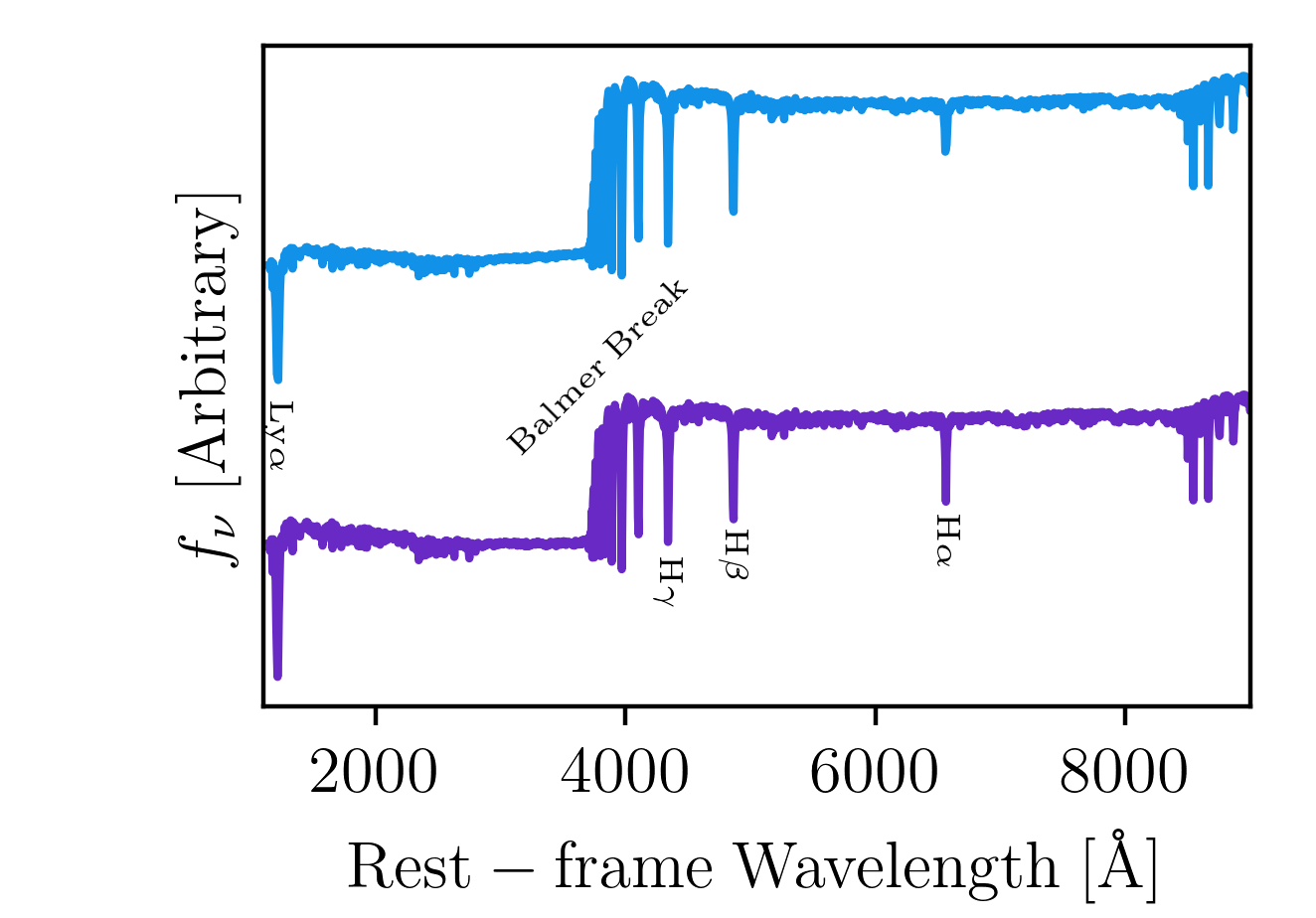}
    \includegraphics[width=0.45\textwidth,trim={0cm 0.0cm 0cm 1.0cm},clip]{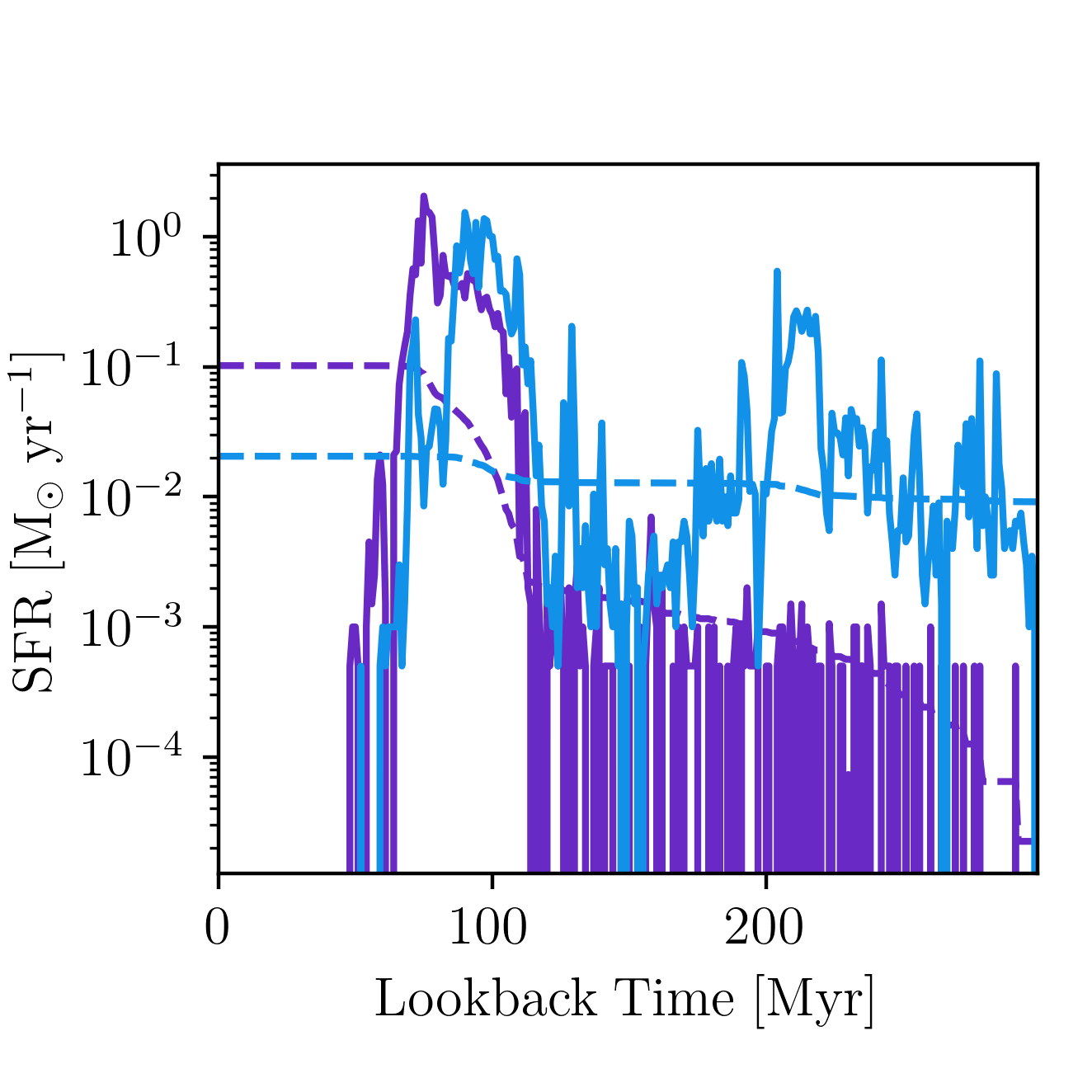}
    \caption{(Top) Spectra of two example mini-quenched galaxies. These spectra come from the efficient star formation and bursty star formation simulations. (Bottom) Star formation histories of the same two galaxies over the past 300 Myr. The dashed lines show the SFR for galaxies of the same mass that are on the star formation main-sequence.}
    \label{fig:miniq_spec}
\end{figure}

Two example mini-quenched galaxy spectra are shown in the top panel of Figure~\ref{fig:miniq_spec}. These two galaxies have halo virial masses of $10^{8.6}$ and $10^9$~M$_{\odot}$, respectively, and gas masses that are $\lesssim1\%$ of the cosmic baryon fraction. The low gas fractions and SFR result in a UV slope close to $-2$ (i.e.\ flat in $f_{\nu}$). Neither nebular emission lines nor the continuum are visible in these spectra. The bottom panel of Figure~\ref{fig:miniq_spec} shows the star formation histories of the two galaxies where a sharp drop in the SFR can be seen $70-80$~Myr ago. The extended period of no star formation is required to reach a UV slope close to $-2$. What is noticeable about these galaxies is that in both, the SFR increased by a factor of $\sim1000$ between 120~Myr ago and $80-90$~Myr ago when the SFR reached its peak. Thus to form a mini-quenched galaxy in our simulations, we seem to require both a very rapid rise in star formation and a rapid decline due to stellar feedback expelling the gas.

In simulations using a similar star formation algorithm, \cite{Katz2023_bursty} showed gravitational instabilities within the galaxy (rather than mergers) are the main drivers for bursts of star formation well above the main sequence. In strongly turbulent environments, our turbulent criterion holds up star formation until the gas gets extremely dense, precipitating a catastrophic starburst when it eventually does. A detailed analysis of the physics that leads to stochastic star formation in the different {\small MEGATRON} simulations will be presented in future work.

One of the primary differences between the mini-quenched galaxies in the {\small MEGATRON} simulations and those observed by JWST is that the simulated analogs tend to be much fainter in the UV. In the simulations, these objects tend to have UV magnitudes in the range of $-15$ to $-16$ and stellar masses $\lesssim10^{7.5}\ {\rm M_{\odot}}$, consistent with results from the {\small SERRA} simulations \citep{Gelli2025}, and the gravitationally lensed object presented in \cite{Strait2023}. In contrast, the object observed by \cite{Looser2024} is much more massive with an estimated stellar mass of $\sim10^{8.5}\ {\rm M_{\odot}}$.

Mini-quenched galaxies are extremely rare in our simulations. Because our simulations do not probe many massive objects at high redshift, it is possible that our sample size is not large enough to produce brighter mini-quenched objects. \cite{Gelli2024} speculated that SN feedback alone cannot quench the system in a way to reproduce the observed SED and posited that outflows from strong radiation pressure may help. Our simulations do include direct UV radiation pressure as well as multi-scattered radiation pressure on dust, but radiation pressure seems to play a minimal role. \cite{Dome2024} also found that other simulations (e.g., {\small IllustrisTNG} and {\small VELA}) fail to reproduce the SED of the \cite{Looser2024} galaxy. The physical nature of the massive mini-quenched galaxies thus remains an open problem.

\subsection{Balmer Break and Post-Starburst Galaxies}
The mini-quenched galaxies represent an extreme tail of the distribution of galaxies with aging stellar populations. Indeed they are clearly offset in our UMAP decomposition. A less extreme population of aging or post-starburst galaxies that have downturns in their SFR are identified by their Balmer break and weaker emission lines. A stacked spectra of these objects is shown in orange in Figure~\ref{fig:umap}. The UMAP decomposition identifies these objects as forming a smooth transition with the normal starforming galaxy population. 

Even in the pre-JWST era, certain redshifts allowed for the photometric identification of the Balmer break \citep[e.g.,][]{Hashimoto2018,RB2020}. While some simulations can predict strong Balmer breaks at high redshift \citep[e.g.,][]{Katz2019_BB,Wilkins2024_BB}, this is not the case for all \citep{Binggeli2019}.

\begin{figure}
    \centering
    \includegraphics[width=0.45\textwidth,trim={0cm 0.0cm 0cm 1.0cm},clip]{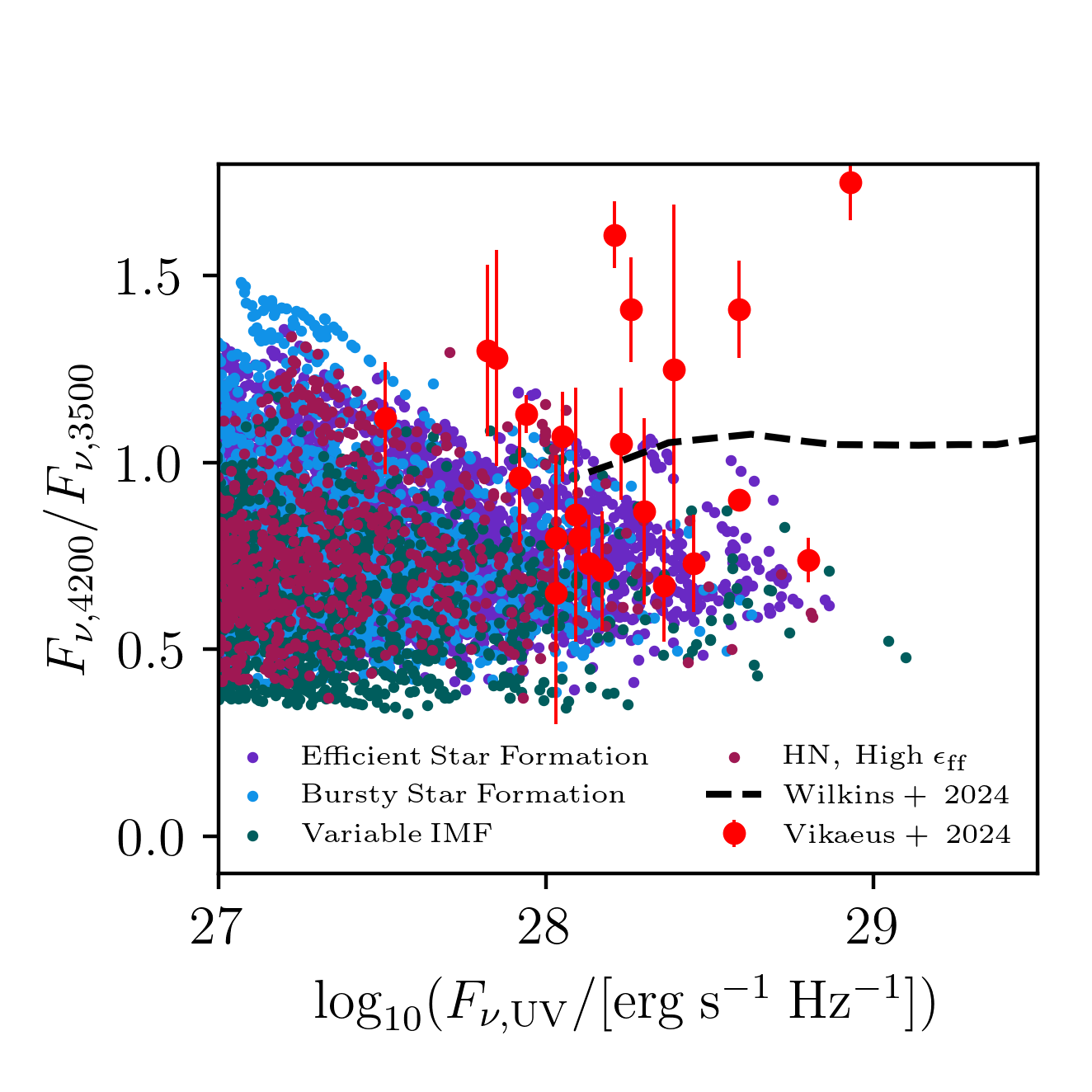}
    \caption{Balmer break strength as a function of UV continuum luminosity. For comparison, we show the sample of Balmer breaks measured for high-redshift JWST galaxies from \protect\cite{Vikaeus2024} as well as the $z=9$ unattenuated predictions from the FLARES simulations \protect\citep{Wilkins2024_BB}.}
    \label{fig:balmer_break}
\end{figure}

We show the Balmer break strength measured as $F_{\rm \nu,4200}/F_{\rm \nu,3500}$ as a function of UV continuum luminosity for {\small MEGATRON} galaxies in Figure~\ref{fig:balmer_break}. For UV fluxes $\lesssim10^{28}\ {\rm erg\ s^{-1}\ Hz^{-1}}$, the {\small MEGATRON} simulations capture the diversity seen by JWST \citep{Vikaeus2024}. We do not find any very bright galaxies with strong Balmer breaks, likely due to the limited volume of the simulation. The brightest objects in our volume are all galaxies that deviate high on the star-formation main-sequence and have spectra with steep UV slopes and Balmer jumps.

Some of the Balmer break galaxies observed by JWST also exhibit strong emission lines \citep{Kuruvanthodi2024}. At lower redshifts, similar features are seen in shocked post-starburst galaxies \citep{Alatalo2016}. Shocks are not always well resolved in our simulations due to the finite spatial resolution, and we do not explicitly post-process the simulation to account for unresolved shocks (see e.g., \citealt{Hirschmann2023}); however, we do find a subset of galaxies that exhibit Balmer breaks with post-starburst features and shock-like emission line ratios. 

\begin{figure}
    \centering
    \includegraphics[width=0.45\textwidth,trim={0cm 0.0cm 0cm 1.0cm},clip]{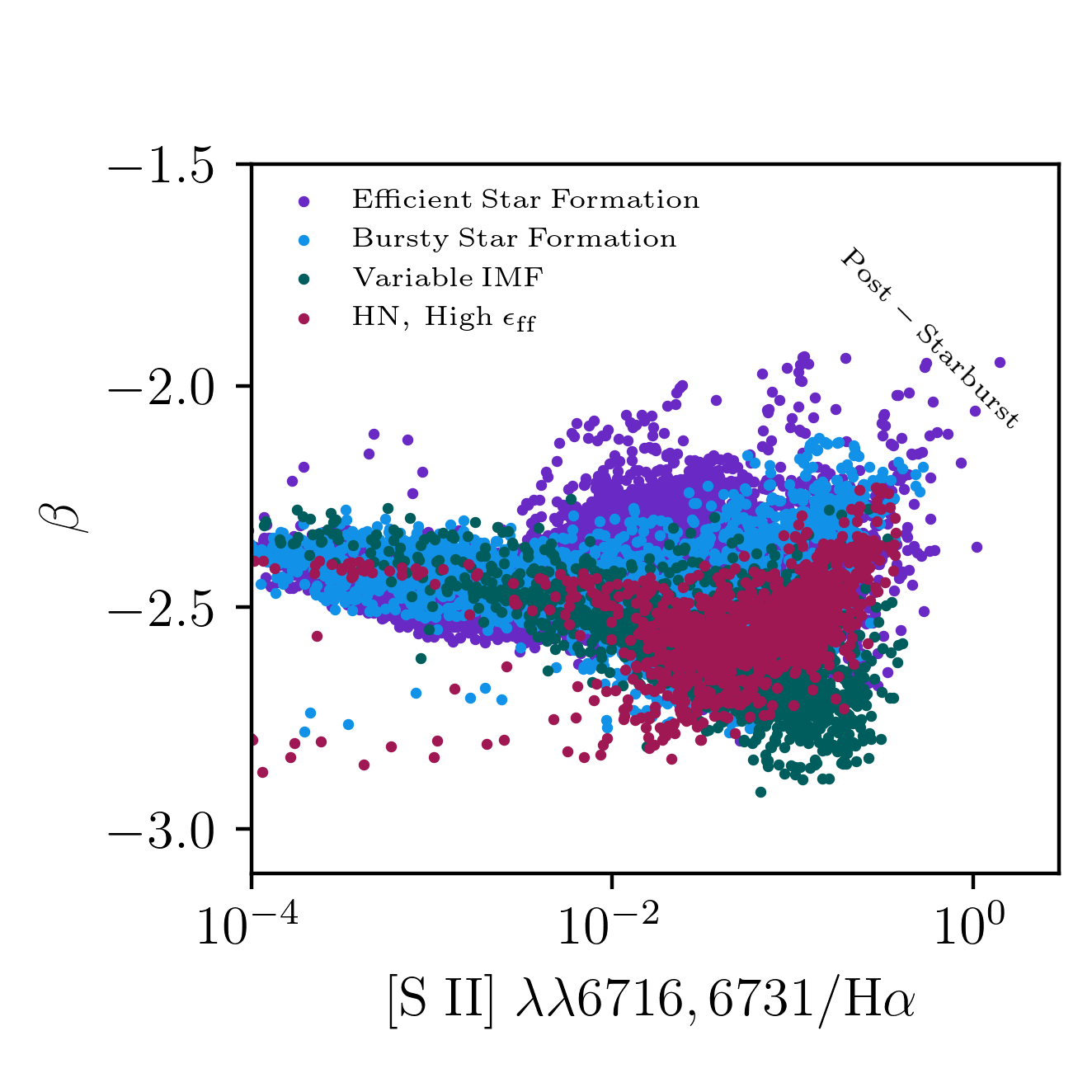}
    \caption{UV slope ($\beta$) as a function of [S~{\small II}]/H$\alpha$ for galaxies with UV magnitudes brighter than $-15$. Galaxies with the reddest (greatest) slopes and the highest [S~{\small II}]/H$\alpha$ ratio represent a post-starburst population. This region is labelled qualitatively.}
    \label{fig:ha_s2_beta}
\end{figure}

One way to search for post-starburst galaxies is to identify systems with strong low ionization state emission lines and UV slopes indicative of aging stellar populations. In Figure~\ref{fig:ha_s2_beta} we show UV slope as a function of ${\rm [S\ {\small II}]~\lambda\lambda6716,6731/H\alpha}$ for galaxies with UV magnitudes $<-15$. ${\rm [S\ {\small II}]~\lambda\lambda6716,6731/H\alpha}$ is a good tracer of post-starburst galaxies because [S~{\small II}] emission is more excited relative to H$\alpha$ in diffuse gas, when the ionization parameter is low, and by shocks. All such conditions are more prevalent with respect to the more highly-ionized gas after a starburst. Particularly in the efficient star formation simulation, there is a small population of galaxies that has $\beta\gtrsim-2$ and ${\rm [S~{\small II}]~\lambda\lambda6716,6731/H\alpha}\gtrsim0.5$. Such a high ${\rm [S~{\small II}]~\lambda\lambda6716,6731/H\alpha}$ ratio is indicative of diffuse gas emission and/or shocks, while a UV slope close to $-2$, in the absence of dust, indicates an aging stellar population.

\begin{figure}
    \centering
    \includegraphics[width=0.45\textwidth,trim={0cm 0.0cm 0cm 0.0cm},clip]{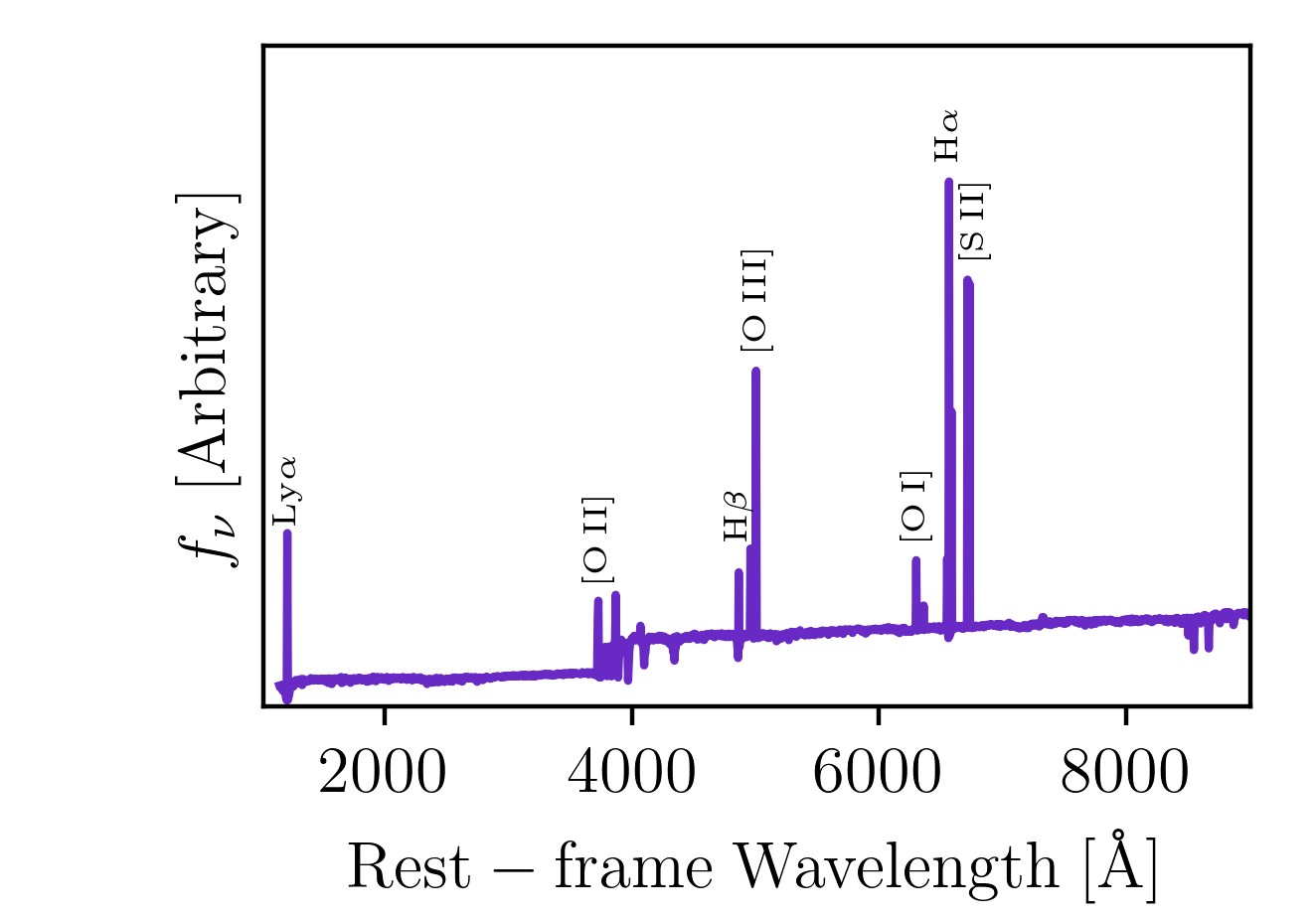}
    \includegraphics[width=0.45\textwidth,trim={0cm 0.0cm 0cm 1.0cm},clip]{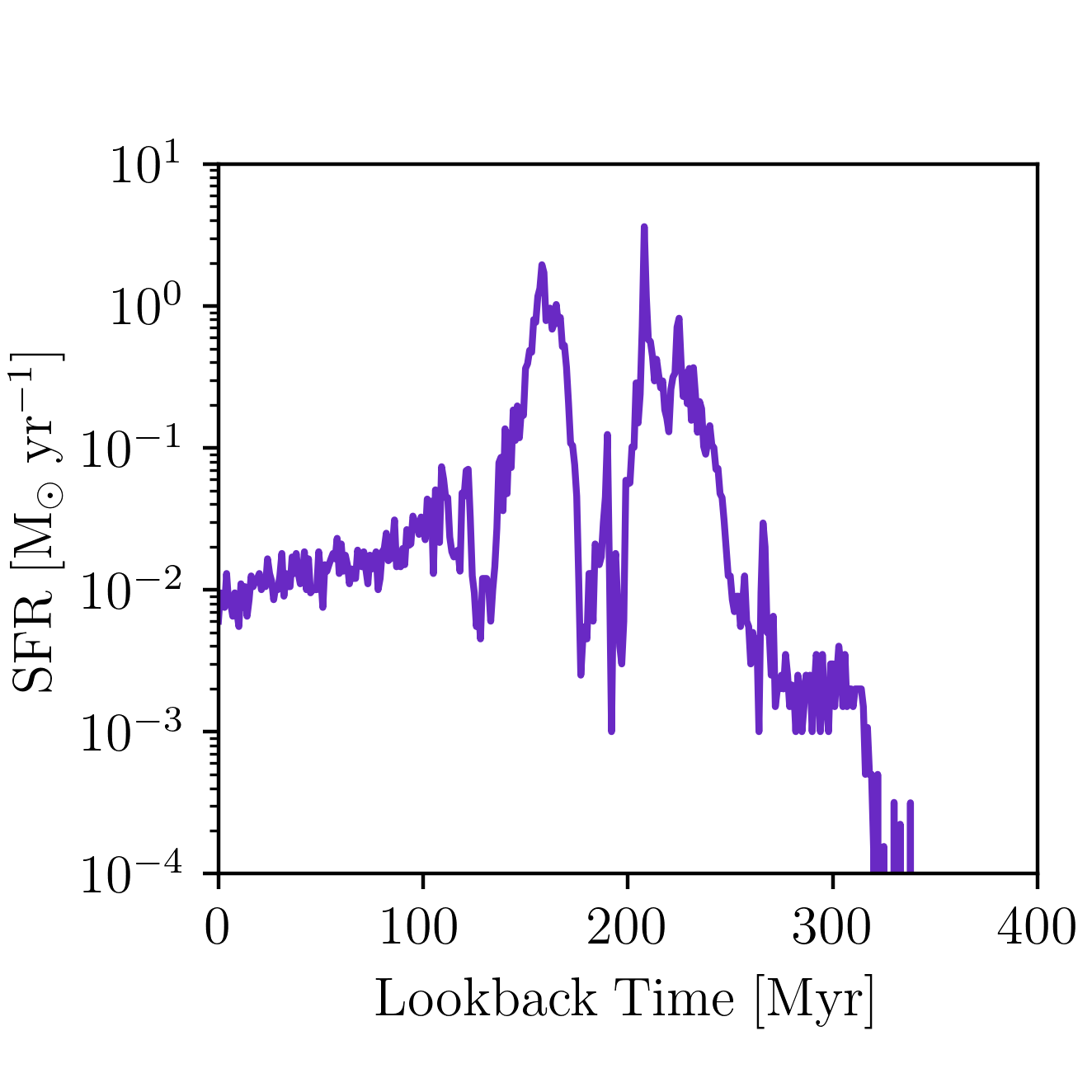}
    \caption{Example spectrum (top) and star formation history (bottom) of a post-starburst galaxy in the bursty star formation simulation.}
    \label{fig:psb_s2}
\end{figure}

The spectrum of the galaxy with the highest ${\rm [S~{\small II}]~\lambda\lambda6716,6731/H\alpha}$ ratio and the reddest UV slope is shown in the top panel of Figure~\ref{fig:psb_s2}. There are indeed very strong [S~{\small II}] and [O~{\small I}] lines as well as [N~{\small II}] and [O~{\small II}], and a clear Balmer break. The SFH of this galaxy is very peculiar (as shown in the bottom panel). The system underwent two extreme bursts of star formation $\sim160$ and $~220$~Myr ago reaching SFRs $\sim200$ times the current value. Subsequently, the SFR has had a relatively steady decline. The old stellar populations are necessary for driving the Balmer break while the current star formation is key for producing the small amount of [O~{\small III}] emission seen in the galaxy. The halo hosting this object is relatively massive, with a virial mass nearing $10^9\ {\rm M_{\odot}}$ and we find no other objects with such peculiar spectra in any of the other simulations.

\vspace{1cm}
\section{Discussion}
We have presented an overview of the {\small MEGATRON} suite of cosmological radiation hydrodynamics simulations that focus on the Lagrange region around a Milky Way-mass environment at high redshift. This suite of simulations is the first to couple a detailed non-equilibrium thermochemistry network of primordial species, metals, and molecules with high enough resolution to predict the intrinsic spectra of galaxies including their stellar and nebular emission. 

Direct comparisons with observations remain the primary tool for constraining the physics of galaxy formation. For example, a key milestone of numerical simulations was the ability to reproduce the Hubble sequence of galaxy morphology at $z=0$ \citep[e.g.,][]{Vogelsberger2014,Schaye2015,Dubois2016} and these simulations were remarkably successful in reproducing numerous other observed or inferred characteristics of galaxies such as the stellar mass function. However, historically at high redshift, besides the UV luminosity function few observational constraints existed on the detailed properties of galaxies in the epoch of reionization. This effort was primarily driven by ALMA \citep[e.g.,][]{Pentericci2016,Bradac2017,Hashimoto2019,Harikane2020,Carniani2020} or the occasional UV line detection from large ground-based facilities \citep[e.g.,][]{Stark2017,Mainali2017}. Such observations were time-consuming for very small samples of galaxies. 

The successful launch of JWST has revolutionized the field of high-redshift astronomy from an observational perspective. This presents an opportunity to test predictions from simulations in a previously unexplored manner and represents new data to calibrate underlying physical models of simulations. One of the key technological advantages of JWST is its spectral capabilities that probe the properties of the high-redshift ISM. Just like the morphological diversity presents a strong constraint on the physics of galaxy formation, with JWST, we can now leverage spectroscopic information. However, few large-scale simulations even attempt to resolve individual molecular clouds \citep[e.g.,][]{Vogelsberger2014,Schaye2015,Dubois2016,Pillepich2018,Lovell2021} and those that do \citep[e.g.,][]{Oshea2015,Ceverino2017,Rosdahl2018,Ma2018,Trebitsch2021,Pallottini2022,Kannan2025} neglect the physics that gives rise to all of the strong emission lines seen in JWST spectra. For these reasons, recent high-resolution simulations use detailed and computationally expensive post-processing to model galaxy spectra. As discussed in \cite{spdrv1}, numerous different, but physically motivated assumptions are often made in post-processing that can limit the fidelity of predicted spectra. Hence, attempting to predict the diversity of galaxy spectra seen in the early Universe by JWST under non-equilibrium conditions, represents one of the primary scientific motivations for the {\small MEGATRON} project. As shown in Figure~\ref{fig:umap}, much of the spectral diversity results from the complex star formation histories driven by gas accretion, mergers, dynamical instabilities, and feedback processes present at high redshift. The modelling details of our simulation, however, remain important. For example, high EW galaxies cannot be produced without very bursty star formation histories. This requires sufficiently high resolution and a star formation model that forms the star particles efficiently in dense gas. Certain emission line ratios, e.g.\ the IR [O {\small III}]/[C {\small II}] cannot currently be reproduced for all chemical yield models (Matsumoto et al. {\it in prep.}). Likewise, non-equilibrium effects can also modify the observed spectra (Katz et al. {\it accepted.}). The qualitative agreement between theory and observation represents only a first step for constraining models of galaxy formation. A more detailed comparison on spectral diagnostics is presented in \cite{Choustikov2026-meg} where we expand on the results from \cite{Katz2024_meg} and demonstrate how observed spectra can be used as a key test of subgrid models.

The results from our simulations are subject to numerous caveats. We refrain from a detailed discussion of the limitations of the numerical methods, in particular those related to finite spatial and mass resolution, subgrid models, and unresolved Strömgren spheres, as these are discussed at length in the context of the {\small MEGATRON} model in \cite{Katz2024_meg}. A key point is that in our calibration simulations, bulk galaxy properties such as stellar mass are converged when going to both lower and higher spatial and mass resolution; however, because ISM and emission properties are density dependent --- e.g.\ luminosity of emission lines $\propto n^2$, while the critical density of certain emission lines is only reached at high enough resolution --- these are not necessarily converged. Likewise, we acknowledge the exclusion of magnetic fields and cosmic rays, which represent an important pressure term and heating/ionization source, respectively in the ISM. These will be addressed in upcoming work. Similarly, we do not attempt to model massive black holes, primarily due to the uncertainties in their seeding/formation channels, how they accrete, their feedback, and their SEDs, despite their unexpectedly high abundance and mass ratios compared to the stellar content of galaxies at high redshift \citep[e.g.,][]{Maiolino2024b,Matthee2024b}. Rather, we highlight uncertainties in our modeling that can have the largest impacts on the observed spectral properties of galaxies. 

First, stellar SEDs, particularly in the ionizing regime are highly uncertain at low metallicity. For example, switching SEDs from single to binary stars and drastically change the reionization history \citep{Ma2016,Rosdahl2018}. Within the context of our modeling, excess ionizing photons or a change in spectral hardness impacts both the EW and ratios of strong emission lines as well as the contribution of the nebular continuum to the total SED. The stellar IMF is also highly uncertain at high redshift and while we varied the upper mass slope in the variable IMF model, we have not considered significant changes to the upper mass limit. Atmospheric models for such massive, metal poor stars are scant (see e.g., \citealt{Martins2020}) and we are thus limited by their availability. However, in the supermassive regime, the luminosities can reach $\sim10^9\ L_{\odot}$ and can dominate the SEDs of the lower-mass galaxies at high redshift. Furthermore, we have not considered deviations from solar abundance patterns, despite metal abundances of the star particles being highly non-solar. $\alpha$-enhancements have been considered in population synthesis models \citep[e.g.,][]{Byrne2022,Park2024} and recently \cite{Katz2024_He} discussed the importance of He enhancements for high-redshift star clusters, but none of these effects are considered in this work. Moreover, X-ray binaries are known to be more important at low metallicity \citep[e.g][]{Mapelli2010,Kaaret2011,Basu-Zych2013,Douna2015,Saxena2021} which would be very relevant for the systems simulated here, and certain JWST observations are better explained when the effects of X-ray binaries are accounted for \citep{Katz2023_jwst}. As a final point, we do not model the spectra of Pop.~II stars individually, but rather we assume an average SED. In reality, especially in the dwarf galaxy regime, there may be local fluctuations in the SED properties due to IMF sampling that may be important \citep{Stanway2023} and are not captured here.  

Second, large uncertainties remain in the chemical yields \citep{Buck2021} and their depletion onto dust. These uncertainties are particularly important in the context of our work because gas cooling is highly sensitive to presence of individual species and their depletion rates, as are the observed metal emission lines. Cooling not only impacts the efficiency by which gas can collapse and form stars or how fast shocks instigated by SN can cool, but most of the strong metal emission lines have emissivities that are exponentially sensitive to electron temperature. Small changes in the cooling can have an outsized impact on the predicted emission. While we are confident that for a given set of abundances, our cooling model is accurate\footnote{Or at the very least is able to reproduce more detailed photionization codes such as CLOUDY.}, if the abundance patterns that our adopted yields predict are not consistent with real galaxies, we may be unable to reproduce some of the spectral characteristics at high redshift. 

Third, little is known about Pop.~III stars because they have never been directly observed. We have adopted a plausible model that has been used previously \citep{Wise2012,Kimm2017}; however, without observational guidance, little progress can be made on their detailed properties. This means that our predictions on the observable properties of the Pop.~III-Pop.~II transition may be highly uncertain. If JWST does not provide more constraints on Pop.~III stars, the ELT may present as a promising alternative \citep{Grisdale2021}.

\section{Conclusions}
With the caveats discussed in the previous section in mind, we highlight our primary conclusions. 
\begin{enumerate}

\item Much of the spectral diversity seen by JWST naturally emerges within the context of a $\Lambda$CDM cosmology and reasonable choices for subgrid models. This is despite the fact that our initial conditions represent a highly biased environment and result from highly varied and complex star formation histories. Nevertheless, significant work remains, both observationally to obtain spectroscopic samples of high-redshift galaxies with a well defined selection function, and the theoretically to not only reproduce the diversity, but also the proportions of each class of spectra. Furthermore, our simulations still struggle to produce certain galaxies like massive mini-quenched systems and other peculiar objects including the high-redshift nitrogen emitters \citep[e.g.][]{Bunker2023,Castellano2024,Naidu2025} and Little Red Dots \citep{Matthee2024}. 

\item While individual haloes rapidly transition from Pop.~III to Pop.~II star formation, across the entire Lagrange region, Pop.~III star formation continues to the lowest redshift ($z=8.5$) currently reached by our simulations. The Pop.~III spectra tend to be redder than those of Pop.~II galaxies due to the strong nebular contribution and high residual ISM gas densities. Most Pop.~III galaxies are too UV faint to be observed, even with optimistic assumptions about strong gravitational lensing. We identify a (new) class of haloes without any stars on the main-sequence where primordial and metal cooling radiation dominate their spectra. The brightest examples of such galaxies appear immediately after a Pop.~III star formation event.

\item The most extreme emission line galaxies form when the instantaneous star formation rates of galaxies spike to values $\gtrsim100\times$ what they were averaged over the previous 50~Myr. $\sim 50\%$ of the simulated galaxies fit the observational definition of being an EELG, which is qualitatively consistent with observations that demonstrate their increased occurrence towards high redshift. Most of our simulations struggle to reproduce the highest [O~{\small III}]~EWs seen in observations; however, we find that the inclusion of hypernova and a top-heavy IMF may help remedy this potential disagreement between simulations and observations.

\item Independent of star formation or feedback model, our simulations easily reproduce the observed star formation main-sequence. This is due to the fact that the majority of the low-mass galaxies in both the simulations and observations seemed to have assembled most of their stellar mass within a 100~Myr time period. Therefore the scatter in this relation and comparing with SFR on shorter timescales will provide a stronger constraint on the physics of high-redshift galaxy formation.

\item Our simulations do not produce bright galaxies with strong Balmer breaks and massive mini-quenched galaxies. Part of this discrepancy may be due to the limited volume of the simulations and thus our feedback model and subgrid physics requires further testing in more extreme environments. Such spectra are much more common in our simulations for faint galaxies that exhibit more stochastic star formation histories.

\end{enumerate}

Here we have focused primarily on high-redshift galaxy formation with the {\small MEGATRON} simulations. Hence we have only addressed two of the four scientific goals of the project. The connection with low redshift via near-field cosmology and the CGM towards Cosmic Noon are explored in companion papers \citep{Rey2025tmp,Cadiou2025tmp}. Furthermore, data products from the {\small MEGATRON} simulations will be released to the wider community in the hope that it aids in both interpreting observations and helps constrain the key physics of early galaxy formation.

\section*{Acknowledgments}
This work made extensive use of the dp265, dp016, dp373, and dp379 projects on the DiRAC ecosystem. HK is particularly grateful to Christopher Mountford and Alastair Basden for support on DIaL3 and Cosma8, respectively. HK and the {\small MEGATRON} team are especially thankful for the support on Glamdring provided by Jonathan Patterson. This work used the DiRAC@Durham facility managed by the Institute for Computational Cosmology on behalf of the STFC DiRAC HPC Facility (\url{www.dirac.ac.uk}). The equipment was funded by BEIS capital funding via STFC capital grants {\small ST/P002293/1}, {\small ST/R002371/1} and {\small ST/S002502/1}, Durham University and STFC operations grant {\small ST/R000832/1}. This work also used the DiRAC Data Intensive service at Leicester, operated by the University of Leicester IT Services, which forms part of the STFC DiRAC HPC Facility. The equipment was funded by BEIS capital funding via STFC capital grants {\small ST/K000373/1} and {\small ST/R002363/1} and STFC DiRAC Operations grant {\small ST/R001014/1}. This work was performed using resources provided by the Cambridge Service for Data Driven Discovery (CSD3) operated by the University of Cambridge Research Computing Service (www.csd3.cam.ac.uk), provided by Dell EMC and Intel using Tier-2 funding from the Engineering and Physical Sciences Research Council (capital grant EP/T022159/1), and DiRAC funding from the Science and Technology Facilities Council (www.dirac.ac.uk). DiRAC is part of the National e-Infrastructure. This work has made use of the Infinity Cluster hosted by Institut d'Astrophysique de Paris. We thank Stephane Rouberol for running smoothly this cluster for us.

The material in this manuscript is based upon work supported by NASA under award No. 80NSSC25K7009. HK acknowledges support from FACCTS. AS and AJC acknowledge funding from the “FirstGalaxies” Advanced Grant from the European Research Council (ERC) under the European Union’s Horizon 2020 research and innovation programme (Grant agreement No.789056). TK is supported by the National Research Foundation of Korea (RS-2022-NR070872 and RS-2025-00516961) and by the Yonsei Fellowship, funded by Lee Youn Jae. AS acknowledges support from the Science and Technology Facilities Council (STFC) for a PhD studentship. FRM is supported by the Kavli Institute for Cosmological physics at the University of Chicago through an endowment from the Kavli Foundation and its founder Fred Kavli. GCJ acknowledges support by the Science and Technology Facilities Council (STFC), by the ERC through Advanced Grant 695671 ``QUENCH'', and by the UKRI Frontier Research grant RISEandFALL. KM acknowledges the Flemish Fund for Scientific Research (FWO-Vlaanderen), Grant number 1169822N. M.S. acknowledges the support from the Swiss National Science Foundation under Grant No. P500PT\_214488. NC acknowledges support from the Science and Technology Facilities Council (STFC) for a PhD studentship. OA acknowledges support from the Knut and Alice Wallenberg Foundation, the Swedish Research Council (grant 2019-04659), the Swedish National Space Agency (SNSA Dnr 2023-00164), and the LMK foundation. The authors also acknowledge financial support from Oriel College’s Research Fund. 

The authors thank Romain Teyssier and Leo Michel-Dansac for both developing and open-sourcing {\small RAMSES} and {\small RASCAS}, respectively. We thank the developers and maintainers of \textsc{pynbody} (\citealt{Pontzen2013,Pontzen2022}), \textsc{NumPy} (\citealt{vanderWalt2011, Harris2020}), \textsc{SciPy} (\citealt{Virtanen2020}), \textsc{jupyter} (\citealt{RaganKelley2014}), \textsc{matplotlib} (\citealt{Hunter2007}), the Astrophysics Data Service and the arXiv pre-print repository for providing open-source software and services that were used extensively in this work.

\bibliographystyle{mn2e}
\bibliography{example}

\begin{thebibliography}{289}
\expandafter\ifx\csname natexlab\endcsname\relax\def\natexlab#1{#1}\fi

\bibitem[{{Agertz} \& {Kravtsov}(2015)}]{Agertz2015}
{Agertz} O., {Kravtsov} A.~V., 2015, \apj, 804, 18

\bibitem[{{Agertz} {et~al}\mbox{.}(2013){Agertz}, {Kravtsov}, {Leitner}, \&
  {Gnedin}}]{Agertz2013}
{Agertz} O., {Kravtsov} A.~V., {Leitner} S.~N., {Gnedin} N.~Y., 2013, \apj,
  770, 25

\bibitem[{{Agertz} {et~al}\mbox{.}(2021){Agertz}, {Renaud}, {Feltzing}, {Read},
  {Ryde}, {Andersson}, {Rey}, {Bensby}, \& {Feuillet}}]{Agertz2021}
{Agertz} O. {et~al.}, 2021, \mnras, 503, 5826

\bibitem[{{Alatalo} {et~al}\mbox{.}(2016){Alatalo}, {Cales}, {Rich},
  {Appleton}, {Kewley}, {Lacy}, {Lanz}, {Medling}, \& {Nyland}}]{Alatalo2016}
{Alatalo} K. {et~al.}, 2016, \apjs, 224, 38

\bibitem[{{Aldrovandi} \& {Pequignot}(1973)}]{Aldrovandi1973}
{Aldrovandi} S.~M.~V., {Pequignot} D., 1973, \aap, 25, 137

\bibitem[{{Amor{\'\i}n} {et~al}\mbox{.}(2015){Amor{\'\i}n},
  {P{\'e}rez-Montero}, {Contini}, {V{\'\i}lchez}, {Bolzonella}, {Tasca},
  {Lamareille}, {Zamorani}, {Maier}, {Carollo}, {Kneib}, {Le F{\`e}vre},
  {Lilly}, {Mainieri}, {Renzini}, {Scodeggio}, {Bardelli}, {Bongiorno},
  {Caputi}, {Cucciati}, {de la Torre}, {de Ravel}, {Franzetti}, {Garilli},
  {Iovino}, {Kampczyk}, {Knobel}, {Kova{\v{c}}}, {Le Borgne}, {Le Brun},
  {Mignoli}, {Pell{\`o}}, {Peng}, {Presotto}, {Ricciardelli}, {Silverman},
  {Tanaka}, {Tresse}, {Vergani}, \& {Zucca}}]{Amorin2015}
{Amor{\'\i}n} R. {et~al.}, 2015, \aap, 578, A105

\bibitem[{{Arata} {et~al}\mbox{.}(2020){Arata}, {Yajima}, {Nagamine}, {Abe}, \&
  {Khochfar}}]{Arata2020}
{Arata} S., {Yajima} H., {Nagamine} K., {Abe} M., {Khochfar} S., 2020, \mnras,
  498, 5541

\bibitem[{{Arnaud} \& {Raymond}(1992)}]{Arnaud1992}
{Arnaud} M., {Raymond} J., 1992, \apj, 398, 394

\bibitem[{{Arnaud} \& {Rothenflug}(1985)}]{Arnaud1985}
{Arnaud} M., {Rothenflug} R., 1985, \aaps, 60, 425

\bibitem[{{Asplund} {et~al}\mbox{.}(2009){Asplund}, {Grevesse}, {Sauval}, \&
  {Scott}}]{Asplund2009}
{Asplund} M., {Grevesse} N., {Sauval} A.~J., {Scott} P., 2009, \araa, 47, 481

\bibitem[{{Baczynski}, {Glover} \& {Klessen}(2015){Baczynski}, {Glover}, \&
  {Klessen}}]{Baczynski2015}
{Baczynski} C., {Glover} S.~C.~O., {Klessen} R.~S., 2015, \mnras, 454, 380

\bibitem[{{Badnell}(2006)}]{Badnell2006}
{Badnell} N.~R., 2006, \apjs, 167, 334

\bibitem[{{Badnell} {et~al}\mbox{.}(2003){Badnell}, {O'Mullane}, {Summers},
  {Altun}, {Bautista}, {Colgan}, {Gorczyca}, {Mitnik}, {Pindzola}, \&
  {Zatsarinny}}]{Badnell2003}
{Badnell} N.~R. {et~al.}, 2003, \aap, 406, 1151

\bibitem[{{Bakes} \& {Tielens}(1994)}]{Bakes1994}
{Bakes} E.~L.~O., {Tielens} A.~G.~G.~M., 1994, \apj, 427, 822

\bibitem[{{Barrag{\'a}n} {et~al}\mbox{.}(2006){Barrag{\'a}n}, {Errea},
  {M{\'e}ndez}, {Rabad{\'a}n}, \& {Riera}}]{Barragan2006}
{Barrag{\'a}n} P., {Errea} L.~F., {M{\'e}ndez} L., {Rabad{\'a}n} I., {Riera}
  A., 2006, \apj, 636, 544

\bibitem[{{Barrow} {et~al}\mbox{.}(2017){Barrow}, {Wise}, {Norman}, {O'Shea},
  \& {Xu}}]{Barrow2017}
{Barrow} K. S.~S., {Wise} J.~H., {Norman} M.~L., {O'Shea} B.~W., {Xu} H., 2017,
  \mnras, 469, 4863

\bibitem[{{Basu-Zych} {et~al}\mbox{.}(2013){Basu-Zych}, {Lehmer},
  {Hornschemeier}, {Gon{\c{c}}alves}, {Fragos}, {Heckman}, {Overzier}, {Ptak},
  \& {Schiminovich}}]{Basu-Zych2013}
{Basu-Zych} A.~R. {et~al.}, 2013, \apj, 774, 152

\bibitem[{{Behroozi} {et~al}\mbox{.}(2019){Behroozi}, {Wechsler}, {Hearin}, \&
  {Conroy}}]{Behroozi2019}
{Behroozi} P., {Wechsler} R.~H., {Hearin} A.~P., {Conroy} C., 2019, \mnras,
  488, 3143

\bibitem[{{Behroozi}, {Wechsler} \& {Wu}(2013){Behroozi}, {Wechsler}, \&
  {Wu}}]{Behroozi2013}
{Behroozi} P.~S., {Wechsler} R.~H., {Wu} H.-Y., 2013, \apj, 762, 109

\bibitem[{{Bhagwat} {et~al}\mbox{.}(2024){Bhagwat}, {Napolitano}, {Pentericci},
  {Ciardi}, \& {Costa}}]{Bhagwat2024}
{Bhagwat} A., {Napolitano} L., {Pentericci} L., {Ciardi} B., {Costa} T., 2024,
  arXiv e-prints, arXiv:2408.16063

\bibitem[{{Bialy} \& {Sternberg}(2019)}]{Bialy2019}
{Bialy} S., {Sternberg} A., 2019, \apj, 881, 160

\bibitem[{{Binggeli} {et~al}\mbox{.}(2019){Binggeli}, {Zackrisson}, {Ma},
  {Inoue}, {Vikaeus}, {Hashimoto}, {Mawatari}, {Shimizu}, \&
  {Ceverino}}]{Binggeli2019}
{Binggeli} C. {et~al.}, 2019, \mnras, 489, 3827

\bibitem[{{Black}(1981)}]{Black1981}
{Black} J.~H., 1981, \mnras, 197, 553

\bibitem[{{Black} \& {Dalgarno}(1977)}]{Black1977}
{Black} J.~H., {Dalgarno} A., 1977, \apjs, 34, 405

\bibitem[{{Bouwens} {et~al}\mbox{.}(2010){Bouwens}, {Illingworth}, {Oesch},
  {Trenti}, {Stiavelli}, {Carollo}, {Franx}, {van Dokkum}, {Labb{\'e}}, \&
  {Magee}}]{Bouwens2010}
{Bouwens} R.~J. {et~al.}, 2010, \apjl, 708, L69

\bibitem[{{Boyett} {et~al}\mbox{.}(2024){Boyett}, {Bunker}, {Curtis-Lake},
  {Chevallard}, {Cameron}, {Jones}, {Saxena}, {Charlot}, {Curti}, {Wallace},
  {Arribas}, {Carniani}, {Willott}, {Alberts}, {Eisenstein}, {Hainline},
  {Hausen}, {Johnson}, {Rieke}, {Robertson}, {Stark}, {Tacchella}, {Williams},
  {Chen}, {Egami}, {Endsley}, {Kumari}, {Laseter}, {Looser}, {Maseda},
  {Scholtz}, {Shivaei}, {Simmonds}, {Smit}, {{\"U}bler}, \&
  {Witstok}}]{Boyett2024}
{Boyett} K. {et~al.}, 2024, \mnras, 535, 1796

\bibitem[{{Brada{\v{c}}} {et~al}\mbox{.}(2017){Brada{\v{c}}}, {Garcia-Appadoo},
  {Huang}, {Vallini}, {Quinn Finney}, {Hoag}, {Lemaux}, {Borello Schmidt},
  {Treu}, {Carilli}, {Dijkstra}, {Ferrara}, {Fontana}, {Jones}, {Ryan}, {Wagg},
  \& {Gonzalez}}]{Bradac2017}
{Brada{\v{c}}} M. {et~al.}, 2017, \apjl, 836, L2

\bibitem[{{Brauer} {et~al}\mbox{.}(2025){Brauer}, {Emerick}, {Mead}, {Ji},
  {Wise}, {Bryan}, {Mac Low}, {C{\^o}t{\'e}}, {Andersson}, \&
  {Frebel}}]{Brauer2025}
{Brauer} K. {et~al.}, 2025, \apj, 980, 41

\bibitem[{{Bromm} {et~al}\mbox{.}(2001){Bromm}, {Ferrara}, {Coppi}, \&
  {Larson}}]{Bromm2001}
{Bromm} V., {Ferrara} A., {Coppi} P.~S., {Larson} R.~B., 2001, \mnras, 328, 969

\bibitem[{{Buck} {et~al}\mbox{.}(2021){Buck}, {Rybizki}, {Buder}, {Obreja},
  {Macci{\`o}}, {Pfrommer}, {Steinmetz}, \& {Ness}}]{Buck2021}
{Buck} T., {Rybizki} J., {Buder} S., {Obreja} A., {Macci{\`o}} A.~V.,
  {Pfrommer} C., {Steinmetz} M., {Ness} M., 2021, \mnras, 508, 3365

\bibitem[{{Bunker} {et~al}\mbox{.}(2024){Bunker}, {Cameron}, {Curtis-Lake},
  {Jakobsen}, {Carniani}, {Curti}, {Witstok}, {Maiolino}, {D'Eugenio},
  {Looser}, {Willott}, {Bonaventura}, {Hainline}, {{\"U}bler}, {Willmer},
  {Saxena}, {Smit}, {Alberts}, {Arribas}, {Baker}, {Baum}, {Bhatawdekar},
  {Bowler}, {Boyett}, {Charlot}, {Chen}, {Chevallard}, {Circosta}, {DeCoursey},
  {de Graaff}, {Egami}, {Eisenstein}, {Endsley}, {Ferruit}, {Giardino},
  {Hausen}, {Helton}, {Hviding}, {Ji}, {Johnson}, {Jones}, {Kumari}, {Laseter},
  {L{\"u}tzgendorf}, {Maseda}, {Nelson}, {Parlanti}, {Perna}, {Rauscher},
  {Rawle}, {Rix}, {Rieke}, {Robertson}, {Rodr{\'\i}guez Del Pino}, {Sandles},
  {Scholtz}, {Sharpe}, {Skarbinski}, {Stark}, {Sun}, {Tacchella}, {Topping},
  {Villanueva}, {Wallace}, {Williams}, \& {Woodrum}}]{Bunker2024}
{Bunker} A.~J. {et~al.}, 2024, \aap, 690, A288

\bibitem[{{Bunker} {et~al}\mbox{.}(2023){Bunker}, {Saxena}, {Cameron},
  {Willott}, {Curtis-Lake}, {Jakobsen}, {Carniani}, {Smit}, {Maiolino},
  {Witstok}, {Curti}, {D'Eugenio}, {Jones}, {Ferruit}, {Arribas}, {Charlot},
  {Chevallard}, {Giardino}, {de Graaff}, {Looser}, {L{\"u}tzgendorf}, {Maseda},
  {Rawle}, {Rix}, {Del Pino}, {Alberts}, {Egami}, {Eisenstein}, {Endsley},
  {Hainline}, {Hausen}, {Johnson}, {Rieke}, {Rieke}, {Robertson}, {Shivaei},
  {Stark}, {Sun}, {Tacchella}, {Tang}, {Williams}, {Willmer}, {Baker}, {Baum},
  {Bhatawdekar}, {Bowler}, {Boyett}, {Chen}, {Circosta}, {Helton}, {Ji},
  {Kumari}, {Lyu}, {Nelson}, {Parlanti}, {Perna}, {Sandles}, {Scholtz},
  {Suess}, {Topping}, {{\"U}bler}, {Wallace}, \& {Whitler}}]{Bunker2023}
{Bunker} A.~J. {et~al.}, 2023, \aap, 677, A88

\bibitem[{{Burton}, {Hollenbach} \& {Tielens}(1990){Burton}, {Hollenbach}, \&
  {Tielens}}]{Burton1990}
{Burton} M.~G., {Hollenbach} D.~J., {Tielens} A.~G.~G.~M., 1990, \apj, 365, 620

\bibitem[{{Byler} {et~al}\mbox{.}(2017){Byler}, {Dalcanton}, {Conroy}, \&
  {Johnson}}]{Byler2017}
{Byler} N., {Dalcanton} J.~J., {Conroy} C., {Johnson} B.~D., 2017, \apj, 840,
  44

\bibitem[{{Byrne} {et~al}\mbox{.}(2022){Byrne}, {Stanway}, {Eldridge},
  {McSwiney}, \& {Townsend}}]{Byrne2022}
{Byrne} C.~M., {Stanway} E.~R., {Eldridge} J.~J., {McSwiney} L., {Townsend}
  O.~T., 2022, \mnras, 512, 5329

\bibitem[{Cadiou, Dubois \& Pichon(2019)Cadiou, Dubois, \&
  Pichon}]{cadiouAccurateTracerParticles2019}
Cadiou C., Dubois Y., Pichon C., 2019, Astronomy \& Astrophysics, 621, A96

\bibitem[{{Cadiou} {et~al}\mbox{.}(2025){Cadiou}, {Katz}, {Rey}, {Agertz},
  {Blaizot}, {Cameron}, {Choustikov}, {Devriendt}, {Hauk}, {Jones}, {Kimm},
  {Laseter}, {Martin-Alvarez}, {Matsumoto}, {Nyhagen}, {Pearce},
  {Rodr{\'\i}guez Montero}, {Rosdahl}, {Rufo Pastor}, {Sanati}, {Saxena},
  {Slyz}, {Stiskalek}, {Storck}, \& {Yee}}]{Cadiou2025tmp}
{Cadiou} C. {et~al.}, 2025, arXiv e-prints, arXiv:2510.05667

\bibitem[{{Cameron} {et~al}\mbox{.}(2023{\natexlab{a}}){Cameron}, {Katz},
  {Rey}, \& {Saxena}}]{Cameron2023_nitrogen}
{Cameron} A.~J., {Katz} H., {Rey} M.~P., {Saxena} A., 2023{\natexlab{a}},
  \mnras, 523, 3516

\bibitem[{{Cameron} {et~al}\mbox{.}(2023{\natexlab{b}}){Cameron}, {Katz},
  {Witten}, {Saxena}, {Laporte}, \& {Bunker}}]{Cameron2023_NDG}
{Cameron} A.~J., {Katz} H., {Witten} C., {Saxena} A., {Laporte} N., {Bunker}
  A.~J., 2023{\natexlab{b}}, arXiv e-prints, arXiv:2311.02051

\bibitem[{{Cameron} {et~al}\mbox{.}(2024){Cameron}, {Katz}, {Witten}, {Saxena},
  {Laporte}, \& {Bunker}}]{Cameron2024}
{Cameron} A.~J., {Katz} H., {Witten} C., {Saxena} A., {Laporte} N., {Bunker}
  A.~J., 2024, \mnras, 534, 523

\bibitem[{{Cameron} {et~al}\mbox{.}(2023{\natexlab{c}}){Cameron}, {Saxena},
  {Bunker}, {D'Eugenio}, {Carniani}, {Maiolino}, {Curtis-Lake}, {Ferruit},
  {Jakobsen}, {Arribas}, {Bonaventura}, {Charlot}, {Chevallard}, {Curti},
  {Looser}, {Maseda}, {Rawle}, {Rodr{\'\i}guez Del Pino}, {Smit}, {{\"U}bler},
  {Willott}, {Witstok}, {Egami}, {Eisenstein}, {Johnson}, {Hainline}, {Rieke},
  {Robertson}, {Stark}, {Tacchella}, {Williams}, {Willmer}, {Bhatawdekar},
  {Bowler}, {Boyett}, {Circosta}, {Helton}, {Jones}, {Kumari}, {Ji}, {Nelson},
  {Parlanti}, {Sandles}, {Scholtz}, \& {Sun}}]{Cameron2023}
{Cameron} A.~J. {et~al.}, 2023{\natexlab{c}}, \aap, 677, A115

\bibitem[{{Cardamone} {et~al}\mbox{.}(2009){Cardamone}, {Schawinski}, {Sarzi},
  {Bamford}, {Bennert}, {Urry}, {Lintott}, {Keel}, {Parejko}, {Nichol},
  {Thomas}, {Andreescu}, {Murray}, {Raddick}, {Slosar}, {Szalay}, \&
  {Vandenberg}}]{Cardamone2009}
{Cardamone} C. {et~al.}, 2009, \mnras, 399, 1191

\bibitem[{{Carniani} {et~al}\mbox{.}(2020){Carniani}, {Ferrara}, {Maiolino},
  {Castellano}, {Gallerani}, {Fontana}, {Kohandel}, {Lupi}, {Pallottini},
  {Pentericci}, {Vallini}, \& {Vanzella}}]{Carniani2020}
{Carniani} S. {et~al.}, 2020, \mnras, 499, 5136

\bibitem[{{Carniani} {et~al}\mbox{.}(2024){Carniani}, {Hainline}, {D'Eugenio},
  {Eisenstein}, {Jakobsen}, {Witstok}, {Johnson}, {Chevallard}, {Maiolino},
  {Helton}, {Willott}, {Robertson}, {Alberts}, {Arribas}, {Baker},
  {Bhatawdekar}, {Boyett}, {Bunker}, {Cameron}, {Cargile}, {Charlot}, {Curti},
  {Curtis-Lake}, {Egami}, {Giardino}, {Isaak}, {Ji}, {Jones}, {Kumari},
  {Maseda}, {Parlanti}, {P{\'e}rez-Gonz{\'a}lez}, {Rawle}, {Rieke}, {Rieke},
  {Del Pino}, {Saxena}, {Scholtz}, {Smit}, {Sun}, {Tacchella}, {{\"U}bler},
  {Venturi}, {Williams}, \& {Willmer}}]{Carniani2024}
{Carniani} S. {et~al.}, 2024, \nat, 633, 318

\bibitem[{{Castellano} {et~al}\mbox{.}(2025){Castellano}, {Fontana}, {Merlin},
  {Santini}, {Napolitano}, {Menci}, {Calabr{\`o}}, {Paris}, {Pentericci},
  {Zavala}, {Dickinson}, {Finkelstein}, {Treu}, {Amorin}, {Arrabal Haro},
  {Bergamini}, {Bisigello}, {Daddi}, {Dayal}, {Dekel}, {Ferrara}, {Fortuni},
  {Gandolfi}, {Giavalisco}, {Grillo}, {Guida}, {Hathi}, {Holwerda},
  {Koekemoer}, {Kokorev}, {Li}, {Llerena}, {Lucas}, {Mascia}, {Metha},
  {Morishita}, {Nanayakkara}, {Pacucci}, {P{\'e}rez-Gonz{\'a}lez},
  {Roberts-Borsani}, {Rodighiero}, {Rosati}, {Salazar}, {Schneider},
  {Somerville}, {Taylor}, {Trenti}, {Trinca}, {Wang}, {Watson}, {Yang}, \&
  {Yung}}]{Castellano2025}
{Castellano} M. {et~al.}, 2025, arXiv e-prints, arXiv:2504.05893

\bibitem[{{Castellano} {et~al}\mbox{.}(2024){Castellano}, {Napolitano},
  {Fontana}, {Roberts-Borsani}, {Treu}, {Vanzella}, {Zavala}, {Arrabal Haro},
  {Calabr{\`o}}, {Llerena}, {Mascia}, {Merlin}, {Paris}, {Pentericci},
  {Santini}, {Bakx}, {Bergamini}, {Cupani}, {Dickinson}, {Filippenko},
  {Glazebrook}, {Grillo}, {Kelly}, {Malkan}, {Mason}, {Morishita},
  {Nanayakkara}, {Rosati}, {Sani}, {Wang}, \& {Yoon}}]{Castellano2024}
{Castellano} M. {et~al.}, 2024, \apj, 972, 143

\bibitem[{{Cen}(1992)}]{Cen1992}
{Cen} R., 1992, \apjs, 78, 341

\bibitem[{{Ceverino}, {Glover} \& {Klessen}(2017){Ceverino}, {Glover}, \&
  {Klessen}}]{Ceverino2017}
{Ceverino} D., {Glover} S. C.~O., {Klessen} R.~S., 2017, \mnras, 470, 2791

\bibitem[{{Ceverino} {et~al}\mbox{.}(2021){Ceverino}, {Hirschmann}, {Klessen},
  {Glover}, {Charlot}, \& {Feltre}}]{Ceverino2021}
{Ceverino} D., {Hirschmann} M., {Klessen} R.~S., {Glover} S. C.~O., {Charlot}
  S., {Feltre} A., 2021, \mnras, 504, 4472

\bibitem[{{Chan} {et~al}\mbox{.}(2025){Chan}, {Richings}, {Theuns}, {Liu},
  {Schaller}, \& {Ivkovic}}]{Chan2025}
{Chan} T.~K., {Richings} A.~J., {Theuns} T., {Liu} Y., {Schaller} M., {Ivkovic}
  M., 2025, arXiv e-prints, arXiv:2508.13277

\bibitem[{{Chemerynska} {et~al}\mbox{.}(2025){Chemerynska}, {Atek}, {Furtak},
  {Chisholm}, {Endsley}, {Kokorev}, {Rosdahl}, {Blaizot}, {Adamo}, {Bouwens},
  {Fujimoto}, {Korber}, {Mason}, {McQuinn}, {Mu{\~n}oz}, {Natarajan}, {Nelson},
  {Oesch}, {Pan}, {Richard}, {Saldana-Lopez}, {Volonteri}, {Zitrin}, {Berg},
  {Claeyssens}, {Dessauges-Zavadsky}, {Jecmen}, {Labb{\'e}}, {Naidu}, \&
  {Trebitsch}}]{Chemerynska2025}
{Chemerynska} I. {et~al.}, 2025, arXiv e-prints, arXiv:2509.24881

\bibitem[{{Chemerynska} {et~al}\mbox{.}(2023){Chemerynska}, {Atek}, {Furtak},
  {Zitrin}, {Greene}, {Dayal}, {Weibel}, {Kokorev}, {Goulding}, {Williams},
  {Nanayakkara}, {Bezanson}, {Brammer}, {Cutler}, {Labbe}, {Leja}, {Pan},
  {Price}, {Wang}, {Weaver}, \& {Whitaker}}]{Chemerynska2023}
{Chemerynska} I. {et~al.}, 2023, arXiv e-prints, arXiv:2312.05030

\bibitem[{{Choustikov} {et~al}\mbox{.}(2026){Choustikov}, {Katz}, {Cameron},
  {Saxena}, {Devriendt}, {Slyz}, {Rey}, {Cadiou}, {Blaizot}, {Kimm}, {Laseter},
  {Matsumoto}, \& {Rosdahl}}]{Choustikov2026-meg}
{Choustikov} N. {et~al.}, 2026, The Open Journal of Astrophysics, 9, 58199

\bibitem[{{Clark} {et~al}\mbox{.}(2011){Clark}, {Glover}, {Klessen}, \&
  {Bromm}}]{Clark2011}
{Clark} P.~C., {Glover} S. C.~O., {Klessen} R.~S., {Bromm} V., 2011, \apj, 727,
  110

\bibitem[{{Commer{\c{c}}on}, {Debout} \& {Teyssier}(2014){Commer{\c{c}}on},
  {Debout}, \& {Teyssier}}]{Commercon2014}
{Commer{\c{c}}on} B., {Debout} V., {Teyssier} R., 2014, \aap, 563, A11

\bibitem[{{Cullen} {et~al}\mbox{.}(2025){Cullen}, {Carnall}, {Scholte},
  {McLeod}, {McLure}, {Arellano-C{\'o}rdova}, {Stanton}, {Donnan}, {Dunlop},
  {Shapley}, {Barrufet}, {Begley}, {Bondestam}, {Cirasuolo}, {Leung},
  {Pollock}, \& {Stevenson}}]{Cullen2025}
{Cullen} F. {et~al.}, 2025, arXiv e-prints, arXiv:2501.11099

\bibitem[{{Cullen} {et~al}\mbox{.}(2024){Cullen}, {McLeod}, {McLure}, {Dunlop},
  {Donnan}, {Carnall}, {Keating}, {Magee}, {Arellano-Cordova}, {Bowler},
  {Begley}, {Flury}, {Hamadouche}, \& {Stanton}}]{Cullen2024}
{Cullen} F. {et~al.}, 2024, \mnras, 531, 997

\bibitem[{{Davis} {et~al}\mbox{.}(2024){Davis}, {Trump}, {Simons}, {McGrath},
  {Wilkins}, {Arrabal Haro}, {Bagley}, {Dickinson}, {Fern{\'a}ndez},
  {Amor{\'\i}n}, {Backhaus}, {Cleri}, {Llerena}, {Brunker}, {Barro},
  {Bisigello}, {Brooks}, {Costantin}, {de La Vega}, {Dekel}, {Finkelstein},
  {Hathi}, {Hirschmann}, {Kartaltepe}, {Koekemoer}, {Lucas}, {Papovich},
  {P{\'e}rez-Gonz{\'a}lez}, {Pirzkal}, {Rodighiero}, {Rose}, {Yung}, \& {Ceers
  Collaborators}}]{Davis2024}
{Davis} K. {et~al.}, 2024, \apj, 974, 42

\bibitem[{{de Graaff} {et~al}\mbox{.}(2024){de Graaff}, {Setton}, {Brammer},
  {Cutler}, {Suess}, {Labbe}, {Leja}, {Weibel}, {Maseda}, {Whitaker},
  {Bezanson}, {Boogaard}, {Cleri}, {De Lucia}, {Franx}, {Greene}, {Hirschmann},
  {Matthee}, {McConachie}, {Naidu}, {Oesch}, {Price}, {Rix}, {Valentino},
  {Wang}, \& {Williams}}]{deGraaff2024}
{de Graaff} A. {et~al.}, 2024, arXiv e-prints, arXiv:2404.05683

\bibitem[{{Dere} {et~al}\mbox{.}(2019){Dere}, {Del Zanna}, {Young}, {Landi}, \&
  {Sutherland}}]{Dere2019}
{Dere} K.~P., {Del Zanna} G., {Young} P.~R., {Landi} E., {Sutherland} R.~S.,
  2019, \apjs, 241, 22

\bibitem[{{D'Eugenio} {et~al}\mbox{.}(2025){D'Eugenio}, {Cameron}, {Scholtz},
  {Carniani}, {Willott}, {Curtis-Lake}, {Bunker}, {Parlanti}, {Maiolino},
  {Willmer}, {Jakobsen}, {Robertson}, {Johnson}, {Tacchella}, {Cargile},
  {Rawle}, {Arribas}, {Chevallard}, {Curti}, {Egami}, {Eisenstein}, {Kumari},
  {Looser}, {Rieke}, {Rodr{\'\i}guez Del Pino}, {Saxena}, {{\"U}bler},
  {Venturi}, {Witstok}, {Baker}, {Bhatawdekar}, {Bonaventura}, {Boyett},
  {Charlot}, {Danhaive}, {Hainline}, {Hausen}, {Helton}, {Ji}, {Ji}, {Jones},
  {Juod{\v{z}}balis}, {Maseda}, {P{\'e}rez-Gonz{\'a}lez}, {Perna},
  {Pusk{\'a}s}, {Shivaei}, {Silcock}, {Simmonds}, {Smit}, {Sun}, {Villanueva},
  {Williams}, \& {Zhu}}]{DEugenio2025}
{D'Eugenio} F. {et~al.}, 2025, \apjs, 277, 4

\bibitem[{{Dijkstra}(2009)}]{Dijkstra2009}
{Dijkstra} M., 2009, \apj, 690, 82

\bibitem[{{Dome} {et~al}\mbox{.}(2024){Dome}, {Tacchella}, {Fialkov},
  {Ceverino}, {Dekel}, {Ginzburg}, {Lapiner}, \& {Looser}}]{Dome2024}
{Dome} T., {Tacchella} S., {Fialkov} A., {Ceverino} D., {Dekel} A., {Ginzburg}
  O., {Lapiner} S., {Looser} T.~J., 2024, \mnras, 527, 2139

\bibitem[{{Dopita} {et~al}\mbox{.}(2000){Dopita}, {Kewley}, {Heisler}, \&
  {Sutherland}}]{Dopita2000}
{Dopita} M.~A., {Kewley} L.~J., {Heisler} C.~A., {Sutherland} R.~S., 2000,
  \apj, 542, 224

\bibitem[{{Douna} {et~al}\mbox{.}(2015){Douna}, {Pellizza}, {Mirabel}, \&
  {Pedrosa}}]{Douna2015}
{Douna} V.~M., {Pellizza} L.~J., {Mirabel} I.~F., {Pedrosa} S.~E., 2015, \aap,
  579, A44

\bibitem[{{Draine}(2011)}]{Draine2011}
{Draine} B.~T., 2011, {Physics of the Interstellar and Intergalactic Medium}

\bibitem[{{Draine} \& {Bertoldi}(1996)}]{Draine1996}
{Draine} B.~T., {Bertoldi} F., 1996, \apj, 468, 269

\bibitem[{{Dubois} {et~al}\mbox{.}(2016){Dubois}, {Peirani}, {Pichon},
  {Devriendt}, {Gavazzi}, {Welker}, \& {Volonteri}}]{Dubois2016}
{Dubois} Y., {Peirani} S., {Pichon} C., {Devriendt} J., {Gavazzi} R., {Welker}
  C., {Volonteri} M., 2016, \mnras, 463, 3948

\bibitem[{{Dunlop} {et~al}\mbox{.}(2013){Dunlop}, {Rogers}, {McLure}, {Ellis},
  {Robertson}, {Koekemoer}, {Dayal}, {Curtis-Lake}, {Wild}, {Charlot},
  {Bowler}, {Schenker}, {Ouchi}, {Ono}, {Cirasuolo}, {Furlanetto}, {Stark},
  {Targett}, \& {Schneider}}]{Dunlop2013}
{Dunlop} J.~S. {et~al.}, 2013, \mnras, 432, 3520

\bibitem[{{Eldridge} {et~al}\mbox{.}(2017){Eldridge}, {Stanway}, {Xiao},
  {McClelland}, {Taylor}, {Ng}, {Greis}, \& {Bray}}]{Eldridge2017}
{Eldridge} J.~J., {Stanway} E.~R., {Xiao} L., {McClelland} L.~A.~S., {Taylor}
  G., {Ng} M., {Greis} S.~M.~L., {Bray} J.~C., 2017, \pasa, 34, e058

\bibitem[{{Federrath} \& {Klessen}(2012)}]{Federrath2012}
{Federrath} C., {Klessen} R.~S., 2012, \apj, 761, 156

\bibitem[{{Ferland} {et~al}\mbox{.}(2017){Ferland}, {Chatzikos}, {Guzm{\'a}n},
  {Lykins}, {van Hoof}, {Williams}, {Abel}, {Badnell}, {Keenan}, {Porter}, \&
  {Stancil}}]{Ferland2017}
{Ferland} G.~J. {et~al.}, 2017, \rmxaa, 53, 385

\bibitem[{{Field}, {Goldsmith} \& {Habing}(1969){Field}, {Goldsmith}, \&
  {Habing}}]{Field1969}
{Field} G.~B., {Goldsmith} D.~W., {Habing} H.~J., 1969, \apjl, 155, L149

\bibitem[{{Finkelstein} {et~al}\mbox{.}(2023){Finkelstein}, {Leung}, {Bagley},
  {Dickinson}, {Ferguson}, {Papovich}, {Akins}, {Arrabal Haro}, {Dave},
  {Dekel}, {Kartaltepe}, {Kocevski}, {Koekemoer}, {Pirzkal}, {Somerville},
  {Yung}, {Amorin}, {Backhaus}, {Behroozi}, {Bisigello}, {Bromm}, {Casey},
  {Chavez Ortiz}, {Cheng}, {Chworowsky}, {Cleri}, {Cooper}, {Davis}, {de la
  Vega}, {Elbaz}, {Franco}, {Fontana}, {Fujimoto}, {Giavalisco}, {Grogin},
  {Holwerda}, {Huertas-Company}, {Hirschmann}, {Iyer}, {Jogee}, {Jung},
  {Larson}, {Lucas}, {Mobasher}, {Morales}, {Morley}, {Mukherjee},
  {Perez-Gonzalez}, {Ravindranath}, {Rodighiero}, {Rowland}, {Tacchella},
  {Taylor}, {Trump}, \& {Wilkins}}]{Finkelstein2023_b}
{Finkelstein} S.~L. {et~al.}, 2023, arXiv e-prints, arXiv:2311.04279

\bibitem[{{Fujimoto} {et~al}\mbox{.}(2025){Fujimoto}, {Naidu}, {Chisholm},
  {Atek}, {Endsley}, {Kokorev}, {Furtak}, {Pan}, {Liu}, {Bromm}, {Venditti},
  {Visbal}, {Sarmento}, {Weibel}, {Oesch}, {Brammer}, {Schaerer}, {Adamo},
  {Berg}, {Bezanson}, {Chemerynska}, {Claeyssens}, {Dessauges-Zavadsky},
  {Frebel}, {Korber}, {Labbe}, {Marques-Chaves}, {Matthee}, {McQuinn},
  {Mu{\~n}oz}, {Natarajan}, {Saldana-Lopez}, {Suess}, {Volonteri}, \&
  {Zitrin}}]{Fujimoto2025}
{Fujimoto} S. {et~al.}, 2025, arXiv e-prints, arXiv:2501.11678

\bibitem[{{Fumagalli} {et~al}\mbox{.}(2012){Fumagalli}, {Patel}, {Franx},
  {Brammer}, {van Dokkum}, {da Cunha}, {Kriek}, {Lundgren}, {Momcheva}, {Rix},
  {Schmidt}, {Skelton}, {Whitaker}, {Labbe}, \& {Nelson}}]{Fumagalli2012}
{Fumagalli} M. {et~al.}, 2012, \apjl, 757, L22

\bibitem[{{Furtak} {et~al}\mbox{.}(2023){Furtak}, {Zitrin}, {Weaver}, {Atek},
  {Bezanson}, {Labb{\'e}}, {Whitaker}, {Leja}, {Price}, {Brammer}, {Wang},
  {Marchesini}, {Pan}, {Dayal}, {van Dokkum}, {Feldmann}, {Fujimoto}, {Franx},
  {Khullar}, {Nelson}, \& {Mowla}}]{Furtak2023}
{Furtak} L.~J. {et~al.}, 2023, \mnras, 523, 4568

\bibitem[{{Garcia}, {Ricotti} \& {Sugimura}(2025){Garcia}, {Ricotti}, \&
  {Sugimura}}]{Garcia2025O}
{Garcia} F. A.~B., {Ricotti} M., {Sugimura} K., 2025, The Open Journal of
  Astrophysics, 8, 146

\bibitem[{{Gardner} {et~al}\mbox{.}(2023){Gardner}, {Mather}, {Abbott},
  {Abell}, {Abernathy}, {Abney}, {Abraham}, {Abraham}, {Abul-Huda}, {Acton},
  {Adams}, {Adams}, {Adler}, {Adriaensen}, {Aguilar}, {Ahmed}, {Ahmed},
  {Ahmed}, {Albat}, {Albert}, {Alberts}, {Aldridge}, {Allen}, {Allen},
  {Altenburg}, {Altunc}, {Alvarez}, {{\'A}lvarez-M{\'a}rquez}, {Alves de
  Oliveira}, {Ambrose}, {Anandakrishnan}, {Andersen}, {Anderson}, {Anderson},
  {Anderson}, {Anderson}, {Aprea}, {Archer}, {Arenberg}, {Argyriou}, {Arribas},
  {Artigau}, {Arvai}, {Atcheson}, {Atkinson}, {Averbukh}, {Aymergen},
  {Bacinski}, {Baggett}, {Bagnasco}, {Baker}, {Balzano}, {Banks}, {Baran},
  {Barker}, {Barrett}, {Barringer}, {Barto}, {Bast}, {Baudoz}, {Baum},
  {Beatty}, {Beaulieu}, {Bechtold}, {Beck}, {Beddard}, {Beichman}, {Bellagama},
  {Bely}, {Berger}, {Bergeron}, {Bernier}, {Bertch}, {Beskow}, {Betz},
  {Biagetti}, {Birkmann}, {Bjorklund}, {Blackwood}, {Blazek}, {Blossfeld},
  {Bluth}, {Boccaletti}, {Boegner}, {Bohlin}, {Boia}, {B{\"o}ker},
  {Bonaventura}, {Bond}, {Bosley}, {Boucarut}, {Bouchet}, {Bouwman}, {Bower},
  {Bowers}, {Bowers}, {Boyce}, {Boyer}, {Boyer}, {Boyer}, {Boyer}, {Bradley},
  {Brady}, {Brandl}, {Brannen}, {Breda}, {Bremmer}, {Brennan}, {Bresnahan},
  {Bright}, {Broiles}, {Bromenschenkel}, {Brooks}, {Brooks}, {Brown}, {Brown},
  {Brown}, {Bruce}, {Bryson}, {Bujanda}, {Bullock}, {Bunker}, {Bureo}, {Burt},
  {Bush}, {Bushouse}, {Bussman}, {Cabaud}, {Cale}, {Calhoon}, {Calvani},
  {Canipe}, {Caputo}, {Cara}, {Carey}, {Case}, {Cesari}, {Cetorelli}, {Chance},
  {Chandler}, {Chaney}, {Chapman}, {Charlot}, {Chayer}, {Cheezum}, {Chen},
  {Chen}, {Cherinka}, {Chichester}, {Chilton}, {Chittiraibalan}, {Clampin},
  {Clark}, {Clark}, {Clark}, {Claybrooks}, {Cleveland}, {Cohen}, {Cohen},
  {Col{\'o}n}, {Coleman}, {Colina}, {Comber}, {Comeau}, {Comer}, {Conde Reis},
  {Connolly}, {Conroy}, {Contos}, {Contreras}, {Cook}, {Cooper}, {Cooper},
  {Correia}, {Correnti}, {Cossou}, {Costanza}, {Coulais}, {Cox}, {Coyle},
  {Cracraft}, {Crew}, {Curtis}, {Cusveller}, {Da Costa Maciel}, {Dailey},
  {Daugeron}, {Davidson}, {Davies}, {Davis}, {Davis}, {Day}, {de Chambure}, {de
  Jong}, {De Marchi}, {Dean}, {Decker}, {Delisa}, {Dell}, \&
  {Dellagatta}}]{Gardner2023}
{Gardner} J.~P. {et~al.}, 2023, \pasp, 135, 068001

\bibitem[{{Gardner} {et~al}\mbox{.}(2006){Gardner}, {Mather}, {Clampin},
  {Doyon}, {Greenhouse}, {Hammel}, {Hutchings}, {Jakobsen}, {Lilly}, {Long},
  {Lunine}, {McCaughrean}, {Mountain}, {Nella}, {Rieke}, {Rieke}, {Rix},
  {Smith}, {Sonneborn}, {Stiavelli}, {Stockman}, {Windhorst}, \&
  {Wright}}]{Gardner2006}
{Gardner} J.~P. {et~al.}, 2006, \ssr, 123, 485

\bibitem[{{Garg} {et~al}\mbox{.}(2024){Garg}, {Narayanan}, {Sanders},
  {Dav{\'e}}, {Popping}, {Shapley}, {Stark}, \& {Trump}}]{Garg2024}
{Garg} P., {Narayanan} D., {Sanders} R.~L., {Dav{\'e}} R., {Popping} G.,
  {Shapley} A.~E., {Stark} D.~P., {Trump} J.~R., 2024, \apj, 972, 113

\bibitem[{{Gelli} {et~al}\mbox{.}(2025){Gelli}, {Pallottini}, {Salvadori},
  {Ferrara}, {Mason}, {Carniani}, \& {Ginolfi}}]{Gelli2025}
{Gelli} V., {Pallottini} A., {Salvadori} S., {Ferrara} A., {Mason} C.,
  {Carniani} S., {Ginolfi} M., 2025, arXiv e-prints, arXiv:2501.16418

\bibitem[{{Gelli} {et~al}\mbox{.}(2024){Gelli}, {Salvadori}, {Ferrara}, \&
  {Pallottini}}]{Gelli2024}
{Gelli} V., {Salvadori} S., {Ferrara} A., {Pallottini} A., 2024, \apj, 964, 76

\bibitem[{{Glazebrook} {et~al}\mbox{.}(2024){Glazebrook}, {Nanayakkara},
  {Schreiber}, {Lagos}, {Kawinwanichakij}, {Jacobs}, {Chittenden}, {Brammer},
  {Kacprzak}, {Labbe}, {Marchesini}, {Marsan}, {Oesch}, {Papovich}, {Remus},
  {Tran}, {Esdaile}, \& {Chandro-Gomez}}]{Glazebrook2024}
{Glazebrook} K. {et~al.}, 2024, \nat, 628, 277

\bibitem[{{Glover} \& {Abel}(2008)}]{Glover2008}
{Glover} S.~C.~O., {Abel} T., 2008, \mnras, 388, 1627

\bibitem[{{Glover} \& {Clark}(2012)}]{Glover2012}
{Glover} S. C.~O., {Clark} P.~C., 2012, \mnras, 421, 116

\bibitem[{{Glover} {et~al}\mbox{.}(2010){Glover}, {Federrath}, {Mac Low}, \&
  {Klessen}}]{Glover2010}
{Glover} S.~C.~O., {Federrath} C., {Mac Low} M.~M., {Klessen} R.~S., 2010,
  \mnras, 404, 2

\bibitem[{{Gnat} \& {Sternberg}(2007)}]{Gnat2007}
{Gnat} O., {Sternberg} A., 2007, \apjs, 168, 213

\bibitem[{{Gnedin}, {Tassis} \& {Kravtsov}(2009){Gnedin}, {Tassis}, \&
  {Kravtsov}}]{Gnedin2009}
{Gnedin} N.~Y., {Tassis} K., {Kravtsov} A.~V., 2009, \apj, 697, 55

\bibitem[{{Gr{\"a}fener} \& {Vink}(2015)}]{Grafener2015}
{Gr{\"a}fener} G., {Vink} J.~S., 2015, \aap, 578, L2

\bibitem[{{Gray} \& {Scannapieco}(2017)}]{Grey2017}
{Gray} W.~J., {Scannapieco} E., 2017, \apj, 849, 132

\bibitem[{{Grevesse} {et~al}\mbox{.}(2010){Grevesse}, {Asplund}, {Sauval}, \&
  {Scott}}]{Grevesse2010}
{Grevesse} N., {Asplund} M., {Sauval} A.~J., {Scott} P., 2010, \apss, 328, 179

\bibitem[{{Grisdale} {et~al}\mbox{.}(2021){Grisdale}, {Thatte}, {Devriendt},
  {Pereira-Santaella}, {Slyz}, {Kimm}, {Dubois}, \& {Yi}}]{Grisdale2021}
{Grisdale} K., {Thatte} N., {Devriendt} J., {Pereira-Santaella} M., {Slyz} A.,
  {Kimm} T., {Dubois} Y., {Yi} S.~K., 2021, \mnras, 501, 5517

\bibitem[{Guillet \& Teyssier(2011)}]{guillet_SimpleMultigridScheme_2011}
Guillet T., Teyssier R., 2011, Journal of Computational Physics, 230, 4756

\bibitem[{{Haiman}, {Thoul} \& {Loeb}(1996){Haiman}, {Thoul}, \&
  {Loeb}}]{Haiman1996}
{Haiman} Z., {Thoul} A.~A., {Loeb} A., 1996, \apj, 464, 523

\bibitem[{{Harikane} {et~al}\mbox{.}(2024){Harikane}, {Nakajima}, {Ouchi},
  {Umeda}, {Isobe}, {Ono}, {Xu}, \& {Zhang}}]{Harikane2024}
{Harikane} Y., {Nakajima} K., {Ouchi} M., {Umeda} H., {Isobe} Y., {Ono} Y.,
  {Xu} Y., {Zhang} Y., 2024, \apj, 960, 56

\bibitem[{{Harikane} {et~al}\mbox{.}(2020){Harikane}, {Ouchi}, {Inoue},
  {Matsuoka}, {Tamura}, {Bakx}, {Fujimoto}, {Moriwaki}, {Ono}, {Nagao},
  {Tadaki}, {Kojima}, {Shibuya}, {Egami}, {Ferrara}, {Gallerani}, {Hashimoto},
  {Kohno}, {Matsuda}, {Matsuo}, {Pallottini}, {Sugahara}, \&
  {Vallini}}]{Harikane2020}
{Harikane} Y. {et~al.}, 2020, \apj, 896, 93

\bibitem[{Harris {et~al}\mbox{.}(2020)Harris, Millman, van~der Walt, Gommers,
  Virtanen, Cournapeau, Wieser, Taylor, Berg, Smith, Kern, Picus, Hoyer, van
  Kerkwijk, Brett, Haldane, del R{\'{i}}o, Wiebe, Peterson,
  G{\'{e}}rard-Marchant, Sheppard, Reddy, Weckesser, Abbasi, Gohlke, \&
  Oliphant}]{Harris2020}
Harris C.~R. {et~al.}, 2020, Nature, 585, 357

\bibitem[{{Hashimoto} {et~al}\mbox{.}(2019){Hashimoto}, {Inoue}, {Mawatari},
  {Tamura}, {Matsuo}, {Furusawa}, {Harikane}, {Shibuya}, {Knudsen}, {Kohno},
  {Ono}, {Zackrisson}, {Okamoto}, {Kashikawa}, {Oesch}, {Ouchi}, {Ota},
  {Shimizu}, {Taniguchi}, {Umehata}, \& {Watson}}]{Hashimoto2019}
{Hashimoto} T. {et~al.}, 2019, \pasj, 71, 71

\bibitem[{{Hashimoto} {et~al}\mbox{.}(2018){Hashimoto}, {Laporte}, {Mawatari},
  {Ellis}, {Inoue}, {Zackrisson}, {Roberts-Borsani}, {Zheng}, {Tamura},
  {Bauer}, {Fletcher}, {Harikane}, {Hatsukade}, {Hayatsu}, {Matsuda}, {Matsuo},
  {Okamoto}, {Ouchi}, {Pell{\'o}}, {Rydberg}, {Shimizu}, {Taniguchi},
  {Umehata}, \& {Yoshida}}]{Hashimoto2018}
{Hashimoto} T. {et~al.}, 2018, \nat, 557, 392

\bibitem[{{Heays}, {Bosman} \& {van Dishoeck}(2017){Heays}, {Bosman}, \& {van
  Dishoeck}}]{Heays2017}
{Heays} A.~N., {Bosman} A.~D., {van Dishoeck} E.~F., 2017, \aap, 602, A105

\bibitem[{{Heger} \& {Woosley}(2002)}]{Heger2002}
{Heger} A., {Woosley} S.~E., 2002, \apj, 567, 532

\bibitem[{{Heintz} {et~al}\mbox{.}(2024){Heintz}, {Brammer}, {Watson}, {Oesch},
  {Keating}, {Hayes}, {Abdurro'uf}, {Arellano-C{\'o}rdova}, {Carnall},
  {Christiansen}, {Cullen}, {Dav{\'e}}, {Dayal}, {Ferrara}, {Finlator},
  {Fynbo}, {Flury}, {Gelli}, {Gillman}, {Gottumukkala}, {Gould}, {Greve},
  {Hardin}, {Y. -Y Hsiao}, {Hutter}, {Jakobsson}, {Killi}, {Khosravaninezhad},
  {Laursen}, {Lee}, {Magdis}, {Matthee}, {Naidu}, {Narayanan}, {Pollock},
  {Prescott}, {Rusakov}, {Shuntov}, {Sneppen}, {Smit}, {Tanvir}, {Terp},
  {Toft}, {Valentino}, {Vijayan}, {Weaver}, {Wise}, \& {Witstok}}]{Heintz2024}
{Heintz} K.~E. {et~al.}, 2024, arXiv e-prints, arXiv:2404.02211

\bibitem[{{Hirai} {et~al}\mbox{.}(2025){Hirai}, {Saitoh}, {Fujii}, {Kaneko}, \&
  {Beers}}]{Hirai2025}
{Hirai} Y., {Saitoh} T.~R., {Fujii} M.~S., {Kaneko} K., {Beers} T.~C., 2025,
  \apjl, 980, L25

\bibitem[{{Hirano} {et~al}\mbox{.}(2014){Hirano}, {Hosokawa}, {Yoshida},
  {Umeda}, {Omukai}, {Chiaki}, \& {Yorke}}]{Hirano2014}
{Hirano} S., {Hosokawa} T., {Yoshida} N., {Umeda} H., {Omukai} K., {Chiaki} G.,
  {Yorke} H.~W., 2014, \apj, 781, 60

\bibitem[{{Hirschmann} {et~al}\mbox{.}(2023){Hirschmann}, {Charlot}, {Feltre},
  {Curtis-Lake}, {Somerville}, {Chevallard}, {Choi}, {Nelson}, {Morisset},
  {Plat}, \& {Vidal-Garcia}}]{Hirschmann2023}
{Hirschmann} M. {et~al.}, 2023, \mnras, 526, 3610

\bibitem[{{Hirschmann}, {Charlot} \& {Somerville}(2023){Hirschmann}, {Charlot},
  \& {Somerville}}]{Hirschmann2023b}
{Hirschmann} M., {Charlot} S., {Somerville} R.~S., 2023, \mnras, 526, 3504

\bibitem[{{Hollenbach} \& {McKee}(1989)}]{Hollenbach1989}
{Hollenbach} D., {McKee} C.~F., 1989, \apj, 342, 306

\bibitem[{{Hu} {et~al}\mbox{.}(2024){Hu}, {Papovich}, {Dickinson}, {Kennicutt},
  {Shen}, {Amor{\'\i}n}, {Arrabal Haro}, {Bagley}, {Bhatawdekar}, {Cleri},
  {Cole}, {Dekel}, {de la Vega}, {Finkelstein}, {Grogin}, {Hathi},
  {Hirschmann}, {Holwerda}, {Hutchison}, {Jung}, {Koekemoer}, {Kartaltepe},
  {Lucas}, {Llerena}, {Mascia}, {Mobasher}, {Napolitano}, {Newman},
  {Pentericci}, {P{\'e}rez-Gonz{\'a}lez}, {Trump}, {Wilkins}, \&
  {Yung}}]{Hu2024}
{Hu} W. {et~al.}, 2024, \apj, 971, 21

\bibitem[{{Hui} \& {Gnedin}(1997)}]{Hui1997}
{Hui} L., {Gnedin} N.~Y., 1997, \mnras, 292, 27

\bibitem[{Hunter(2007)}]{Hunter2007}
Hunter J.~D., 2007, Computing in Science \& Engineering, 9, 90

\bibitem[{{Isobe} {et~al}\mbox{.}(2023){Isobe}, {Ouchi}, {Tominaga},
  {Watanabe}, {Nakajima}, {Umeda}, {Yajima}, {Harikane}, {Fukushima}, {Xu},
  {Ono}, \& {Zhang}}]{Isobe2023_nitrogen}
{Isobe} Y. {et~al.}, 2023, \apj, 959, 100

\bibitem[{{Izotov} {et~al}\mbox{.}(2021){Izotov}, {Guseva}, {Fricke}, {Henkel},
  {Schaerer}, \& {Thuan}}]{Izotov2021}
{Izotov} Y.~I., {Guseva} N.~G., {Fricke} K.~J., {Henkel} C., {Schaerer} D.,
  {Thuan} T.~X., 2021, \aap, 646, A138

\bibitem[{{Izotov}, {Guseva} \& {Thuan}(2011){Izotov}, {Guseva}, \&
  {Thuan}}]{Izotov2011}
{Izotov} Y.~I., {Guseva} N.~G., {Thuan} T.~X., 2011, \apj, 728, 161

\bibitem[{{Jaacks} {et~al}\mbox{.}(2018){Jaacks}, {Thompson}, {Finkelstein}, \&
  {Bromm}}]{Jaacks2018}
{Jaacks} J., {Thompson} R., {Finkelstein} S.~L., {Bromm} V., 2018, \mnras, 475,
  4396

\bibitem[{{Jappsen} {et~al}\mbox{.}(2007){Jappsen}, {Glover}, {Klessen}, \&
  {Mac Low}}]{Jappsen2007}
{Jappsen} A.~K., {Glover} S.~C.~O., {Klessen} R.~S., {Mac Low} M.~M., 2007,
  \apj, 660, 1332

\bibitem[{{Jin}, {Kewley} \& {Sutherland}(2022){Jin}, {Kewley}, \&
  {Sutherland}}]{Jin2022}
{Jin} Y., {Kewley} L.~J., {Sutherland} R.~S., 2022, \apjl, 934, L8

\bibitem[{{Joshi} {et~al}\mbox{.}(2025){Joshi}, {Pontzen}, {Agertz}, {Rey},
  {Read}, \& {Pillepich}}]{Joshi2025}
{Joshi} G.~D., {Pontzen} A., {Agertz} O., {Rey} M.~P., {Read} J., {Pillepich}
  A., 2025, \mnras, 537, 3792

\bibitem[{{Kaaret}, {Schmitt} \& {Gorski}(2011){Kaaret}, {Schmitt}, \&
  {Gorski}}]{Kaaret2011}
{Kaaret} P., {Schmitt} J., {Gorski} M., 2011, \apj, 741, 10

\bibitem[{{Kannan} {et~al}\mbox{.}(2022){Kannan}, {Garaldi}, {Smith}, {Pakmor},
  {Springel}, {Vogelsberger}, \& {Hernquist}}]{Kannan2022}
{Kannan} R., {Garaldi} E., {Smith} A., {Pakmor} R., {Springel} V.,
  {Vogelsberger} M., {Hernquist} L., 2022, \mnras, 511, 4005

\bibitem[{{Kannan} {et~al}\mbox{.}(2025){Kannan}, {Puchwein}, {Smith},
  {Borrow}, {Garaldi}, {Keating}, {Vogelsberger}, {Zier}, {McClymont}, {Shen},
  {Popovic}, {Tacchella}, {Hernquist}, \& {Springel}}]{Kannan2025}
{Kannan} R. {et~al.}, 2025, arXiv e-prints, arXiv:2502.20437

\bibitem[{{Katz}(2022)}]{RTZ}
{Katz} H., 2022, \mnras, 512, 348

\bibitem[{{Katz} {et~al}\mbox{.}(2024{\natexlab{a}}){Katz}, {Cameron},
  {Saxena}, {Barrufet}, {Choustikov}, {Cleri}, {de Graaff}, {Ellis}, {Fosbury},
  {Heintz}, {Maseda}, {Matthee}, {McConchie}, \& {Oesch}}]{Katz2024_BJ}
{Katz} H. {et~al.}, 2024{\natexlab{a}}, arXiv e-prints, arXiv:2408.03189

\bibitem[{{Katz} {et~al}\mbox{.}(2019{\natexlab{a}}){Katz}, {Galligan}, {Kimm},
  {Rosdahl}, {Haehnelt}, {Blaizot}, {Devriendt}, {Slyz}, {Laporte}, \&
  {Ellis}}]{Katz2019_ism}
{Katz} H. {et~al.}, 2019{\natexlab{a}}, \mnras, 487, 5902

\bibitem[{{Katz} {et~al}\mbox{.}(2024{\natexlab{b}}){Katz}, {Ji}, {Telford}, \&
  {Senchyna}}]{Katz2024_He}
{Katz} H., {Ji} A.~P., {Telford} G., {Senchyna} P., 2024{\natexlab{b}}, The
  Open Journal of Astrophysics, 7, 106

\bibitem[{{Katz} {et~al}\mbox{.}(2023{\natexlab{a}}){Katz}, {Kimm}, {Ellis},
  {Devriendt}, \& {Slyz}}]{Katz2023_Pop3}
{Katz} H., {Kimm} T., {Ellis} R.~S., {Devriendt} J., {Slyz} A.,
  2023{\natexlab{a}}, \mnras, 524, 351

\bibitem[{{Katz} {et~al}\mbox{.}(2019{\natexlab{b}}){Katz}, {Laporte}, {Ellis},
  {Devriendt}, \& {Slyz}}]{Katz2019_BB}
{Katz} H., {Laporte} N., {Ellis} R.~S., {Devriendt} J., {Slyz} A.,
  2019{\natexlab{b}}, \mnras, 484, 4054

\bibitem[{{Katz} {et~al}\mbox{.}(2022){Katz}, {Liu}, {Kimm}, {Rey},
  {Andersson}, {Cameron}, {Rodriguez-Montero}, {Agertz}, {Devriendt}, \&
  {Slyz}}]{Katz2022_prism}
{Katz} H. {et~al.}, 2022, arXiv e-prints, arXiv:2211.04626

\bibitem[{{Katz} {et~al}\mbox{.}(2024{\natexlab{c}}){Katz}, {Rey}, {Cadiou},
  {Kimm}, \& {Agertz}}]{Katz2024_meg}
{Katz} H., {Rey} M.~P., {Cadiou} C., {Kimm} T., {Agertz} O.,
  2024{\natexlab{c}}, arXiv e-prints, arXiv:2411.07282

\bibitem[{{Katz} {et~al}\mbox{.}(2023{\natexlab{b}}){Katz}, {Rosdahl}, {Kimm},
  {Blaizot}, {Choustikov}, {Farcy}, {Garel}, {Haehnelt}, {Michel-Dansac}, \&
  {Ocvirk}}]{spdrv1}
{Katz} H. {et~al.}, 2023{\natexlab{b}}, The Open Journal of Astrophysics, 6, 44

\bibitem[{{Katz} {et~al}\mbox{.}(2023{\natexlab{c}}){Katz}, {Saxena},
  {Cameron}, {Carniani}, {Bunker}, {Arribas}, {Bhatawdekar}, {Bowler},
  {Boyett}, {Cresci}, {Curtis-Lake}, {D'Eugenio}, {Kumari}, {Looser},
  {Maiolino}, {{\"U}bler}, {Willott}, \& {Witstok}}]{Katz2023_jwst}
{Katz} H. {et~al.}, 2023{\natexlab{c}}, \mnras, 518, 592

\bibitem[{{Katz} {et~al}\mbox{.}(2023{\natexlab{d}}){Katz}, {Saxena},
  {Rosdahl}, {Kimm}, {Blaizot}, {Garel}, {Michel-Dansac}, {Haehnelt}, {Ellis},
  {Pentericci}, {Devriendt}, \& {Slyz}}]{Katz2023_bursty}
{Katz} H. {et~al.}, 2023{\natexlab{d}}, \mnras, 518, 270

\bibitem[{{Kewley}, {Nicholls} \& {Sutherland}(2019){Kewley}, {Nicholls}, \&
  {Sutherland}}]{Kewley2019}
{Kewley} L.~J., {Nicholls} D.~C., {Sutherland} R.~S., 2019, \araa, 57, 511

\bibitem[{{Khatri}, {Romano-D{\'\i}az} \& {Porciani}(2025){Khatri},
  {Romano-D{\'\i}az}, \& {Porciani}}]{Khatri2025}
{Khatri} P., {Romano-D{\'\i}az} E., {Porciani} C., 2025, \aap, 697, A174

\bibitem[{{Khostovan} {et~al}\mbox{.}(2024){Khostovan}, {Malhotra}, {Rhoads},
  {Sobral}, {Harish}, {Tilvi}, {Coughlin}, \& {Rezaee}}]{Khostovan2024}
{Khostovan} A.~A., {Malhotra} S., {Rhoads} J.~E., {Sobral} D., {Harish} S.,
  {Tilvi} V., {Coughlin} A., {Rezaee} S., 2024, \mnras, 535, 2903

\bibitem[{{Kim} \& {Ostriker}(2015)}]{Kim2015}
{Kim} C.-G., {Ostriker} E.~C., 2015, \apj, 802, 99

\bibitem[{{Kim} {et~al}\mbox{.}(2023){Kim}, {Gong}, {Kim}, \&
  {Ostriker}}]{Kim2023}
{Kim} J.-G., {Gong} M., {Kim} C.-G., {Ostriker} E.~C., 2023, \apjs, 264, 10

\bibitem[{{Kimm} {et~al}\mbox{.}(2017){Kimm}, {Katz}, {Haehnelt}, {Rosdahl},
  {Devriendt}, \& {Slyz}}]{Kimm2017}
{Kimm} T., {Katz} H., {Haehnelt} M., {Rosdahl} J., {Devriendt} J., {Slyz} A.,
  2017, \mnras, 466, 4826

\bibitem[{{Kingdon} \& {Ferland}(1996)}]{Kingdon1996}
{Kingdon} J.~B., {Ferland} G.~J., 1996, \apjs, 106, 205

\bibitem[{{Kobayashi} {et~al}\mbox{.}(2006){Kobayashi}, {Umeda}, {Nomoto},
  {Tominaga}, \& {Ohkubo}}]{Kobayashi2006}
{Kobayashi} C., {Umeda} H., {Nomoto} K., {Tominaga} N., {Ohkubo} T., 2006,
  \apj, 653, 1145

\bibitem[{{Kokorev} {et~al}\mbox{.}(2024){Kokorev}, {Atek}, {Chisholm},
  {Endsley}, {Chemerynska}, {Mu{\~n}oz}, {Furtak}, {Pan}, {Berg}, {Fujimoto},
  {Oesch}, {Weibel}, {Adamo}, {Blaizot}, {Bouwens}, {Dessauges-Zavadsky},
  {Khullar}, {Korber}, {Goovaerts}, {Jecmen}, {Labb{\'e}}, {Leclercq},
  {Marques-Chaves}, {Mason}, {McQuinn}, {Naidu}, {Natarajan}, {Nelson},
  {Rosdahl}, {Saldana-Lopez}, {Schaerer}, {Trebitsch}, {Volonteri}, \&
  {Zitrin}}]{Kokorev2024}
{Kokorev} V. {et~al.}, 2024, arXiv e-prints, arXiv:2411.13640

\bibitem[{{Koyama} \& {Inutsuka}(2000)}]{Koyama2000}
{Koyama} H., {Inutsuka} S.-I., 2000, \apj, 532, 980

\bibitem[{{Kroupa}(2001)}]{Kroupa2001}
{Kroupa} P., 2001, \mnras, 322, 231

\bibitem[{Kumar \& Chen(2025)}]{Kumar2025}
Kumar S., Chen H.-W., 2025

\bibitem[{{Kuruvanthodi} {et~al}\mbox{.}(2024){Kuruvanthodi}, {Schaerer},
  {Marques-Chaves}, {Korber}, {Weibel}, {Oesch}, \&
  {Roberts-Borsani}}]{Kuruvanthodi2024}
{Kuruvanthodi} A., {Schaerer} D., {Marques-Chaves} R., {Korber} D., {Weibel}
  A., {Oesch} P.~A., {Roberts-Borsani} G., 2024, \aap, 691, A310

\bibitem[{{Larkin}, {Gerasimov} \& {Burgasser}(2023){Larkin}, {Gerasimov}, \&
  {Burgasser}}]{Larkin2023}
{Larkin} M.~M., {Gerasimov} R., {Burgasser} A.~J., 2023, \aj, 165, 2

\bibitem[{{Laseter} {et~al}\mbox{.}(2024){Laseter}, {Maseda}, {Curti},
  {Maiolino}, {D'Eugenio}, {Cameron}, {Looser}, {Arribas}, {Baker},
  {Bhatawdekar}, {Boyett}, {Bunker}, {Carniani}, {Charlot}, {Chevallard},
  {Curtis-lake}, {Egami}, {Eisenstein}, {Hainline}, {Hausen}, {Ji}, {Kumari},
  {Perna}, {Rawle}, {Rix}, {Robertson}, {Rodr{\'\i}guez Del Pino}, {Sandles},
  {Scholtz}, {Smit}, {Tacchella}, {{\"U}bler}, {Williams}, {Willott}, \&
  {Witstok}}]{Laseter2024}
{Laseter} I.~H. {et~al.}, 2024, \aap, 681, A70

\bibitem[{{Leitherer} {et~al}\mbox{.}(1999){Leitherer}, {Schaerer}, {Goldader},
  {Delgado}, {Robert}, {Kune}, {de Mello}, {Devost}, \&
  {Heckman}}]{Leitherer1999}
{Leitherer} C. {et~al.}, 1999, \apjs, 123, 3

\bibitem[{{Leung} {et~al}\mbox{.}(2023){Leung}, {Bagley}, {Finkelstein},
  {Ferguson}, {Koekemoer}, {P{\'e}rez-Gonz{\'a}lez}, {Morales}, {Kocevski},
  {Yang}, {Somerville}, {Wilkins}, {Yung}, {Fujimoto}, {Larson}, {Papovich},
  {Pirzkal}, {Berg}, {Lotz}, {Castellano}, {Ch{\'a}vez Ortiz}, {Cheng},
  {Dickinson}, {Giavalisco}, {Hathi}, {Hutchison}, {Jung}, {Kartaltepe},
  {Natarajan}, \& {Rothberg}}]{Leung2023}
{Leung} G. C.~K. {et~al.}, 2023, \apjl, 954, L46

\bibitem[{{Levermore}(1984)}]{Levermore1984}
{Levermore} C.~D., 1984, \jqsrt, 31, 149

\bibitem[{{Limongi} \& {Chieffi}(2018)}]{Limongi2018}
{Limongi} M., {Chieffi} A., 2018, \apjs, 237, 13

\bibitem[{{Llerena} {et~al}\mbox{.}(2024){Llerena}, {Amor{\'\i}n},
  {Pentericci}, {Arrabal Haro}, {Backhaus}, {Bagley}, {Calabr{\`o}}, {Cleri},
  {Davis}, {Dickinson}, {Finkelstein}, {Gawiser}, {Grogin}, {Hathi},
  {Hirschmann}, {Kartaltepe}, {Koekemoer}, {McGrath}, {Mobasher}, {Napolitano},
  {Papovich}, {Pirzkal}, {Trump}, {Wilkins}, \& {Yung}}]{Llerena2024}
{Llerena} M. {et~al.}, 2024, \aap, 691, A59

\bibitem[{{Looser} {et~al}\mbox{.}(2024){Looser}, {D'Eugenio}, {Maiolino},
  {Witstok}, {Sandles}, {Curtis-Lake}, {Chevallard}, {Tacchella}, {Johnson},
  {Baker}, {Suess}, {Carniani}, {Ferruit}, {Arribas}, {Bonaventura}, {Bunker},
  {Cameron}, {Charlot}, {Curti}, {de Graaff}, {Maseda}, {Rawle}, {Rix}, {Del
  Pino}, {Smit}, {{\"U}bler}, {Willott}, {Alberts}, {Egami}, {Eisenstein},
  {Endsley}, {Hausen}, {Rieke}, {Robertson}, {Shivaei}, {Williams}, {Boyett},
  {Chen}, {Ji}, {Jones}, {Kumari}, {Nelson}, {Perna}, {Saxena}, \&
  {Scholtz}}]{Looser2024}
{Looser} T.~J. {et~al.}, 2024, \nat, 629, 53

\bibitem[{{Lovell} {et~al}\mbox{.}(2021){Lovell}, {Vijayan}, {Thomas},
  {Wilkins}, {Barnes}, {Irodotou}, \& {Roper}}]{Lovell2021}
{Lovell} C.~C., {Vijayan} A.~P., {Thomas} P.~A., {Wilkins} S.~M., {Barnes}
  D.~J., {Irodotou} D., {Roper} W., 2021, \mnras, 500, 2127

\bibitem[{{Lupi} {et~al}\mbox{.}(2020){Lupi}, {Pallottini}, {Ferrara},
  {Bovino}, {Carniani}, \& {Vallini}}]{Lupi2020}
{Lupi} A., {Pallottini} A., {Ferrara} A., {Bovino} S., {Carniani} S., {Vallini}
  L., 2020, \mnras, 496, 5160

\bibitem[{{Luridiana}, {Morisset} \& {Shaw}(2015){Luridiana}, {Morisset}, \&
  {Shaw}}]{Luridiana2015}
{Luridiana} V., {Morisset} C., {Shaw} R.~A., 2015, \aap, 573, A42

\bibitem[{{Ma} {et~al}\mbox{.}(2018){Ma}, {Hopkins}, {Garrison-Kimmel},
  {Faucher-Gigu{\`e}re}, {Quataert}, {Boylan-Kolchin}, {Hayward}, {Feldmann},
  \& {Kere{\v{s}}}}]{Ma2018}
{Ma} X. {et~al.}, 2018, \mnras, 478, 1694

\bibitem[{{Ma} {et~al}\mbox{.}(2016){Ma}, {Hopkins}, {Kasen}, {Quataert},
  {Faucher-Gigu{\`e}re}, {Kere{\v{s}}}, {Murray}, \& {Strom}}]{Ma2016}
{Ma} X., {Hopkins} P.~F., {Kasen} D., {Quataert} E., {Faucher-Gigu{\`e}re}
  C.-A., {Kere{\v{s}}} D., {Murray} N., {Strom} A., 2016, \mnras, 459, 3614

\bibitem[{{Mainali} {et~al}\mbox{.}(2017){Mainali}, {Kollmeier}, {Stark},
  {Simcoe}, {Walth}, {Newman}, \& {Miller}}]{Mainali2017}
{Mainali} R., {Kollmeier} J.~A., {Stark} D.~P., {Simcoe} R.~A., {Walth} G.,
  {Newman} A.~B., {Miller} D.~R., 2017, \apjl, 836, L14

\bibitem[{{Maiolino} {et~al}\mbox{.}(2024){Maiolino}, {Scholtz}, {Curtis-Lake},
  {Carniani}, {Baker}, {de Graaff}, {Tacchella}, {{\"U}bler}, {D'Eugenio},
  {Witstok}, {Curti}, {Arribas}, {Bunker}, {Charlot}, {Chevallard},
  {Eisenstein}, {Egami}, {Ji}, {Jones}, {Lyu}, {Rawle}, {Robertson},
  {Rujopakarn}, {Perna}, {Sun}, {Venturi}, {Williams}, \&
  {Willott}}]{Maiolino2024b}
{Maiolino} R. {et~al.}, 2024, \aap, 691, A145

\bibitem[{{Maoz} \& {Graur}(2017)}]{Maoz2017}
{Maoz} D., {Graur} O., 2017, \apj, 848, 25

\bibitem[{{Mapelli} {et~al}\mbox{.}(2010){Mapelli}, {Ripamonti}, {Zampieri},
  {Colpi}, \& {Bressan}}]{Mapelli2010}
{Mapelli} M., {Ripamonti} E., {Zampieri} L., {Colpi} M., {Bressan} A., 2010,
  \mnras, 408, 234

\bibitem[{{Marks} {et~al}\mbox{.}(2012){Marks}, {Kroupa}, {Dabringhausen}, \&
  {Pawlowski}}]{Marks2012}
{Marks} M., {Kroupa} P., {Dabringhausen} J., {Pawlowski} M.~S., 2012, \mnras,
  422, 2246

\bibitem[{{M{\'a}rmol-Queralt{\'o}}
  {et~al}\mbox{.}(2016){M{\'a}rmol-Queralt{\'o}}, {McLure}, {Cullen}, {Dunlop},
  {Fontana}, \& {McLeod}}]{MQ2016}
{M{\'a}rmol-Queralt{\'o}} E., {McLure} R.~J., {Cullen} F., {Dunlop} J.~S.,
  {Fontana} A., {McLeod} D.~J., 2016, \mnras, 460, 3587

\bibitem[{{Martins} {et~al}\mbox{.}(2020){Martins}, {Schaerer},
  {Haemmerl{\'e}}, \& {Charbonnel}}]{Martins2020}
{Martins} F., {Schaerer} D., {Haemmerl{\'e}} L., {Charbonnel} C., 2020, \aap,
  633, A9

\bibitem[{{Matthee} {et~al}\mbox{.}(2024{\natexlab{a}}){Matthee}, {Naidu},
  {Brammer}, {Chisholm}, {Eilers}, {Goulding}, {Greene}, {Kashino}, {Labbe},
  {Lilly}, {Mackenzie}, {Oesch}, {Weibel}, {Wuyts}, {Xiao}, {Bordoloi},
  {Bouwens}, {van Dokkum}, {Illingworth}, {Kramarenko}, {Maseda}, {Mason},
  {Meyer}, {Nelson}, {Reddy}, {Shivaei}, {Simcoe}, \& {Yue}}]{Matthee2024}
{Matthee} J. {et~al.}, 2024{\natexlab{a}}, \apj, 963, 129

\bibitem[{{Matthee} {et~al}\mbox{.}(2024{\natexlab{b}}){Matthee}, {Naidu},
  {Kotiwale}, {Furtak}, {Kramarenko}, {Mackenzie}, {Greene}, {Adamo},
  {Bouwens}, {Di Cesare}, {Eilers}, {de Graaff}, {Heintz}, {Kashino}, {Maseda},
  {Tacchella}, \& {Torralba}}]{Matthee2024b}
{Matthee} J. {et~al.}, 2024{\natexlab{b}}, arXiv e-prints, arXiv:2412.02846

\bibitem[{{McClymont} {et~al}\mbox{.}(2024){McClymont}, {Tacchella}, {Smith},
  {Kannan}, {Maiolino}, {Belfiore}, {Hernquist}, {Li}, \&
  {Vogelsberger}}]{McClymont2024}
{McClymont} W. {et~al.}, 2024, \mnras, 532, 2016

\bibitem[{{McInnes}, {Healy} \& {Melville}(2018){McInnes}, {Healy}, \&
  {Melville}}]{umap2}
{McInnes} L., {Healy} J., {Melville} J., 2018, ArXiv e-prints

\bibitem[{McInnes {et~al}\mbox{.}(2018)McInnes, Healy, Saul, \&
  Grossberger}]{umap1e}
McInnes L., Healy J., Saul N., Grossberger L., 2018, The Journal of Open Source
  Software, 3, 861

\bibitem[{{McKee} \& {Ostriker}(1977)}]{McKee1977}
{McKee} C.~F., {Ostriker} J.~P., 1977, \apj, 218, 148

\bibitem[{{McKee} {et~al}\mbox{.}(1982){McKee}, {Storey}, {Watson}, \&
  {Green}}]{McKee1982}
{McKee} C.~F., {Storey} J.~W.~V., {Watson} D.~M., {Green} S., 1982, \apj, 259,
  647

\bibitem[{{McKinney} {et~al}\mbox{.}(2025){McKinney}, {Cooper}, {Casey},
  {Mu{\~n}oz}, {Akins}, {Lambrides}, \& {Long}}]{McKinney2025}
{McKinney} J., {Cooper} O.~R., {Casey} C.~M., {Mu{\~n}oz} J.~B., {Akins} H.,
  {Lambrides} E., {Long} A.~S., 2025, \apjl, 985, L21

\bibitem[{{Michel-Dansac} {et~al}\mbox{.}(2020){Michel-Dansac}, {Blaizot},
  {Garel}, {Verhamme}, {Kimm}, \& {Trebitsch}}]{Leo2020}
{Michel-Dansac} L., {Blaizot} J., {Garel} T., {Verhamme} A., {Kimm} T.,
  {Trebitsch} M., 2020, \aap, 635, A154

\bibitem[{{Mingozzi} {et~al}\mbox{.}(2022){Mingozzi}, {James},
  {Arellano-C{\'o}rdova}, {Berg}, {Senchyna}, {Chisholm}, {Brinchmann},
  {Aloisi}, {Amor{\'\i}n}, {Charlot}, {Feltre}, {Hayes}, {Heckman}, {Henry},
  {Hernandez}, {Kumari}, {Leitherer}, {Llerena}, {Martin}, {Nanayakkara},
  {Ravindranath}, {Skillman}, {Sugahara}, {Wofford}, \& {Xu}}]{Mingozzi2022}
{Mingozzi} M. {et~al.}, 2022, \apj, 939, 110

\bibitem[{{Naidu} {et~al}\mbox{.}(2025){Naidu}, {Oesch}, {Brammer}, {Weibel},
  {Li}, {Matthee}, {Chisholm}, {Pollock}, {Heintz}, {Johnson}, {Shen},
  {Hviding}, {Leja}, {Tacchella}, {Ganguly}, {Witten}, {Atek}, {Belli}, {Bose},
  {Bouwens}, {Dayal}, {Decarli}, {de Graaff}, {Fudamoto}, {Giovinazzo},
  {Greene}, {Illingworth}, {Inoue}, {Kane}, {Labbe}, {Leonova},
  {Marques-Chaves}, {Meyer}, {Nelson}, {Roberts-Borsani}, {Schaerer}, {Simcoe},
  {Stefanon}, {Sugahara}, {Toft}, {van der Wel}, {van Dokkum}, {Walter},
  {Watson}, {Weaver}, \& {Whitaker}}]{Naidu2025}
{Naidu} R.~P. {et~al.}, 2025, arXiv e-prints, arXiv:2505.11263

\bibitem[{{Nakajima} {et~al}\mbox{.}(2025){Nakajima}, {Ouchi}, {Harikane},
  {Vanzella}, {Ono}, {Isobe}, {Nishigaki}, {Tsujimoto}, {Nakamura}, {Xu},
  {Umeda}, \& {Zhang}}]{Nakajima2025b}
{Nakajima} K. {et~al.}, 2025, arXiv e-prints, arXiv:2506.11846

\bibitem[{{Nakazato} {et~al}\mbox{.}(2026){Nakazato}, {Sugimura}, {Inoue}, \&
  {Ricotti}}]{Nakazato2026}
{Nakazato} Y., {Sugimura} K., {Inoue} A.~K., {Ricotti} M., 2026, \apj, 998, 26

\bibitem[{{Nakazato}, {Yoshida} \& {Ceverino}(2023){Nakazato}, {Yoshida}, \&
  {Ceverino}}]{Nakazato2023}
{Nakazato} Y., {Yoshida} N., {Ceverino} D., 2023, \apj, 953, 140

\bibitem[{{Nelson} {et~al}\mbox{.}(2018){Nelson}, {Pillepich}, {Springel},
  {Weinberger}, {Hernquist}, {Pakmor}, {Genel}, {Torrey}, {Vogelsberger},
  {Kauffmann}, {Marinacci}, \& {Naiman}}]{Nelson2018}
{Nelson} D. {et~al.}, 2018, \mnras, 475, 624

\bibitem[{{Nelson} \& {Langer}(1997)}]{Nelson1997}
{Nelson} R.~P., {Langer} W.~D., 1997, \apj, 482, 796

\bibitem[{{Nomoto}, {Kobayashi} \& {Tominaga}(2013){Nomoto}, {Kobayashi}, \&
  {Tominaga}}]{Nomoto2013}
{Nomoto} K., {Kobayashi} C., {Tominaga} N., 2013, \araa, 51, 457

\bibitem[{{Nomoto} {et~al}\mbox{.}(2006){Nomoto}, {Tominaga}, {Umeda},
  {Kobayashi}, \& {Maeda}}]{Nomoto2006}
{Nomoto} K., {Tominaga} N., {Umeda} H., {Kobayashi} C., {Maeda} K., 2006,
  \nphysa, 777, 424

\bibitem[{{Nyhagen} {et~al}\mbox{.}(2024){Nyhagen}, {Schimek}, {Cicone},
  {Decataldo}, \& {Shen}}]{Nyhagen2024}
{Nyhagen} C.~T., {Schimek} A., {Cicone} C., {Decataldo} D., {Shen} S., 2024,
  arXiv e-prints, arXiv:2410.18471

\bibitem[{{Oh}, {Haiman} \& {Rees}(2001){Oh}, {Haiman}, \& {Rees}}]{Oh2001}
{Oh} S.~P., {Haiman} Z., {Rees} M.~J., 2001, \apj, 553, 73

\bibitem[{{Omukai} {et~al}\mbox{.}(2005){Omukai}, {Tsuribe}, {Schneider}, \&
  {Ferrara}}]{Omukai2005}
{Omukai} K., {Tsuribe} T., {Schneider} R., {Ferrara} A., 2005, \apj, 626, 627

\bibitem[{{Oppenheimer} \& {Schaye}(2013)}]{Oppenheimer2013}
{Oppenheimer} B.~D., {Schaye} J., 2013, \mnras, 434, 1043

\bibitem[{{O'Shea} {et~al}\mbox{.}(2015){O'Shea}, {Wise}, {Xu}, \&
  {Norman}}]{Oshea2015}
{O'Shea} B.~W., {Wise} J.~H., {Xu} H., {Norman} M.~L., 2015, \apjl, 807, L12

\bibitem[{{Osterbrock} \& {Ferland}(2006)}]{Osterbrock2006}
{Osterbrock} D.~E., {Ferland} G.~J., 2006, {Astrophysics of gaseous nebulae and
  active galactic nuclei}

\bibitem[{{Padoan} \& {Nordlund}(2011)}]{Padoan2011}
{Padoan} P., {Nordlund} {\r{A}}., 2011, \apj, 730, 40

\bibitem[{{Pallottini} {et~al}\mbox{.}(2022){Pallottini}, {Ferrara},
  {Gallerani}, {Behrens}, {Kohandel}, {Carniani}, {Vallini}, {Salvadori},
  {Gelli}, {Sommovigo}, {D'Odorico}, {Di Mascia}, \&
  {Pizzati}}]{Pallottini2022}
{Pallottini} A. {et~al.}, 2022, \mnras, 513, 5621

\bibitem[{{Pallottini} {et~al}\mbox{.}(2014){Pallottini}, {Ferrara},
  {Gallerani}, {Salvadori}, \& {D'Odorico}}]{Pallottini2014}
{Pallottini} A., {Ferrara} A., {Gallerani} S., {Salvadori} S., {D'Odorico} V.,
  2014, \mnras, 440, 2498

\bibitem[{{Park} {et~al}\mbox{.}(2024){Park}, {Conroy}, {Johnson}, {Leja},
  {Dotter}, \& {Cargile}}]{Park2024}
{Park} M., {Conroy} C., {Johnson} B.~D., {Leja} J., {Dotter} A., {Cargile}
  P.~A., 2024, arXiv e-prints, arXiv:2410.21375

\bibitem[{{Pentericci} {et~al}\mbox{.}(2016){Pentericci}, {Carniani},
  {Castellano}, {Fontana}, {Maiolino}, {Guaita}, {Vanzella}, {Grazian},
  {Santini}, {Yan}, {Cristiani}, {Conselice}, {Giavalisco}, {Hathi}, \&
  {Koekemoer}}]{Pentericci2016}
{Pentericci} L. {et~al.}, 2016, \apjl, 829, L11

\bibitem[{{Pillepich} {et~al}\mbox{.}(2018){Pillepich}, {Nelson}, {Hernquist},
  {Springel}, {Pakmor}, {Torrey}, {Weinberger}, {Genel}, {Naiman}, {Marinacci},
  \& {Vogelsberger}}]{Pillepich2018}
{Pillepich} A. {et~al.}, 2018, \mnras, 475, 648

\bibitem[{{Planck Collaboration} {et~al}\mbox{.}(2016){Planck Collaboration},
  {Ade}, {Aghanim}, {Arnaud}, {Ashdown}, {Aumont}, {Baccigalupi}, {Banday},
  {Barreiro}, {Bartlett}, {Bartolo}, {Battaner}, {Battye}, {Benabed},
  {Beno{\^\i}t}, {Benoit-L{\'e}vy}, {Bernard}, {Bersanelli}, {Bielewicz},
  {Bock}, {Bonaldi}, {Bonavera}, {Bond}, {Borrill}, {Bouchet}, {Boulanger},
  {Bucher}, {Burigana}, {Butler}, {Calabrese}, {Cardoso}, {Catalano},
  {Challinor}, {Chamballu}, {Chary}, {Chiang}, {Chluba}, {Christensen},
  {Church}, {Clements}, {Colombi}, {Colombo}, {Combet}, {Coulais}, {Crill},
  {Curto}, {Cuttaia}, {Danese}, {Davies}, {Davis}, {de Bernardis}, {de Rosa},
  {de Zotti}, {Delabrouille}, {D{\'e}sert}, {Di Valentino}, {Dickinson},
  {Diego}, {Dolag}, {Dole}, {Donzelli}, {Dor{\'e}}, {Douspis}, {Ducout},
  {Dunkley}, {Dupac}, {Efstathiou}, {Elsner}, {En{\ss}lin}, {Eriksen},
  {Farhang}, {Fergusson}, {Finelli}, {Forni}, {Frailis}, {Fraisse},
  {Franceschi}, {Frejsel}, {Galeotta}, {Galli}, {Ganga}, {Gauthier}, {Gerbino},
  {Ghosh}, {Giard}, {Giraud-H{\'e}raud}, {Giusarma}, {Gjerl{\o}w},
  {Gonz{\'a}lez-Nuevo}, {G{\'o}rski}, {Gratton}, {Gregorio}, {Gruppuso},
  {Gudmundsson}, {Hamann}, {Hansen}, {Hanson}, {Harrison}, {Helou},
  {Henrot-Versill{\'e}}, {Hern{\'a}ndez-Monteagudo}, {Herranz}, {Hildebrandt},
  {Hivon}, {Hobson}, {Holmes}, {Hornstrup}, {Hovest}, {Huang}, {Huffenberger},
  {Hurier}, {Jaffe}, {Jaffe}, {Jones}, {Juvela}, {Keih{\"a}nen}, {Keskitalo},
  {Kisner}, {Kneissl}, {Knoche}, {Knox}, {Kunz}, {Kurki-Suonio}, {Lagache},
  {L{\"a}hteenm{\"a}ki}, {Lamarre}, {Lasenby}, {Lattanzi}, {Lawrence}, {Leahy},
  {Leonardi}, {Lesgourgues}, {Levrier}, {Lewis}, {Liguori}, {Lilje},
  {Linden-V{\o}rnle}, {L{\'o}pez-Caniego}, {Lubin}, {Mac{\'\i}as-P{\'e}rez},
  {Maggio}, {Maino}, {Mandolesi}, {Mangilli}, {Marchini}, {Maris}, {Martin},
  {Martinelli}, {Mart{\'\i}nez-Gonz{\'a}lez}, {Masi}, {Matarrese}, {McGehee},
  {Meinhold}, {Melchiorri}, {Melin}, {Mendes}, {Mennella}, {Migliaccio},
  {Millea}, {Mitra}, {Miville-Desch{\^e}nes}, {Moneti}, {Montier}, {Morgante},
  {Mortlock}, {Moss}, {Munshi}, {Murphy}, {Naselsky}, {Nati}, {Natoli},
  {Netterfield}, {N{\o}rgaard-Nielsen}, {Noviello}, {Novikov}, {Novikov},
  {Oxborrow}, {Paci}, {Pagano}, {Pajot}, {Paladini}, {Paoletti}, {Partridge},
  {Pasian}, {Patanchon}, {Pearson}, {Perdereau}, {Perotto}, {Perrotta},
  {Pettorino}, {Piacentini}, {Piat}, {Pierpaoli}, {Pietrobon}, {Plaszczynski},
  {Pointecouteau}, {Polenta}, {Popa}, {Pratt}, {Pr{\'e}zeau}, {Prunet},
  {Puget}, {Rachen}, {Reach}, {Rebolo}, {Reinecke}, {Remazeilles}, {Renault},
  {Renzi}, {Ristorcelli}, {Rocha}, {Rosset}, {Rossetti}, {Roudier},
  {Rouill{\'e} d'Orfeuil}, {Rowan-Robinson}, {Rubi{\~n}o-Mart{\'\i}n},
  {Rusholme}, {Said}, {Salvatelli}, {Salvati}, {Sandri}, {Santos},
  {Savelainen}, {Savini}, {Scott}, {Seiffert}, {Serra}, {Shellard}, {Spencer},
  {Spinelli}, {Stolyarov}, {Stompor}, {Sudiwala}, {Sunyaev}, {Sutton},
  {Suur-Uski}, {Sygnet}, {Tauber}, {Terenzi}, {Toffolatti}, {Tomasi},
  {Tristram}, {Trombetti}, {Tucci}, {Tuovinen}, {T{\"u}rler}, {Umana},
  {Valenziano}, {Valiviita}, {Van Tent}, {Vielva}, {Villa}, {Wade}, {Wandelt},
  {Wehus}, {White}, {White}, {Wilkinson}, {Yvon}, {Zacchei}, \&
  {Zonca}}]{Planck2016Cosmo}
{Planck Collaboration} {et~al.}, 2016, \aap, 594, A13

\bibitem[{{Ploeckinger} {et~al}\mbox{.}(2025){Ploeckinger}, {Richings},
  {Schaye}, {Trayford}, {Schaller}, \& {Chaikin}}]{Ploeckinger2025}
{Ploeckinger} S., {Richings} A.~J., {Schaye} J., {Trayford} J.~W., {Schaller}
  M., {Chaikin} E., 2025, arXiv e-prints, arXiv:2506.15773

\bibitem[{{Pontzen} {et~al}\mbox{.}(2021){Pontzen}, {Rey}, {Cadiou}, {Agertz},
  {Teyssier}, {Read}, \& {Orkney}}]{Pontzen2021}
{Pontzen} A., {Rey} M.~P., {Cadiou} C., {Agertz} O., {Teyssier} R., {Read} J.,
  {Orkney} M. D.~A., 2021, \mnras, 501, 1755

\bibitem[{{Pontzen} {et~al}\mbox{.}(2023){Pontzen}, {Ro{\v{s}}kar}, {Cadiou},
  {Stinson}, {Mastropietro}, {Rey}, {Keller}, {Duffy}, {mkrets}, {Tremmel},
  {Davies}, {Franck}, {Quinn}, {Sarmento}, {Bovy}, {nroth0815}, {Coles}, {Ji},
  {Applebaum}, {Zana}, {Biernacki}, {Herpich}, {mihaimt}, {Woods}, {EthTay},
  {Altay}, {Winkler}, {Shaw}, \& {Moon}}]{Pontzen2022}
{Pontzen} A. {et~al.}, 2023, {pynbody/pynbody: Version 1.5.2}

\bibitem[{{Pontzen} {et~al}\mbox{.}(2013){Pontzen}, {Ro{\v{s}}kar}, {Stinson},
  \& {Woods}}]{Pontzen2013}
{Pontzen} A., {Ro{\v{s}}kar} R., {Stinson} G., {Woods} R., 2013, {pynbody:
  N-Body/SPH analysis for python}. Astrophysics Source Code Library, record
  ascl:1305.002

\bibitem[{{Pontzen} {et~al}\mbox{.}(2017){Pontzen}, {Tremmel}, {Roth},
  {Peiris}, {Saintonge}, {Volonteri}, {Quinn}, \& {Governato}}]{Pontzen2017}
{Pontzen} A., {Tremmel} M., {Roth} N., {Peiris} H.~V., {Saintonge} A.,
  {Volonteri} M., {Quinn} T., {Governato} F., 2017, \mnras, 465, 547

\bibitem[{{Ragan-Kelley} {et~al}\mbox{.}(2014){Ragan-Kelley}, {Perez},
  {Granger}, {Kluyver}, {Ivanov}, {Frederic}, \&
  {Bussonnier}}]{RaganKelley2014}
{Ragan-Kelley} M., {Perez} F., {Granger} B., {Kluyver} T., {Ivanov} P.,
  {Frederic} J., {Bussonnier} M., 2014, in AGU Fall Meeting Abstracts, Vol.
  2014, pp. H44D--07

\bibitem[{{Raiter}, {Schaerer} \& {Fosbury}(2010){Raiter}, {Schaerer}, \&
  {Fosbury}}]{Raiter2010}
{Raiter} A., {Schaerer} D., {Fosbury} R.~A.~E., 2010, \aap, 523, A64

\bibitem[{{Ramambason} {et~al}\mbox{.}(2022){Ramambason}, {Lebouteiller},
  {Bik}, {Richardson}, {Galliano}, {Schaerer}, {Morisset}, {Polles}, {Madden},
  {Chevance}, \& {De Looze}}]{Ramambason2022}
{Ramambason} L. {et~al.}, 2022, \aap, 667, A35

\bibitem[{{Reddy} {et~al}\mbox{.}(2018){Reddy}, {Shapley}, {Sanders}, {Kriek},
  {Coil}, {Shivaei}, {Freeman}, {Mobasher}, {Siana}, {Azadi}, {Fetherolf},
  {Fornasini}, {Leung}, {Price}, {Zick}, \& {Barro}}]{Reddy2018}
{Reddy} N.~A. {et~al.}, 2018, \apj, 869, 92

\bibitem[{{R{\'e}my-Ruyer} {et~al}\mbox{.}(2014){R{\'e}my-Ruyer}, {Madden},
  {Galliano}, {Galametz}, {Takeuchi}, {Asano}, {Zhukovska}, {Lebouteiller},
  {Cormier}, {Jones}, {Bocchio}, {Baes}, {Bendo}, {Boquien}, {Boselli},
  {DeLooze}, {Doublier-Pritchard}, {Hughes}, {Karczewski}, \&
  {Spinoglio}}]{RR2014}
{R{\'e}my-Ruyer} A. {et~al.}, 2014, \aap, 563, A31

\bibitem[{{Rey} {et~al}\mbox{.}(2023){Rey}, {Agertz}, {Starkenburg}, {Renaud},
  {Joshi}, {Pontzen}, {Martin}, {Feuillet}, \& {Read}}]{Rey2023Vintergatan}
{Rey} M.~P. {et~al.}, 2023, \mnras, 521, 995

\bibitem[{{Rey} {et~al}\mbox{.}(2025){Rey}, {Katz}, {Cadiou}, {Sanati},
  {Agertz}, {Blaizot}, {Cameron}, {Choustikov}, {Devriendt}, {Hauk}, {Ji},
  {Jones}, {Kimm}, {Laseter}, {Martin-Alvarez}, {Matsumoto}, {Pearce}, {Revaz},
  {Rodriguez Montero}, {Rosdahl}, {Saxena}, {Slyz}, {Stiskalek}, {Storck},
  {Veenema}, \& {Yee}}]{Rey2025tmp}
{Rey} M.~P. {et~al.}, 2025, arXiv e-prints, arXiv:2510.05232

\bibitem[{{Rey} \& {Pontzen}(2018)}]{Rey2018}
{Rey} M.~P., {Pontzen} A., 2018, \mnras, 474, 45

\bibitem[{{Rey}, {Pontzen} \& {Saintonge}(2019){Rey}, {Pontzen}, \&
  {Saintonge}}]{Rey2019}
{Rey} M.~P., {Pontzen} A., {Saintonge} A., 2019, \mnras, 485, 1906

\bibitem[{{Rey} \& {Starkenburg}(2022)}]{Rey2022}
{Rey} M.~P., {Starkenburg} T.~K., 2022, \mnras, 510, 4208

\bibitem[{{Richings} {et~al}\mbox{.}(2022){Richings}, {Faucher-Gigu{\`e}re},
  {Gurvich}, {Schaye}, \& {Hayward}}]{Richings2022}
{Richings} A.~J., {Faucher-Gigu{\`e}re} C.-A., {Gurvich} A.~B., {Schaye} J.,
  {Hayward} C.~C., 2022, \mnras, 517, 1557

\bibitem[{{Richings}, {Schaye} \& {Oppenheimer}(2014){Richings}, {Schaye}, \&
  {Oppenheimer}}]{Richings2014}
{Richings} A.~J., {Schaye} J., {Oppenheimer} B.~D., 2014, \mnras, 440, 3349

\bibitem[{{Ritter} {et~al}\mbox{.}(2018){Ritter}, {Herwig}, {Jones},
  {Pignatari}, {Fryer}, \& {Hirschi}}]{Ritter2018}
{Ritter} C., {Herwig} F., {Jones} S., {Pignatari} M., {Fryer} C., {Hirschi} R.,
  2018, \mnras, 480, 538

\bibitem[{{Roberts-Borsani} {et~al}\mbox{.}(2024){Roberts-Borsani}, {Treu},
  {Shapley}, {Fontana}, {Pentericci}, {Castellano}, {Morishita}, {Bergamini},
  \& {Rosati}}]{RB2024}
{Roberts-Borsani} G. {et~al.}, 2024, \apj, 976, 193

\bibitem[{{Roberts-Borsani}, {Ellis} \& {Laporte}(2020){Roberts-Borsani},
  {Ellis}, \& {Laporte}}]{RB2020}
{Roberts-Borsani} G.~W., {Ellis} R.~S., {Laporte} N., 2020, \mnras, 497, 3440

\bibitem[{{R{\"o}llig} {et~al}\mbox{.}(2006){R{\"o}llig}, {Ossenkopf},
  {Jeyakumar}, {Stutzki}, \& {Sternberg}}]{Rollig2006}
{R{\"o}llig} M., {Ossenkopf} V., {Jeyakumar} S., {Stutzki} J., {Sternberg} A.,
  2006, \aap, 451, 917

\bibitem[{{Rosdahl} {et~al}\mbox{.}(2013){Rosdahl}, {Blaizot}, {Aubert},
  {Stranex}, \& {Teyssier}}]{Rosdahl2013}
{Rosdahl} J., {Blaizot} J., {Aubert} D., {Stranex} T., {Teyssier} R., 2013,
  \mnras, 436, 2188

\bibitem[{{Rosdahl} {et~al}\mbox{.}(2018){Rosdahl}, {Katz}, {Blaizot}, {Kimm},
  {Michel-Dansac}, {Garel}, {Haehnelt}, {Ocvirk}, \& {Teyssier}}]{Rosdahl2018}
{Rosdahl} J. {et~al.}, 2018, \mnras, 479, 994

\bibitem[{{Rosdahl} \& {Teyssier}(2015)}]{Rosdahl2015}
{Rosdahl} J., {Teyssier} R., 2015, \mnras, 449, 4380

\bibitem[{{Roth}, {Pontzen} \& {Peiris}(2016){Roth}, {Pontzen}, \&
  {Peiris}}]{Roth2016}
{Roth} N., {Pontzen} A., {Peiris} H.~V., 2016, \mnras, 455, 974

\bibitem[{{Sanders} {et~al}\mbox{.}(2023){Sanders}, {Shapley}, {Topping},
  {Reddy}, \& {Brammer}}]{Sanders2023}
{Sanders} R.~L., {Shapley} A.~E., {Topping} M.~W., {Reddy} N.~A., {Brammer}
  G.~B., 2023, \apj, 955, 54

\bibitem[{{Sarmento}, {Scannapieco} \& {Cohen}(2018){Sarmento}, {Scannapieco},
  \& {Cohen}}]{Sarmento2018}
{Sarmento} R., {Scannapieco} E., {Cohen} S., 2018, \apj, 854, 75

\bibitem[{{Saxena} {et~al}\mbox{.}(2024){Saxena}, {Cameron}, {Katz}, {Bunker},
  {Chevallard}, {D'Eugenio}, {Arribas}, {Bhatawdekar}, {Boyett}, {Cargile},
  {Carniani}, {Charlot}, {Curti}, {Curtis-Lake}, {Hainline}, {Ji}, {Johnson},
  {Jones}, {Kumari}, {Laseter}, {Maseda}, {Robertson}, {Simmonds}, {Tacchella},
  {Ubler}, {Williams}, {Willott}, {Witstok}, \& {Zhu}}]{Saxena2024}
{Saxena} A. {et~al.}, 2024, arXiv e-prints, arXiv:2411.14532

\bibitem[{{Saxena} {et~al}\mbox{.}(2021){Saxena}, {Ellis}, {F{\"o}rster},
  {Calabr{\`o}}, {Pentericci}, {Carnall}, {Castellano}, {Cullen}, {Fontana},
  {Franco}, {Fynbo}, {Gargiulo}, {Garilli}, {Hathi}, {McLeod}, {Amor{\'\i}n},
  \& {Zamorani}}]{Saxena2021}
{Saxena} A. {et~al.}, 2021, \mnras, 505, 4798

\bibitem[{{Schaerer}(2002)}]{Schaerer2002}
{Schaerer} D., 2002, \aap, 382, 28

\bibitem[{{Schaerer} {et~al}\mbox{.}(2025){Schaerer}, {Guibert},
  {Marques-Chaves}, \& {Martins}}]{Schaerer2025}
{Schaerer} D., {Guibert} J., {Marques-Chaves} R., {Martins} F., 2025, \aap,
  693, A271

\bibitem[{{Schaerer} {et~al}\mbox{.}(1993){Schaerer}, {Meynet}, {Maeder}, \&
  {Schaller}}]{Schaerer1993}
{Schaerer} D., {Meynet} G., {Maeder} A., {Schaller} G., 1993, \aaps, 98, 523

\bibitem[{{Schaye} {et~al}\mbox{.}(2015){Schaye}, {Crain}, {Bower}, {Furlong},
  {Schaller}, {Theuns}, {Dalla Vecchia}, {Frenk}, {McCarthy}, {Helly},
  {Jenkins}, {Rosas-Guevara}, {White}, {Baes}, {Booth}, {Camps}, {Navarro},
  {Qu}, {Rahmati}, {Sawala}, {Thomas}, \& {Trayford}}]{Schaye2015}
{Schaye} J. {et~al.}, 2015, \mnras, 446, 521

\bibitem[{{Schimek} {et~al}\mbox{.}(2024){Schimek}, {Cicone}, {Shen},
  {Decataldo}, {Klaassen}, \& {Mayer}}]{Schimek2024}
{Schimek} A., {Cicone} C., {Shen} S., {Decataldo} D., {Klaassen} P., {Mayer}
  L., 2024, \aap, 687, L10

\bibitem[{{Schmidt}(1959)}]{Schmidt1959}
{Schmidt} M., 1959, \apj, 129, 243

\bibitem[{{Schneider} {et~al}\mbox{.}(2002){Schneider}, {Ferrara}, {Natarajan},
  \& {Omukai}}]{Schneider2002}
{Schneider} R., {Ferrara} A., {Natarajan} P., {Omukai} K., 2002, \apj, 571, 30

\bibitem[{{Schreiber} {et~al}\mbox{.}(2015){Schreiber}, {Pannella}, {Elbaz},
  {B{\'e}thermin}, {Inami}, {Dickinson}, {Magnelli}, {Wang}, {Aussel}, {Daddi},
  {Juneau}, {Shu}, {Sargent}, {Buat}, {Faber}, {Ferguson}, {Giavalisco},
  {Koekemoer}, {Magdis}, {Morrison}, {Papovich}, {Santini}, \&
  {Scott}}]{Schreiber2015}
{Schreiber} C. {et~al.}, 2015, \aap, 575, A74

\bibitem[{{Seitenzahl} {et~al}\mbox{.}(2013){Seitenzahl},
  {Ciaraldi-Schoolmann}, {R{\"o}pke}, {Fink}, {Hillebrandt}, {Kromer},
  {Pakmor}, {Ruiter}, {Sim}, \& {Taubenberger}}]{Seitenzahl2013}
{Seitenzahl} I.~R. {et~al.}, 2013, \mnras, 429, 1156

\bibitem[{{Senchyna} {et~al}\mbox{.}(2024){Senchyna}, {Plat}, {Stark}, {Rudie},
  {Berg}, {Charlot}, {James}, \& {Mingozzi}}]{Senchyna2024}
{Senchyna} P., {Plat} A., {Stark} D.~P., {Rudie} G.~C., {Berg} D., {Charlot}
  S., {James} B.~L., {Mingozzi} M., 2024, \apj, 966, 92

\bibitem[{{Shapley} {et~al}\mbox{.}(2019){Shapley}, {Sanders}, {Shao}, {Reddy},
  {Kriek}, {Coil}, {Mobasher}, {Siana}, {Shivaei}, {Freeman}, {Azadi}, {Price},
  {Leung}, {Fetherolf}, {de Groot}, {Zick}, {Fornasini}, \&
  {Barro}}]{Shapley2019}
{Shapley} A.~E. {et~al.}, 2019, \apjl, 881, L35

\bibitem[{{Shapley} {et~al}\mbox{.}(2025){Shapley}, {Sanders}, {Topping},
  {Reddy}, {Berg}, {Bouwens}, {Brammer}, {Carnall}, {Cullen}, {Dav{\'e}},
  {Dunlop}, {Ellis}, {F{\"o}rster Schreiber}, {Furlanetto}, {Glazebrook},
  {Illingworth}, {Jones}, {Kriek}, {McLeod}, {McLure}, {Narayanan}, {Oesch},
  {Pahl}, {Pettini}, {Schaerer}, {Stark}, {Steidel}, {Tang}, {Clarke},
  {Donnan}, \& {Kehoe}}]{Shapley2025}
{Shapley} A.~E. {et~al.}, 2025, \apj, 980, 242

\bibitem[{{Shull} \& {van Steenberg}(1982)}]{Shull1982}
{Shull} J.~M., {van Steenberg} M., 1982, \apjs, 48, 95

\bibitem[{{Shuntov} {et~al}\mbox{.}(2025){Shuntov}, {Ilbert}, {Toft},
  {Arango-Toro}, {Akins}, {Casey}, {Franco}, {Harish}, {Kartaltepe},
  {Koekemoer}, {McCracken}, {Paquereau}, {Laigle}, {Bethermin}, {Dubois},
  {Drakos}, {Faisst}, {Gozaliasl}, {Gillman}, {Hayward}, {Hirschmann},
  {Huertas-Company}, {Jespersen}, {Jin}, {Kokorev}, {Lambrides}, {Le Borgne},
  {Liu}, {Magdis}, {Massey}, {McPartland}, {Mercier}, {McCleary}, {McKinney},
  {Oesch}, {Renzini}, {Rhodes}, {Rich}, {Robertson}, {Sanders}, {Trebitsch},
  {Tresse}, {Valentino}, {Vijayan}, {Weaver}, {Weibel}, {Wilkins}, \&
  {Yang}}]{Shuntov2025}
{Shuntov} M. {et~al.}, 2025, \aap, 695, A20

\bibitem[{{Speagle} {et~al}\mbox{.}(2014){Speagle}, {Steinhardt}, {Capak}, \&
  {Silverman}}]{Speagle2014}
{Speagle} J.~S., {Steinhardt} C.~L., {Capak} P.~L., {Silverman} J.~D., 2014,
  \apjs, 214, 15

\bibitem[{{Stacy}, {Bromm} \& {Lee}(2016){Stacy}, {Bromm}, \&
  {Lee}}]{Stacy2016}
{Stacy} A., {Bromm} V., {Lee} A.~T., 2016, \mnras, 462, 1307

\bibitem[{{Stancil} {et~al}\mbox{.}(1999){Stancil}, {Schultz}, {Kimura}, {Gu},
  {Hirsch}, \& {Buenker}}]{Stancil1999}
{Stancil} P.~C., {Schultz} D.~R., {Kimura} M., {Gu} J.~P., {Hirsch} G.,
  {Buenker} R.~J., 1999, \aaps, 140, 225

\bibitem[{{Stanway} \& {Eldridge}(2018)}]{Stanway2018}
{Stanway} E.~R., {Eldridge} J.~J., 2018, \mnras, 479, 75

\bibitem[{{Stanway} \& {Eldridge}(2023)}]{Stanway2023}
{Stanway} E.~R., {Eldridge} J.~J., 2023, \mnras, 522, 4430

\bibitem[{{Stark} {et~al}\mbox{.}(2017){Stark}, {Ellis}, {Charlot},
  {Chevallard}, {Tang}, {Belli}, {Zitrin}, {Mainali}, {Gutkin},
  {Vidal-Garc{\'\i}a}, {Bouwens}, \& {Oesch}}]{Stark2017}
{Stark} D.~P. {et~al.}, 2017, \mnras, 464, 469

\bibitem[{{Steidel} {et~al}\mbox{.}(2016){Steidel}, {Strom}, {Pettini},
  {Rudie}, {Reddy}, \& {Trainor}}]{Steidel2016}
{Steidel} C.~C., {Strom} A.~L., {Pettini} M., {Rudie} G.~C., {Reddy} N.~A.,
  {Trainor} R.~F., 2016, \apj, 826, 159

\bibitem[{{Sternberg} \& {Dalgarno}(1989)}]{Sternberg1989}
{Sternberg} A., {Dalgarno} A., 1989, \apj, 338, 197

\bibitem[{{Stopyra} {et~al}\mbox{.}(2021){Stopyra}, {Pontzen}, {Peiris},
  {Roth}, \& {Rey}}]{Stopyra2021}
{Stopyra} S., {Pontzen} A., {Peiris} H., {Roth} N., {Rey} M.~P., 2021, \apjs,
  252, 28

\bibitem[{{Storck} {et~al}\mbox{.}(2026){Storck}, {Katz}, {Devriendt}, {Slyz},
  {Cadiou}, {Choustikov}, {Rey}, {Saxena}, {Agertz}, \&
  {Kimm}}]{Storck2026-meg}
{Storck} A. {et~al.}, 2026, \mnras, 548, stag529

\bibitem[{{Strait} {et~al}\mbox{.}(2023){Strait}, {Brammer}, {Muzzin},
  {Desprez}, {Asada}, {Abraham}, {Brada{\v{c}}}, {Iyer}, {Martis}, {Mowla},
  {Noirot}, {Sarrouh}, {Sawicki}, {Willott}, {Gould}, {Grindlay}, {Matharu}, \&
  {Rihtar{\v{s}}i{\v{c}}}}]{Strait2023}
{Strait} V. {et~al.}, 2023, \apjl, 949, L23

\bibitem[{{Strom} {et~al}\mbox{.}(2017){Strom}, {Steidel}, {Rudie}, {Trainor},
  {Pettini}, \& {Reddy}}]{Strom2017}
{Strom} A.~L., {Steidel} C.~C., {Rudie} G.~C., {Trainor} R.~F., {Pettini} M.,
  {Reddy} N.~A., 2017, \apj, 836, 164

\bibitem[{{Suess} {et~al}\mbox{.}(2024){Suess}, {Weaver}, {Price}, {Pan},
  {Wang}, {Bezanson}, {Brammer}, {Cutler}, {Labb{\'e}}, {Leja}, {Williams},
  {Whitaker}, {Atek}, {Dayal}, {de Graaff}, {Feldmann}, {Franx}, {Fudamoto},
  {Fujimoto}, {Furtak}, {Goulding}, {Greene}, {Khullar}, {Kokorev}, {Kriek},
  {Lorenz}, {Marchesini}, {Maseda}, {Matthee}, {Miller}, {Mitsuhashi}, {Mowla},
  {Muzzin}, {Naidu}, {Nanayakkara}, {Nelson}, {Oesch}, {Setton}, {Shipley},
  {Smit}, {Spilker}, {van Dokkum}, \& {Zitrin}}]{Suess2024}
{Suess} K.~A. {et~al.}, 2024, \apj, 976, 101

\bibitem[{{Teyssier}(2002)}]{Teyssier2002}
{Teyssier} R., 2002, \aap, 385, 337

\bibitem[{{Tielens}(2005)}]{Tielens2005}
{Tielens} A.~G.~G.~M., 2005, {The Physics and Chemistry of the Interstellar
  Medium}

\bibitem[{{Topping} {et~al}\mbox{.}(2025){Topping}, {Sanders}, {Shapley},
  {Pahl}, {Reddy}, {Stark}, {Berg}, {Clarke}, {Cullen}, {Dunlop}, {Ellis},
  {F{\"o}rster Schreiber}, {Illingworth}, {Jones}, {Narayanan}, {Pettini}, \&
  {Schaerer}}]{Topping2025}
{Topping} M.~W. {et~al.}, 2025, arXiv e-prints, arXiv:2502.08712

\bibitem[{{Topping} {et~al}\mbox{.}(2024){Topping}, {Stark}, {Endsley},
  {Whitler}, {Hainline}, {Johnson}, {Robertson}, {Tacchella}, {Chen},
  {Alberts}, {Baker}, {Bunker}, {Carniani}, {Charlot}, {Chevallard},
  {Curtis-Lake}, {DeCoursey}, {Egami}, {Eisenstein}, {Ji}, {Maiolino},
  {Williams}, {Willmer}, {Willott}, \& {Witstok}}]{Topping2024}
{Topping} M.~W. {et~al.}, 2024, \mnras, 529, 4087

\bibitem[{Toro(2009)}]{toro_RiemannSolversNumerical_2009}
Toro E.~F., 2009, Riemann Solvers and Numerical Methods for Fluid Dynamics: A
  Practical Introduction, 3rd edn. {Springer}, {Berlin}

\bibitem[{{Trebitsch} {et~al}\mbox{.}(2021){Trebitsch}, {Dubois}, {Volonteri},
  {Pfister}, {Cadiou}, {Katz}, {Rosdahl}, {Kimm}, {Pichon}, {Beckmann},
  {Devriendt}, \& {Slyz}}]{Trebitsch2021}
{Trebitsch} M. {et~al.}, 2021, \aap, 653, A154

\bibitem[{{Trussler} {et~al}\mbox{.}(2025){Trussler}, {Conselice}, {Adams},
  {Austin}, {Caruana}, {Harvey}, {Li}, {Lovell}, {Seeyave}, {Vijayan}, \&
  {Wilkins}}]{Trussler2025}
{Trussler} J. A.~A. {et~al.}, 2025, \mnras

\bibitem[{{Trussler} {et~al}\mbox{.}(2023){Trussler}, {Conselice}, {Adams},
  {Maiolino}, {Nakajima}, {Zackrisson}, {Austin}, {Ferreira}, \&
  {Harvey}}]{Trussler2023}
{Trussler} J. A.~A. {et~al.}, 2023, \mnras, 525, 5328

\bibitem[{{Tumlinson} \& {Shull}(2000)}]{Tumlinson2000}
{Tumlinson} J., {Shull} J.~M., 2000, \apjl, 528, L65

\bibitem[{{Tumlinson}, {Shull} \& {Venkatesan}(2003){Tumlinson}, {Shull}, \&
  {Venkatesan}}]{Tumlinson2003}
{Tumlinson} J., {Shull} J.~M., {Venkatesan} A., 2003, \apj, 584, 608

\bibitem[{{Umeda} \& {Nomoto}(2002)}]{Umeda2002}
{Umeda} H., {Nomoto} K., 2002, \apj, 565, 385

\bibitem[{{van der Walt}, {Colbert} \& {Varoquaux}(2011){van der Walt},
  {Colbert}, \& {Varoquaux}}]{vanderWalt2011}
{van der Walt} S., {Colbert} S.~C., {Varoquaux} G., 2011, Computing in Science
  and Engineering, 13, 22

\bibitem[{van
  Leer(1979)}]{vannoordvanvanleerUltimateConservativeDifference1979}
van Leer B., 1979, Journal of Computational Physics, 32, 101

\bibitem[{{Vanzella} {et~al}\mbox{.}(2024){Vanzella}, {Loiacono}, {Messa},
  {Castellano}, {Bergamini}, {Zanella}, {Annibali}, {Sun}, {Dickinson},
  {Adamo}, {Calura}, {Ricotti}, {Rosati}, {Meneghetti}, {Grillo},
  {Brada{\v{c}}}, {Conselice}, {Yan}, {Bolamperti}, {Me{\v{s}}tri{\'c}},
  {Gilli}, {Gronke}, {Willott}, {Sani}, {Acebron}, {Comastri}, {Mignoli},
  {Gruppioni}, {Mercurio}, {Strait}, {Pascale}, {Annunziatella}, {Frye},
  {Bradley}, {Grogin}, {Koekemoer}, {Ravindranath}, {D'Silva}, {Summers},
  {Rihtar{\v{s}}i{\v{c}}}, \& {Windhorst}}]{Vanzella2024}
{Vanzella} E. {et~al.}, 2024, \aap, 691, A251

\bibitem[{{Verner} {et~al}\mbox{.}(1996){Verner}, {Ferland}, {Korista}, \&
  {Yakovlev}}]{Verner1996}
{Verner} D.~A., {Ferland} G.~J., {Korista} K.~T., {Yakovlev} D.~G., 1996, \apj,
  465, 487

\bibitem[{{Vikaeus} {et~al}\mbox{.}(2024){Vikaeus}, {Zackrisson}, {Wilkins},
  {Nabizadeh}, {Kokorev}, {Abdurro'uf}, {Bradley}, {Coe}, {Dayal}, \&
  {Ricotti}}]{Vikaeus2024}
{Vikaeus} A. {et~al.}, 2024, \mnras, 529, 1299

\bibitem[{Virtanen {et~al}\mbox{.}(2020)Virtanen, Gommers, Oliphant, Haberland,
  Reddy, Cournapeau, Burovski, Peterson, Weckesser, Bright, {van der Walt},
  Brett, Wilson, Millman, Mayorov, Nelson, Jones, Kern, Larson, Carey, Polat,
  Feng, Moore, {VanderPlas}, Laxalde, Perktold, Cimrman, Henriksen, Quintero,
  Harris, Archibald, Ribeiro, Pedregosa, {van Mulbregt}, \& {SciPy 1.0
  Contributors}}]{Virtanen2020}
Virtanen P. {et~al.}, 2020, Nature Methods, 17, 261

\bibitem[{{Vogelsberger} {et~al}\mbox{.}(2014){Vogelsberger}, {Genel},
  {Springel}, {Torrey}, {Sijacki}, {Xu}, {Snyder}, {Nelson}, \&
  {Hernquist}}]{Vogelsberger2014}
{Vogelsberger} M. {et~al.}, 2014, \mnras, 444, 1518

\bibitem[{{Vogelsberger} {et~al}\mbox{.}(2020){Vogelsberger}, {Nelson},
  {Pillepich}, {Shen}, {Marinacci}, {Springel}, {Pakmor}, {Tacchella},
  {Weinberger}, {Torrey}, \& {Hernquist}}]{Vogelsberger2020}
{Vogelsberger} M. {et~al.}, 2020, \mnras, 492, 5167

\bibitem[{{Voronov}(1997)}]{Voronov1997}
{Voronov} G.~S., 1997, Atomic Data and Nuclear Data Tables, 65, 1

\bibitem[{{Wang} {et~al}\mbox{.}(2024){Wang}, {Leja}, {Labb{\'e}}, {Bezanson},
  {Whitaker}, {Brammer}, {Furtak}, {Weaver}, {Price}, {Zitrin}, {Atek}, {Coe},
  {Cutler}, {Dayal}, {van Dokkum}, {Feldmann}, {Marchesini}, {Franx},
  {F{\"o}rster Schreiber}, {Fujimoto}, {Geha}, {Glazebrook}, {de Graaff},
  {Greene}, {Juneau}, {Kassin}, {Kriek}, {Khullar}, {Maseda}, {Mowla},
  {Muzzin}, {Nanayakkara}, {Nelson}, {Oesch}, {Pacifici}, {Pan}, {Papovich},
  {Setton}, {Shapley}, {Smit}, {Stefanon}, {Suess}, {Taylor}, \&
  {Williams}}]{Wang2024}
{Wang} B. {et~al.}, 2024, \apjs, 270, 12

\bibitem[{{Weaver} {et~al}\mbox{.}(2024){Weaver}, {Cutler}, {Pan}, {Whitaker},
  {Labb{\'e}}, {Price}, {Bezanson}, {Brammer}, {Marchesini}, {Leja}, {Wang},
  {Furtak}, {Zitrin}, {Atek}, {Chemerynska}, {Coe}, {Dayal}, {van Dokkum},
  {Feldmann}, {F{\"o}rster Schreiber}, {Franx}, {Fujimoto}, {Fudamoto},
  {Glazebrook}, {de Graaff}, {Greene}, {Juneau}, {Kassin}, {Kriek}, {Khullar},
  {Maseda}, {Mowla}, {Muzzin}, {Nanayakkara}, {Nelson}, {Oesch}, {Pacifici},
  {Papovich}, {Setton}, {Shapley}, {Shipley}, {Smit}, {Stefanon}, {Taylor},
  {Weibel}, \& {Williams}}]{Weaver2024}
{Weaver} J.~R. {et~al.}, 2024, \apjs, 270, 7

\bibitem[{{Weingartner} \& {Draine}(2001)}]{Weingartner2001}
{Weingartner} J.~C., {Draine} B.~T., 2001, \apj, 563, 842

\bibitem[{{Wilkins} {et~al}\mbox{.}(2024){Wilkins}, {Lovell}, {Irodotou},
  {Vijayan}, {Vikaeus}, {Zackrisson}, {Caruana}, {Stanway}, {Conselice},
  {Seeyave}, {Roper}, {Chworowsky}, \& {Finkelstein}}]{Wilkins2024_BB}
{Wilkins} S.~M. {et~al.}, 2024, \mnras, 527, 7965

\bibitem[{{Wilkins} {et~al}\mbox{.}(2023){Wilkins}, {Lovell}, {Vijayan},
  {Irodotou}, {Adams}, {Roper}, {Caruana}, {Matthee}, {Seeyave}, {Conselice},
  {P{\'e}rez-Gonz{\'a}lez}, {Turner}, {Donnellan}, {Verma}, \&
  {Trussler}}]{Wilkins2023}
{Wilkins} S.~M. {et~al.}, 2023, \mnras, 522, 4014

\bibitem[{{Willott} {et~al}\mbox{.}(2023){Willott}, {Desprez}, {Asada},
  {Sarrouh}, {Abraham}, {Brada{\v{c}}}, {Brammer}, {Estrada-Carpenter}, {Iyer},
  {Martis}, {Matharu}, {Mowla}, {Muzzin}, {Noirot}, {Sawicki}, {Strait},
  {Rihtar{\v{s}}i{\v{c}}}, \& {Withers}}]{Willott2023}
{Willott} C.~J. {et~al.}, 2023, arXiv e-prints, arXiv:2311.12234

\bibitem[{{Wise} {et~al}\mbox{.}(2012){Wise}, {Turk}, {Norman}, \&
  {Abel}}]{Wise2012}
{Wise} J.~H., {Turk} M.~J., {Norman} M.~L., {Abel} T., 2012, \apj, 745, 50

\bibitem[{{Witstok} {et~al}\mbox{.}(2024){Witstok}, {Jakobsen}, {Maiolino},
  {Helton}, {Johnson}, {Robertson}, {Tacchella}, {Cameron}, {Smit}, {Bunker},
  {Saxena}, {Sun}, {Alberts}, {Arribas}, {Baker}, {Bhatawdekar}, {Boyett},
  {Cargile}, {Carniani}, {Charlot}, {Chevallard}, {Curti}, {Curtis-Lake},
  {D'Eugenio}, {Eisenstein}, {Hainline}, {Jones}, {Kumari}, {Maseda},
  {P{\'e}rez-Gonz{\'a}lez}, {Rinaldi}, {Scholtz}, {{\"U}bler}, {Williams},
  {Willmer}, {Willott}, \& {Zhu}}]{Witstok2024}
{Witstok} J. {et~al.}, 2024, arXiv e-prints, arXiv:2408.16608

\bibitem[{{Wolfire} {et~al}\mbox{.}(1995){Wolfire}, {Hollenbach}, {McKee},
  {Tielens}, \& {Bakes}}]{Wolfire1995}
{Wolfire} M.~G., {Hollenbach} D., {McKee} C.~F., {Tielens} A.~G.~G.~M., {Bakes}
  E.~L.~O., 1995, \apj, 443, 152

\bibitem[{{Wolfire} {et~al}\mbox{.}(2003){Wolfire}, {McKee}, {Hollenbach}, \&
  {Tielens}}]{Wolfire2003}
{Wolfire} M.~G., {McKee} C.~F., {Hollenbach} D., {Tielens} A.~G.~G.~M., 2003,
  \apj, 587, 278

\bibitem[{{Xiao}, {Stanway} \& {Eldridge}(2018){Xiao}, {Stanway}, \&
  {Eldridge}}]{Xiao2018}
{Xiao} L., {Stanway} E.~R., {Eldridge} J.~J., 2018, \mnras, 477, 904

\bibitem[{{Xu} {et~al}\mbox{.}(2016){Xu}, {Norman}, {O'Shea}, \&
  {Wise}}]{Xu2016}
{Xu} H., {Norman} M.~L., {O'Shea} B.~W., {Wise} J.~H., 2016, \apj, 823, 140

\bibitem[{{Yajima} {et~al}\mbox{.}(2022){Yajima}, {Abe}, {Khochfar},
  {Nagamine}, {Inoue}, {Kodama}, {Arata}, {Dalla Vecchia}, {Fukushima},
  {Hashimoto}, {Kashikawa}, {Kubo}, {Li}, {Matsuda}, {Mawatari}, {Ouchi}, \&
  {Umehata}}]{Yajima2022}
{Yajima} H. {et~al.}, 2022, \mnras, 509, 4037

\bibitem[{{Yang} {et~al}\mbox{.}(2017){Yang}, {Malhotra}, {Rhoads}, \&
  {Wang}}]{Yang2017}
{Yang} H., {Malhotra} S., {Rhoads} J.~E., {Wang} J., 2017, \apj, 847, 38

\bibitem[{{Yang} {et~al}\mbox{.}(2023){Yang}, {Lidz}, {Smith}, {Benson}, \&
  {Li}}]{Yang2023}
{Yang} S., {Lidz} A., {Smith} A., {Benson} A., {Li} H., 2023, \mnras, 525, 5989

\bibitem[{{Zel'dovich}(1970)}]{Zeldovich1970}
{Zel'dovich} Y.~B., 1970, \aap, 5, 84

\bibitem[{{Zubko}, {Dwek} \& {Arendt}(2004){Zubko}, {Dwek}, \&
  {Arendt}}]{Zubko2004}
{Zubko} V., {Dwek} E., {Arendt} R.~G., 2004, \apjs, 152, 211

\end{thebibliography}

\appendix
\section{Numerical Methods}
\label{app:gf}
Here we provide a review of the primary numerical methods used for the {\small MEGATRON} simulations. Because the simulation methods are fully described in \cite{Katz2024_meg}, here we only highlight the key galaxy formation physics, initial conditions, and additional developments. 

\subsection{Galaxy Formation Physics}
Our simulations model gravity, radiation hydrodynamics, non-equilibrium thermochemistry, star formation, and feedback in a cosmological context. The gravitational potential is calculated on the AMR grid using a multigrid scheme \citep{guillet_SimpleMultigridScheme_2011}. Hydrodynamics is evolved using a MUSCL-Hancock scheme \citep{vannoordvanvanleerUltimateConservativeDifference1979}, and  the HLLC approximate Riemann solver \citep{toro_RiemannSolversNumerical_2009}, assuming an adiabatic index of $\gamma = 5/3$ to close the relation between gas pressure and internal energy. Multi-frequency radiation transfer (RT) is solved in eight energy bins spanning the IR to the EUV using a two-moment (M1) scheme \citep{Rosdahl2013,Levermore1984}. We adopt a reduced speed of light approximation setting $c_{\rm sim}=0.01 \, c$. Radiation transport is subcycled up to 500 times per hydrodynamic time step \citep{Commercon2014} and thermochemistry is computed during each subcycle. We model radiation pressure in the single scattering limit for UV and optical photons and multiple scatterings for IR photons \citep{Rosdahl2015}. 

Radiation and hydrodynamics are coupled to an 82 species non-equilibrium chemistry solver for primordial species, metals, and molecules. The solver follows H~{\small I}-{\small II}, He~{\small I}-{\small III}, $e^{-}$, C~{\small I}-{\small VI}, N~{\small I}-{\small VII}, O~{\small I}-{\small VIII}, Ne~{\small I}-{\small X}, Mg~{\small I}-{\small X}, Si~{\small I}-{\small XI}, S~{\small I}-{\small XI}, Fe~{\small I}-{\small XI}, H$_2$, and CO. All metal ionization states that are not followed are assumed to be in collisional ionization equilibrium. 

For each atomic species, we account for non-equilibrium ionization, recombination and charge exchange processes. Heating and cooling processes include photoheating, photoelectric heating, H$_2$ formation heating, H$_2$ excitation/dissociation heating, H$_2$ cooling, CO cooling, dust recombination cooling, dust-gas collisional processes, primordial heating/cooling, and metal line cooling. These processes are all modeled in non-equilibrium (see \citealt{Katz2022_prism, Katz2024_meg} for the reaction network). 

The reaction network and atomic data are derived from \cite{Cen1992,Hui1997,Badnell2006,Badnell2003,Aldrovandi1973,Shull1982,Arnaud1985,Arnaud1992,Kingdon1996,Stancil1999,Barragan2006,Voronov1997,Weingartner2001,Verner1996,Glover2010,Gnedin2009,Bialy2019,Glover2008,Baczynski2015,Nelson1997,Glover2012,Heays2017}. The dust-to-gas mass ratio as a function of metallicity is adopted following the empirical trends derived by \cite{RR2014} and we assume a dust composition following the BARE-GR-S model of \cite{Zubko2004}. Depletion of elements onto dust follows \cite{Dopita2000}, and we adopt the solar abundance patterns of \cite{Grevesse2010} when converting to solar metallicity units. Heating and cooling rates are adopted from \cite{Rosdahl2013,Bakes1994,Wolfire1995,Wolfire2003,Sternberg1989,Rollig2006,Black1977,Draine1996,Burton1990,Cen1992,Hui1997,Osterbrock2006,Haiman1996,Black1981,McKee1982,Hollenbach1989,Koyama2000,Draine2011,Bialy2019,Ferland2017}.

Star formation is modeled following a method that considers local turbulent properties of the gas \citep{Padoan2011,Federrath2012,Kimm2017,Rosdahl2018}. Gas cells can form stars if (1) their gas density is $>10~{\rm cm^{-3}}$ and $>200\times$ the mean background, (2) the local turbulent Jeans length is unresolved\footnote{In most simulations, we compare the turbulent Jeans length to $\Delta x$ but in the HN, High $\epsilon_{\rm ff}$ simulation we consider $4\Delta x$ which encourages star formation at lower densities.}, (3) the gas cell represents a local maximum in the density field, and (4) the fluid flow is locally convergent. The star formation rate is then determined using a Schmidt law \citep{Schmidt1959}, where the efficiency per free-fall time varies\footnote{In the HN, High $\epsilon_{\rm ff}$ simulation, the efficiency is always set to 100\%.} based on the local turbulent gas properties following results from high-resolution turbulent box simulations \citep{Padoan2011,Federrath2012}. 

The number of star particles formed is drawn from a Poisson distribution. When the metallicity of the gas is above the critical metallicity\footnote{There is some debate on the exact critical metallicity and it is unlikely to be a single number because physics such as redshift (via the CMB temperature), local radiation field, and dust probably all play a role \citep{Bromm2001,Schneider2002,Omukai2005,Jappsen2007}. We have adopted a very conservative value which assumes that dust cooling at low-metallicity can lead to fragmentation.} ($Z\geq2\times10^{-8}$) Pop.~II star formation occurs. We employ a minimum mass of 500~M$_{\odot}$ for the high-redshift simulations\footnote{In the `HN, High $\epsilon_{\rm ff}$' simulation, the minimum stellar particle mass is 2,000~M$_{\odot}$.} or 4,600~M$_{\odot}$ for the cosmic noon simulations\footnote{Note that the cosmic noon simulations begin with a metallicity floor and Pop.~III star formation is not modeled.}. Below the critical metallicity threshold, Pop.~III star formation is modeled with individual stars drawn from a log-normal distribution with a mean mass of 100~M$_{\odot}$ \citep{Wise2012,Kimm2017}. For all simulations, we assume a Kroupa IMF \citep{Kroupa2001} with a maximum mass of $300~{\rm M_{\odot}}$. For the variable IMF simulation, we assume that the upper mass slope varies with gas density and metallicity following \cite{Marks2012}.

Star particles interact with the gas via the radiation they emit, injecting energy and momentum in the form of feedback, and depositing mass and heavy elements into the gas. Stellar SEDs are adopted from the {\small BPASS v2.2.1} binary stellar evolution model \citep{Eldridge2017,Stanway2018} for Pop.~II stars and \cite{Schaerer2002} for Pop.~III stars\footnote{Note that \cite{Schaerer2002} does not provide the full SED for Pop.~III stars so when computing their spectra, we use the models of \cite{Larkin2023} for the same effective temperature.}. The Variable IMF simulation adopts SEDs that were computed with {\small Starburst99} \citep{Leitherer1999}. 

For Pop.~II stars, we model core-collapse SN, type Ia SN, and stellar winds closely following \cite{Agertz2021,Agertz2015,Agertz2013}, which was calibrated for Milky Way mass galaxies at $z=0$, with only minor modifications. Progenitors of core-collapse SN are stochastically sampled from the IMF following the main-sequence lifetime-mass relation from \cite{Schaerer1993}. Stars with  $8 \leq m_{\star}/{\si{\Msun}} \leq 25$ explode with an energy of $E_{\rm SN} = 10^{51}\ {\rm erg}$. For the Variable IMF and HN, High $\epsilon_{\rm ff}$ simulations, stars more massive than $25\ {\rm M_{\odot}}$ can explode as hypernovae \cite{Nomoto2006}. In this case the energy depends on mass. The fraction of high-mass stars that explode as hypernovae strongly depends on metallicity \citep{Kobayashi2006}. SNIa explode following a delay time distribution of \cite{Maoz2017}, also injecting $10^{51}\ {\rm erg}$. In contrast to SN explosions, energy and momentum injection from stellar winds of lower-mass stars is computed as an IMF average \citep[see][]{Agertz2013}.

Pop.~III star particles either explode as core-collapse SN injecting $E_{\rm SN} = 10^{51}\ {\rm erg}$ for masses in the range $10-20\ {\rm M_{\odot}}$, explode as a hypernova in the mass range $20-40\ {\rm M_{\odot}}$, directly collapse to a black hole in the mass ranges $40-140\ {\rm M_{\odot}}$ and $>300\ {\rm M_{\odot}}$, or explode as a PISN following \cite{Heger2002}.

For both Pop.~III and Pop.~II SN, we check whether the cooling radius is resolved and then either inject energy or momentum. This method is calibrated so that the terminal momentum of SN explosions expected from high-resolution simulations \citep{Kim2015} is recovered independent of resolution and local gas conditions \citep{Katz2024_meg}.

Heavy elements are injected into the ``local oct'' of the star particle for SN (based on stochastic IMF sampling) and stellar winds (IMF averages) based on \citet{Limongi2018,Ritter2018,Seitenzahl2013,Umeda2002,Nomoto2006,Nomoto2013,Heger2002}.

The appeal of such a detailed non-equilibrium thermochemistry and star formation is to naturally predict the (intrinsic) spectra of galaxies including the stellar continuum, nebular continuum, and nebular emission lines. For the stellar continuum, we simply adopt the same SED models that were used in the simulation, while emissivities for gas emission follow \cite{Ferland2017,Luridiana2015,Dere2019}. However, in certain cases, the $\mathrm{H}\,{\textsc{ii}}$ regions around star particles are numerically unresolved. This typically happens for older star particles that have low ionizing output (and thus rarely contribute meaningfully to the spectrum), or very young stars that are embedded in dense gas clouds. When this occurs, the local cell properties become a mix between the neutral and ionized phases and must be corrected. We thus replace the emission of cells with unresolved Strömgren spheres by a spherical {\small CLOUDY} \citep{Ferland2017} calculation adopting the gas properties of the host cell and SED of the star particle. We emphasize that in the spherical Strömgren sphere limit, {\small CLOUDY} and {\small RAMSES-RTZ} predict very similar emission line luminosities (to better than 10\% on average; \citealt{Katz2022_prism}).

We allow the AMR grid to refine on multiple criteria. A quasi-Lagrangian refinement strategy is employed so that a cell is refined into eight children cells when its dark matter or baryonic mass are equal to eight times that on the base grid in the initial conditions. Moreover, we ensure that the local Jeans length is resolved by at least four cells. For simulations that are part of our high-redshift suite, we allow for a constant-comoving spatial resolution such that the grid can refine up to the maximum level at any redshift. This corresponds to a resolution $1.7$~pc~$h^{-1}$ when the first Pop.~III stars form at $z\sim27$ and $5$~pc~$h^{-1}$ at $z=8.5$. For the simulations primarily focused on studying the CGM towards Cosmic Noon, we employ constant physical refinement such that additional levels in the AMR grid are released at a fixed scale factor to maintain close to $24$~pc~$h^{-1}$ resolution at all times. 

We highlight the fact that our current galaxy formation model does not consider black holes or AGN feedback. While AGN are empirically very important at high-redshift \citep[e.g.,][]{Maiolino2024b,Matthee2024b}, significant uncertainties remain on how they are seeded, how they accrete, the shape of their SEDs, and the importance of their feedback. Hence, we are currently unable to model the large class of broad emission line galaxies and peculiar objects such as Little Red Dots. Moreover, if AGN are important in low-mass galaxies, the properties of the ISM may differ due to the enhanced X-ray contribution and modes of energetic feedback. This will have an impact on the observed SEDs --- e.g.\ enhancing emission from O~{\small I}, N~{\small II}, and S~{\small II}.

\subsection{Tracer Particles}

In order to be able to track inflows and outflows of gas (and more generally any kinematics), we sample the Lagrangian history of the gas using the tracer particle implementation of \cite{cadiouAccurateTracerParticles2019}.

Briefly, the passive tracers rely on a Monte-Carlo approach to sample mass fluxes between gas cells as well as from any gas cell onto stars (including Pop.~II and Pop.~III star formation) and from stars back into gas cells (including AGB winds and SNe). Mass fluxes between cells are sampled by moving particles across the cell boundaries with a probability $\Delta m_\mathrm{cell} / m_\mathrm{cell}$, where $m_\mathrm{cell}$ is the old cell mass and $\Delta m_\mathrm{cell}$ is the mass change of the cell as computed by the hydro solver. We also follow star formation in a similar fashion by attaching gas tracers onto newly-formed stars with probability $m_\star/m_\mathrm{cell}$, where $m_\star$ is the mass of the star. Star tracers remain attached to their host star unless they are yielded back to the nearest gas cell in a wind or supernova event. This happens with probability $\Delta m_\star/m_\star$, where $\Delta m_\star$ is the mass lost by the star particle through either wind or supernova ejecta.

We initialize tracers in the zoomed region such that each initial cell contains $m_\mathrm{cell}/m_\mathrm{tracer}$ tracers (on average). Our tracer mass is $m_\mathrm{tracer} = \SI{7.3e4}{\rm M_{\odot}}$ ($150\times$ the stellar particle mass for most high-redshift simulations) for a total of 14,607,238 tracers. Tracer particles record the hydrodynamic properties of each gas cell. We also keep track of the number of times a tracer has been processed through a star and yielded back into the gas phase (either through winds or SN explosion).

\subsection{Halo Finding}
Haloes are extracted from the simulations using the {\small ROCKSTAR} 6D, phase-space, halo finder \citep{Behroozi2013}. We only consider haloes with at least 300 high-resolution DM particles in the friends-of-friends group (before unbinding) and set the virial radius to be that where the enclosed average density is 200 times the mean background. The halo finder is run only on dark matter particles and when computing stellar and gas properties, we consider only those particles and cells that reside within 25\% of the virial radius. This value was determined to maximize the total fraction of star particles assigned to haloes and minimize the overlap between main haloes and subhaloes. We empirically find there is $\lesssim2\%$ overlap in terms of star particles assigned to more than one halo while nearly 95\% of star particles have a host\footnote{Part of the reason 100\% of star particles are not assigned to haloes is because some star particles reside at larger distances than 25\% of the virial radius while others form in haloes with fewer than 300 dark matter particles.}. When computing spectra, we only consider haloes with at least one star particle, and those that are resolved by $>$1,000 dark matter particles and $>$1,000 gas cells, as this constitutes our ``well-resolved'' galaxy sample.

\subsection{Dust Radiation Transport}
Although in this work we focus primary on the intrinsic properties of galaxies (i.e.\ those prior to dust attenuation), absorption and scattering by dust can have significant impacts on the observable properties of high-redshift galaxies. For this reason, we have developed a custom interface\footnote{\url{https://git-cral.univ-lyon1.fr/rascas/rascas/-/tree/megatron?ref_type=heads}} to the public Monte Carlo radiation transport code {\small RASCAS} \citep{Leo2020}. We adopt the same dust model that is used on-the-fly in the simulations that combines the empirical results for the dust-to-gas mass ratio as a function of metallicity \citep{RR2014} with the BARE-GR-S dust composition from \cite{Zubko2004}. An example image combining multiple JWST filters into a single RGB image is shown in Figure~\ref{fig:hero2}. In addition to imaging, {\small RASCAS} also allows
us to compute the dust-attenuated spectrum and a mock IFU along any sight line as also shown in Figure~\ref{fig:hero2} for the massive $z=4$ galaxy from the Cosmic Noon suite.

\end{document}